\documentclass[useAMS,a4paper,usenatbib]{mnras}

\usepackage{newtxtext}

\usepackage[T1]{fontenc}
\usepackage{ae,aecompl}

\usepackage{graphicx}	% Including figure files
\usepackage{float}
\usepackage{caption}
\usepackage{amsmath}	% Advanced maths commands
\usepackage{amssymb}	% Extra maths symbols
\usepackage{listings}
\usepackage{color}
\usepackage{booktabs}
\usepackage{natbib}

\title[Formation of Planetary Populations I]{Formation of Planetary Populations I: Metallicity \& Envelope Opacity Effects}

\author[M. Alessi \& R. Pudritz]{
Matthew Alessi$^{1}$\thanks{E-mail:
alessimj@mcmaster.ca (MA); pudritz@mcmaster.ca (REP)} and
Ralph E. Pudritz$^{1,2}$\footnotemark[1]\\
$^{1}$Department of Physics and Astronomy, McMaster University, Hamilton, ON L8S 4M1, Canada\\
$^{2}$Origins Institute, McMaster University, Hamilton, ON L8S 4M1, Canada\\
}

\date{Accepted XXX. Received YYY; in original form ZZZ}

\pubyear{2017}

\begin{document}
\label{firstpage}
\pagerange{\pageref{firstpage}--\pageref{lastpage}}
\maketitle

% Abstract of the paper
\begin{abstract}
%single paragraph not more than 250 words (200 words for Letters).
%No references should appear in the abstract.
%Emphasis on more physical treatment of planet accretion, reduced parameters.  Then shift to key results - pick the top 3.  What problem does this paper solve?  

We present a comprehensive body of simulations of the formation of exoplanetary populations that incorporate the role of planet traps in slowing planetary migration. The traps we include in our model are the water ice line, the disk heat transition, and the dead zone outer edge. We reduce our model parameter set to two physical parameters: the opacity of the accreting planetary atmospheres ($\kappa_{\rm{env}}$) and a measure of the efficiency of planetary accretion after gap opening ($f_{\rm{max}}$). We perform planet population synthesis calculations based on the initial observed distributions of host star and disk properties - their disk masses, lifetimes, and stellar metallicities. We find the frequency of giant planet formation scales with disk metallicity, in agreement with the observed Jovian planet frequency-metallicity relation. We consider both X-ray and cosmic ray disk ionization models, whose differing ionization rates lead to different dead zone trap locations. In both cases, Jovian planets form in our model out to 2-3 AU, with a distribution at smaller radii dependent on the disk ionization source and the setting of envelope opacity. We find that low values of $\kappa_{\rm{env}}$ (0.001-0.002 cm$^2$ g$^{-1}$) and X-ray disk ionization are necessary to obtain a separation between hot Jupiters near 0.1 AU, and warm Jupiters outside 0.6 AU, a feature present in the data. Our model also produces a large number of super Earths, but the majority are outside of 2 AU. As our model assumes a constant dust to gas ratio, we suggest that radial dust evolution must be taken into account to reproduce the observed super Earth population. %We conclude, therefore, that X-rays are the dominant source of disk ionization, and that low envelope opacities are necessary to obtain a separation of hot and warm Jupiter populations.

\end{abstract}

% Select between one and six entries from the list of approved keywords.
% Don't make up new ones.

%%%%%%%%%%%%%%%%%%%%%%%%%%%%%%%%%%%%%%%%%%%%%%%%%%

%%%%%%%%%%%%%%%%% BODY OF PAPER %%%%%%%%%%%%%%%%%%

\section{Introduction}

%Current outlook on observations. Trends in data and what they suggest. Link to figure of current distribution from exoplanets.org

With the ever-growing sample of nearly 3000 confirmed exoplanets and over 2300 unconfirmed planetary candidates, we are gaining a statistical understanding of the outcomes of planet formation \citep{Borucki2011, Mayor2011, Cassan2012, Batalha2013, Burke2014, Fischer2014, Rowe2014, Morton2016}. The mass semi-major axis distribution contains a tremendous amount of information revealed by observations that can strongly constrain planet formation theories. As was first suggested by \citet{ChiangLaughlin2013}, and also discussed in \citet{HP13, HP14}, \& \citet{Hasegawa2016}, the diagram can be divided into zones that define sub-populations that appear within the distribution. The frequencies by which planets populate different zones offers strong constraints and insight into the process of planet formation. 

\begin{figure}
\centering
\includegraphics[width = 3.5 in]{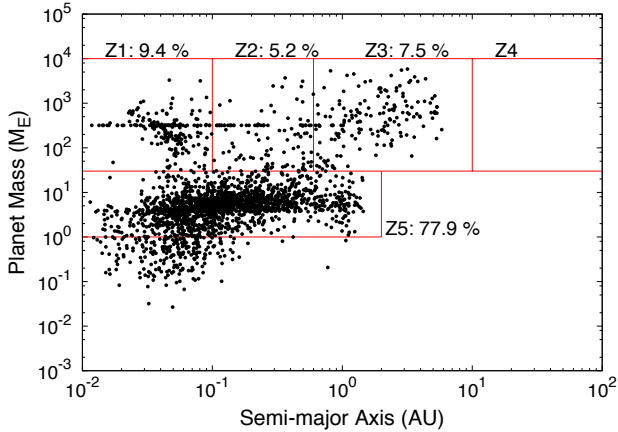}
\caption{The observed planetary mass semi-major axis distribution is shown for G-type host stars, including both confirmed planets and unconfirmed \emph{Kepler} candidates. The diagram can be divided into five zones defining populations seen in the data (first suggested by \citet{ChiangLaughlin2013}). Zone 1 contains hot Jupiters. Zone 2 has warm Jupiters. Zone 3 contains the largest population of gas giants orbiting near or outside 1 AU. Zone 4 contains distant planets. Lastly, zone 5 contains the largest population of planets, super Earths and Neptune-mass planets. This data was retrieved from http://exoplanets.org on October 16, 2017 \citep{Han2014}.}
\label{ObservedDistribution}
\end{figure}

We summarize the observations in figure \ref{ObservedDistribution}, where we show the most current m-a diagram for exoplanetary data. The vast majority of observed planets are comprised of super Earths (masses 1 - 10 M$_\oplus$) and hot Neptunes (masses 10 - 30 M$_\oplus$) orbiting within 2 AU of their host stars, which lie in zone 5 on the distribution. In the Jovian zones, the hot Jupiters (zone 1 planets), whose semi-major axes are within 0.1 AU of their host stars, comprise nearly an equal fraction of observed planets as Jupiters orbiting near or outside 1 AU (zone 3 planets). Conversely, at orbital radii between 0.6-1 AU (zone 2), there is a reduction in the number of Jovian planets compared to the adjacent zones. Lastly, zone 4 contains distant planets, a region of the diagram that is observationally incomplete \citep{Cumming2008, Bryan2016}. 

In this paper, we use the features and trends in the observed mass semi-major axis distribution to constrain the core accretion model of planet formation. The core accretion model predicts Jovian planets to form in a bottom-up manner, starting with planetary cores that are a fraction of an Earth mass accreting solids before becoming massive enough to directly accrete gas from the surrounding protoplanetary disk. This model has been shown to successfully form Jupiter-mass planets in disks with average to long lifetimes of $\gtrsim 3$ Myr \citep{Alibert2005, Lissauer2009, HP12}. 

As was shown in \citet*{IdaLin2008, Mordasini2009, HP11, HP13, Alessi2017}, the natal disk's mass and lifetime are two key parameters that strongly affect the outcome of core accretion calculations. For example, super Earths or hot Neptunes have been shown to arise in cases where the planet's formation timescale greatly exceeds the disk's lifetime, so that it does not have sufficient time to build a massive core that can quickly accrete gas \citep{Alessi2017}. This situation is encountered in low-mass disks whose lower densities give rise to long planet formation timescales, or in disks with short lifetimes. 

%HP14, Mordasini 2014, Metallicities; why this is an important inclusion.
Disk metallicity has also been shown to greatly affect outcomes of core accretion models \citep{IdaLin2004b, Mordasini2012c, HP14}. Disk metallicity affects the global disk dust to gas ratio, and therefore the solid accretion timescale throughout planet formation. This is particularly important for setting the timescale for early stages of core accretion prior to gas accretion. Stellar metallicities range from -0.6 $\leq$ [Fe/H] $\leq$ 0.6 among the sample of G-type stars hosting observed planets \citep{Han2014}. The metallicities of the gaseous disks reflect those of their parent stars, which formed out of this material. The observed planet-metallicity relation shows that the detected gas giant frequency scales with host star metallicity \citep{Fischer2005, Valenti2008, Wang2015}, highlighting the importance of disk metallicity in the framework of planet formation.

The technique of planet population synthesis is a useful method for calculating the outcomes of a core accretion model while considering a range of several input and model parameters. In particular, one can consider the observed distributions of disk lifetimes, masses, and metallicities as priors in a population synthesis calculation and determine the corresponding statistical distribution of planet properties. Planet population synthesis has been used in this manner in many previous works, considering either planetesimal accretion (e.g. \citet{IdaLin2004, IdaLin2008, Mordasini2009, Mordasini2009b} \& \citet{HP13}) or pebble accretion (e.g. \citet*{Bitsch2015, AliDib2017, Ndugu2018}). We will take a similar approach here to account for the observationally constrained ranges of disk parameters on the outcomes of our core accretion model that considers planetesimal accretion.

%Through HP13, HP14, and Hasegawa 2016 discuss their pop. synth with inclusion of planet traps. First models of traps. Link distribution to trapped type-I migration.

A central feature of any theory of planet formation is how to prevent the rapid loss of planetary embryos to the central star due to rapid Type I migration \citep{Alibert2004, IdaLin2008, Mordasini2009}.  A robust solution to this problem is the existence of regions of zero net torque - or planet traps - that arise at various kinds of disk inhomogeneities and transitions. This could arise at the inner edge of a dead zone \citep{Masset2006}.  More generally, planet traps arise in a number of regions throughout the body of protoplanetary disks.

Planet traps have been previously considered in population synthesis models in \citet*{MP2007, HP13, HP14} and \citet{Hasegawa2016}, who found that including multiple traps originating at a range of orbital radii resulted in the formation of different classes of observed planets. The planet traps considered in this model are the water ice line, the heat transition that exists at the boundary between an inner, viscously-heated region of disks and an outer region heated via direct radiation from the host star, and lastly the dead zone's outer edge which separates turbulently active and inactive regions of disks\citep{HP11}. 

%HP14, Mordasini 2014: Accretion in gaseous envelope. Envelope opacities.

The outcomes of core accretion models are sensitive to the calculated gas accretion rates. Gas accretion onto forming planets is set by the Kelvin-Helmholtz timescale, which describes how quickly a forming planet's envelope can cool and contract \citep{Pollack1996}. In semi-analytic core accretion models, the Kelvin-Helmholtz timescale is often written in terms of two poorly constrained parameters which are physically linked to the forming planet's envelope opacity \citet*{Ikoma2000, IdaLin2004}. 

To track a forming planet's envelope opacity throughout its formation self-consistently, one would need to consider the size distribution of grains in the atmosphere as well as the compositions of those grains as they are accreted from the disk onto the planet \citep{Mordasini2014b, Ormel2014, Venturini2016}. Due to the difficulty of such a calculation, the envelope opacity of a forming planet is a somewhat poorly constrained parameter, and results of core accretion calculations are sensitive to the related Kelvin-Helmholtz parameters \citep{Ikoma2000}.

%Goal of paper: Connect observed distributions of exoplanets (mass-period relation) to observed properties of host stars and disks

The goal of this paper is to connect the statistical distribution of planets on the mass semi-major axis diagram to observed properties of host stars and protoplanetary disks. We will consider observationally constrained distributions of disk masses, metallicities, and lifetimes in population synthesis calculations. This will allow us to determine if the core accretion model, including the effects of trapped type-I migration, can reproduce features of the observed mass semi-major axis diagram. 

We also greatly improve the physical model of the accretion process onto planets by reducing earlier highly parameterized treatments to only one - the envelope opacity - which plays the central role in controlling the accretion rate onto the planetary atmospheres.  As was previously considered in \citet{Mordasini2014}, we aim to constrain envelope opacity values by including the parameter in our population synthesis calculations.

The remainder of this paper is organized as follows. In section \ref{Observations}, we summarize the observational constraints on host star and disk properties, motivating the chosen distribution functions for our population synthesis calculations. In section \ref{Model} we outline our model, describing our calculation of physical disk conditions, the time-dependent locations of planet traps, our core accretion model, and lastly the methods used in our population synthesis approach. In section \ref{Results}, we show the planet populations resulting from our calculations that consider a range of envelope opacities and disk metallicities. We discuss our results and contrast them with other models in section \ref{Discussion}. Lastly, in section \ref{Conclusion}, we summarize our key findings.

\section{Distributions of Disk Properties} \label{Observations}

In our planet population synthesis approach we connect the mass-semimajor axis distribution of planets to the range of protoplanetary disk properties. The distributions of disk lifetimes, initial masses, and metallicities are therefore external inputs to our population synthesis model as they depend on processes external to planet formation. Initial disk masses and metallicities are set during the star formation process, and disk lifetimes are linked to the ongoing process of disk photoevaporation driven by ionizing radiation fields from the host stars.

The rate at which disk photoevaporation takes place is dependent on the UV and X-ray luminosity of the host star. Disks surrounding stars with high X-ray and/or UV luminosities will have high photoevaporation rates and correspondingly short disk lifetimes \citep*{Gorti2009, Owen2011, Gorti2015}. In this sense, the disk lifetime distribution shares a physical link with the distribution of X-ray and UV luminosities for disk-hosting stars. Additionally, photoevaporation rates have been shown to depend on disk metallicity \citep{Ercolano2010, Nakatani2017} since metallicity affects the disk structure, and therefore the optical depth to photoevaporating radiation. In particular, these works have shown disk lifetimes resulting from photoevaporation to be longer when disk metallicity is increased. Therefore, the relation between disk lifetime and UV/X-ray luminosity of the host star is more complex than one-to-one. In addition to radiation from the host star, the external radiation field can also affect photoevaporative rates and disk lifetimes of a disk within a star cluster \citep{Clarke2007}.

\begin{figure}
\centering
\includegraphics[width = 3.5 in]{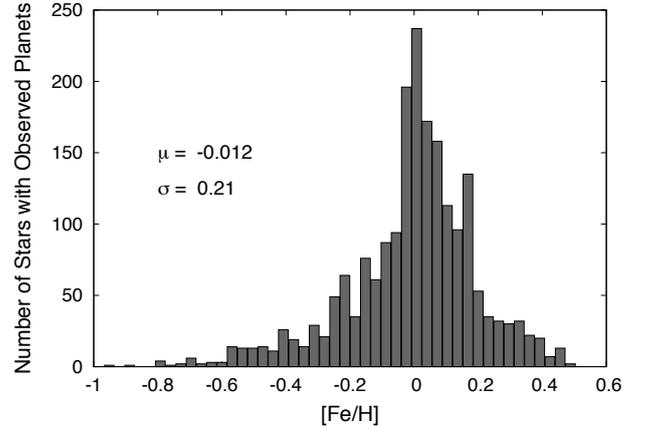}
\caption{The metallicity distribution for G-type stars hosting planets is shown. The mean and standard deviation of this distribution are used to define the parameters in our population synthesis metallicity distribution (see equation \ref{NormalDist}). This data was retrieved from http://exoplanets.org on October 16, 2017 \citep{Han2014}.}
\label{MetallicityHist}
\end{figure}

For the distributions of disk lifetimes, $t_{\rm{LT}}$, and initial masses $f_M$, we use a log-normal distribution,
\begin{equation} P(X|\mu_x,\sigma_x)  \sim \exp\left(-\frac{(\log(X)-\log(\mu_x))^2}{2\sigma_x^2}\right) \;.\label{LogNormal} \end{equation}
In the case of the disk lifetime, we use an average $\mu_{\rm{lt}} = 3$ Myr and standard deviation $\sigma_{\rm{lt}} = 0.222$, which results in 1.8 Myr - 5 Myr corresponding to the $\pm\sigma_{\rm{lt}}$ range of the distribution. For initial disk masses, we use $\mu_{\rm{m}} = 0.1$ and $\sigma_{\rm{m}} = 0.138$, whereby $f_M = 0.073 - 0.137$ are within one sigma of the mean. We note that this range of initial disk masses corresponds to a 1-$\sigma$ range of $\sim$ 0.037 - 0.065 M$_\odot$ after 1 Myr of disk evolution. We choose these parameters to correspond with lifetime and initial disk accretion rate distributions used in \citet{HP13}. Our resulting disk lifetime and initial mass distributions correspond reasonably well with those used in previous population synthesis works such as \citet{IdaLin2008} and \citet{Mordasini2009}.

In figure \ref{MetallicityHist}, we show the distribution of metallicities ([Fe/H]) among G-type stars hosting observed planets. We consider the initial disk metallicity distribution to follow that of the observed host stars, and model the observed data using a normal distribution,
\begin{equation} P(X|\mu_x,\sigma_x) \sim \exp\left(-\frac{(X-\mu_x)^2}{2\sigma_x^2}\right) \;, \label{NormalDist} \end{equation}
with mean $\mu_{\rm{z}} = -0.012$ and $\sigma_{\rm{z}} = 0.21$. 

\begin{figure*}
\centering
\includegraphics[width = .33\textwidth]{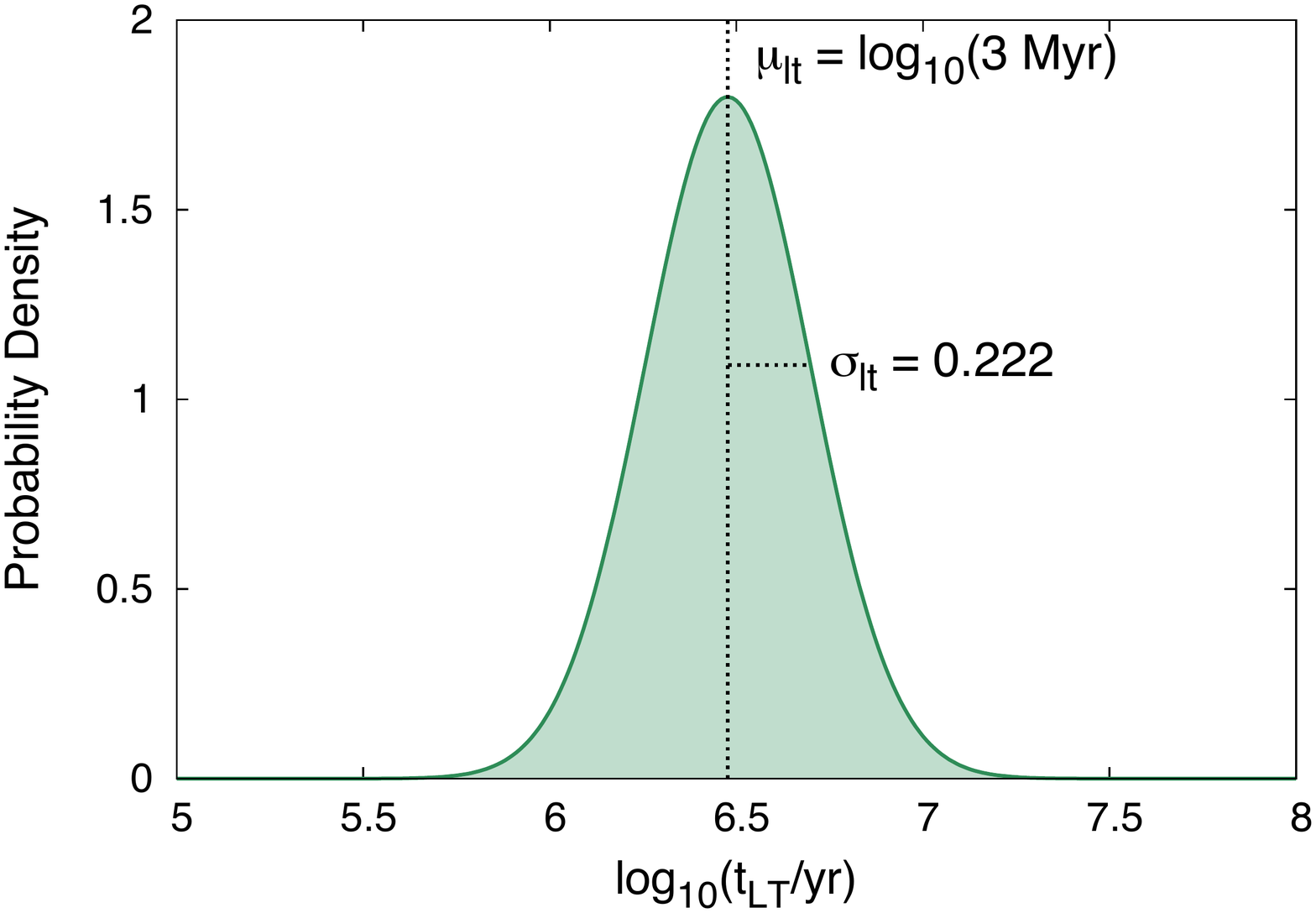} \includegraphics[width = .33\textwidth]{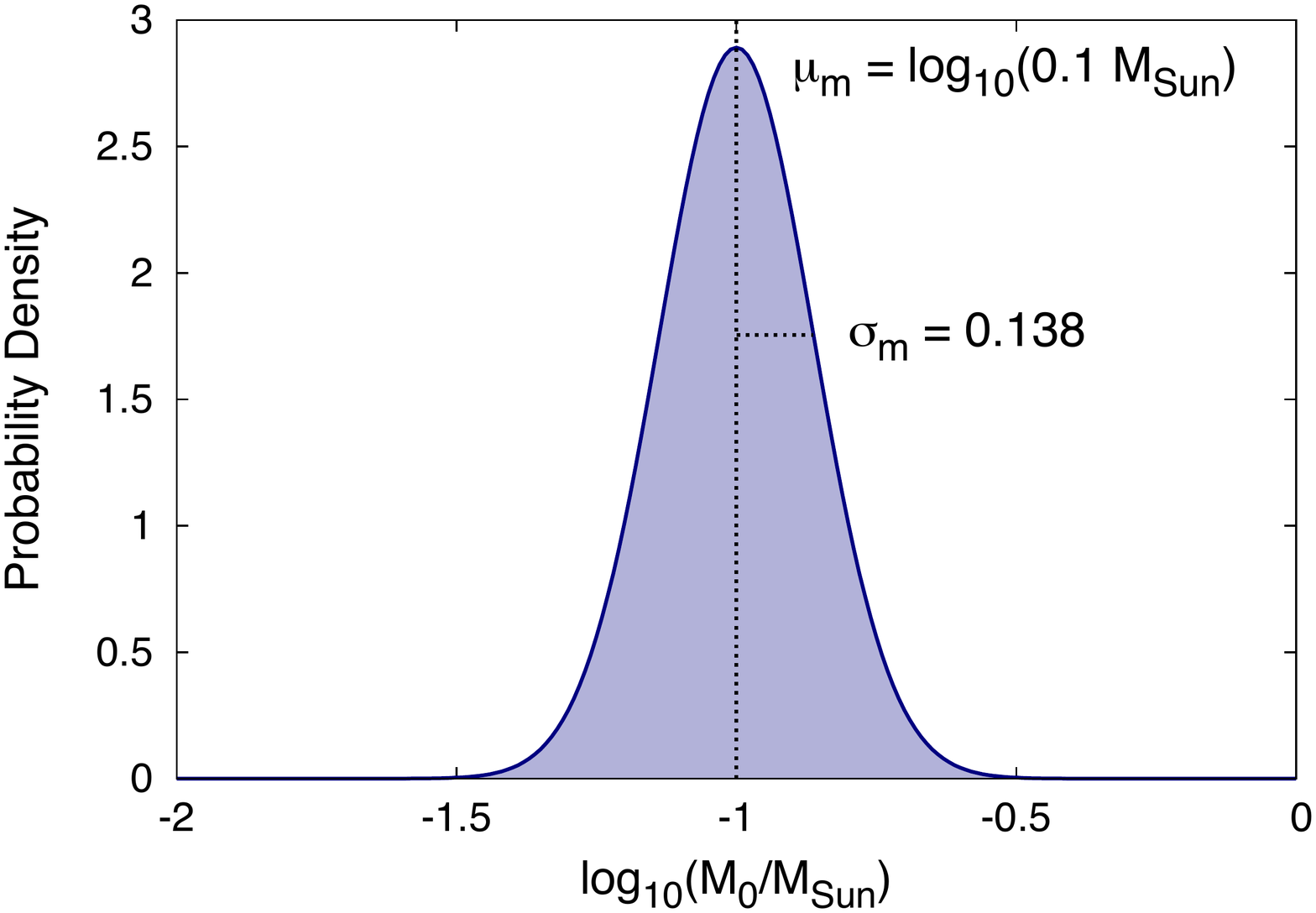}
 \includegraphics[width = .33\textwidth]{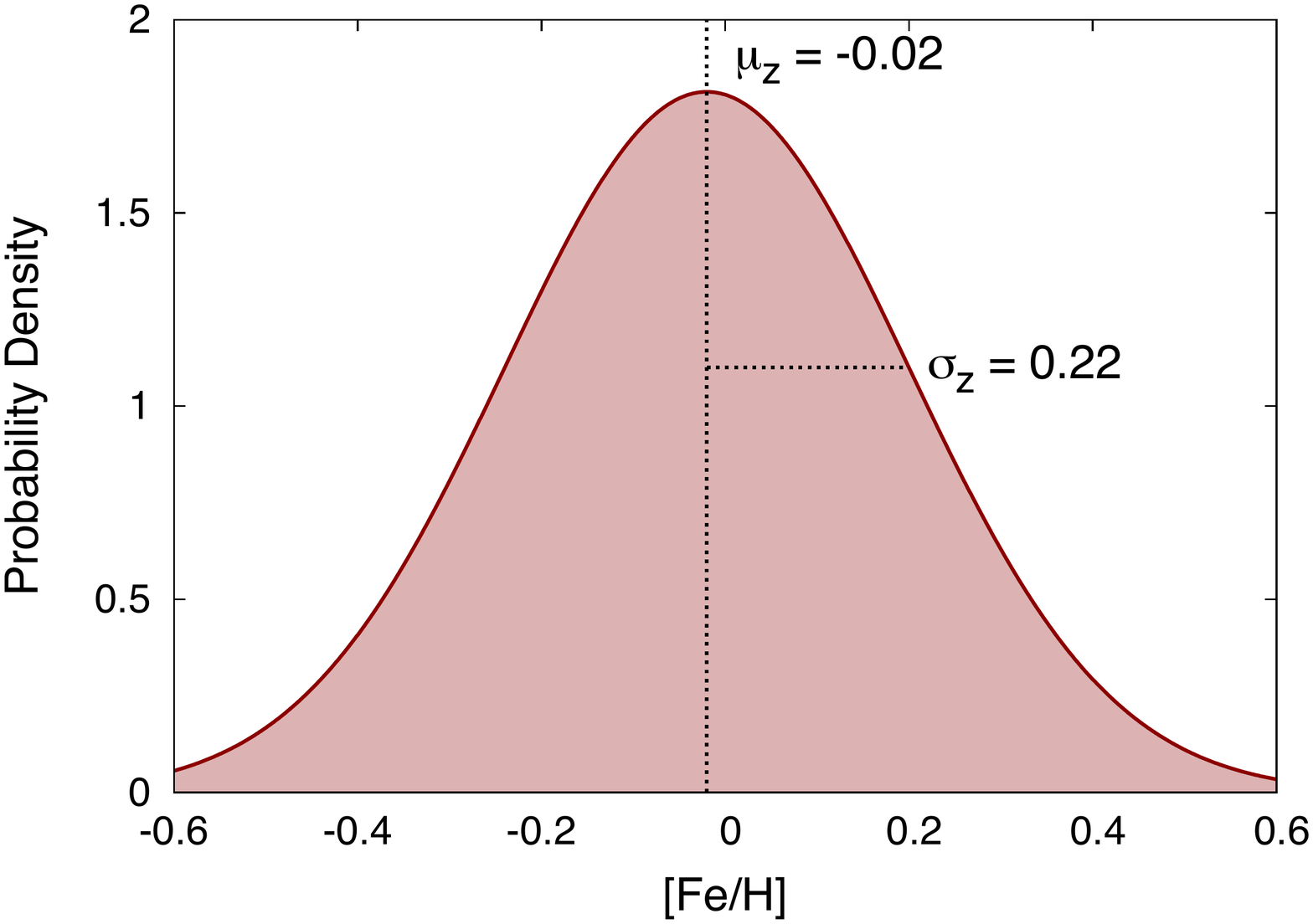}
\caption{Distributions of the four varied quantities in our population synthesis calculations are shown. We use log-normal distributions for disk lifetimes and masses, and a normal distribution for disk metallicities.}
\label{distributions}
\end{figure*}

In figure \ref{distributions}, we plot the distributions of disk parameters stochastically varied in our population synthesis models. The lifetime and initial mass distribution (left and centre panels, respectively) correspond to the distributions described in equation \ref{LogNormal} while the metallicity distribution is described by equation \ref{NormalDist}.

\section{Planet Formation Model} \label{Model}

%Reference APC17 to start and go from there. Main points of disk model, planet formation model, etc.

% main additions to model: envelope opacities (Gas accretion), metallicity (dust-to-gas ratio and chemistry), observational bias (Appendix A)

% Table of parameters and different categories. Outline ones we control vs ones set by observations. Key point of paper is to show that the observed ones have the largest impact on PF as opposed to model parameters we can tune.

Our model combines a calculation of the evolving physical conditions within a protoplanetary disk with a core accretion model of planet formation that includes a prescription for planet formation. For a complete, detailed description of the model used in this work, we refer the reader to \citet{Alessi2017}. In this section, we summarize its key features as well as our implementation of population synthesis, and note any updates to the model presented in \citet{Alessi2017}.

\subsection{Disk Evolution Model} \label{DiskModel}

To calculate disk structure and evolution, we use the 1+1D semi-analytic model presented in \citet{Chambers2009}. An analytic disk model is advantageous for our purposes as it allows for efficient calculation of disk conditions in population synthesis calculations that model the formation of thousands of planets. The \citet{Chambers2009} model provides a self-similar solution to the viscous evolution equation describing the evolving surface density profile $\Sigma(r,t)$ of a protoplanetary disk,
\begin{equation} \frac{\partial \Sigma}{\partial t} = \frac{3}{r}\frac{\partial}{\partial r}\left[r^{1/2}\frac{\partial}{\partial r}\left(r^{1/2} \nu \Sigma\right) \right]\, , \label{ViscousDisk} \end{equation}
where $\nu(r,t)$ is the disk's viscosity. Self-similar solutions to this equation can be obtained for $\alpha$-disk models, where the disk viscosity is written in terms of an effective viscosity coefficient, $\alpha$, defined as \citep{SS1973, LBP1974},
\begin{equation} \nu = \alpha c_s H\;, \end{equation}
where $c_s$ is the local sound speed and $H$ is the disk scale height.

We recall that self-similar solutions for disk evolution require a constant $\alpha$ throughout the disk and we take that value to be $\alpha=10^{-3}$.  However, there are two mechanisms for angular momentum transport throughout the disk corresponding to angular momentum transport through MRI generated turbulence and torques exerted by MHD disk winds,
\begin{equation} \alpha = \alpha_{\rm{turb}} + \alpha_{\rm{wind}}\,. \end{equation}
Models considering angular momentum transport to occur solely via MRI turbulence predict an outer, turbulently active region that has a larger $\alpha_{\rm{turb}} \sim 10^{-3}-10^{-2}$ value than an inner, turbulently inactive regions (the so-called \emph{dead zone}), with $\alpha_{\rm{turb}} \sim 10^{-5}-10^{-4}$ \citep{Gammie1996, MP2003}. However, disk winds have been shown to sustain accretion rates within dead zones \citep{BaiStone2013, Gressel2015, Gressel2015b, Bai2016}, motivating our assumption of a globally constant disk $\alpha$ in spite of a radially dependent $\alpha_{\rm{turb}}$ component.

Our model accounts for disk evolution to occur through both the generalized ``viscous'' accretion and photoevaporation, whereby UV and X-ray radiation from the protostar disperses disk material throughout its evolution \citep*{Pascucci2009, Owen2011}. The time-dependent disk accretion rate used in this work,
\begin{equation} \dot{M}(t) = \frac{\dot{M}_0}{(1 + t/\tau_{\rm{vis}})^{19/16}} \exp\left(-\frac{t - \tau_{\rm{int}}}{t_{\rm{LT}}}\right)\;, \label{ViscousAccretion}\end{equation}
includes a viscous evolution term multiplied by an exponentially-decreasing factor that models the effect of photoevaporation on our disk's viscous evolution. In equation \ref{ViscousAccretion}, $\tau_{\rm{vis}}$ is the disk's viscous timescale, $\dot{M}_0$ is the accretion rate at the initial time $\tau_{\rm{int}} = 10^5$ years, and $t_{\rm{LT}}$ is the disk's lifetime. 

We note that equation \ref{ViscousAccretion} is an improvement of the model presented in \citet{Alessi2017}, as we now set the $e$-folding timescale of disks equal to their disk lifetimes, thereby reducing our set of parameters by one. With this change all of the disks in our populations have accretion rates that are reduced by a factor of $e$ at the end of their lifetimes as driven by photoevaporation in addition to their viscous evolution that is handled by the \citet{Chambers2009} model. Following our discussion in section \ref{Observations}, disk lifetimes are linked to photoevaporation driven by UV and X-ray luminosities of their host stars, and with this change to equation \ref{ViscousAccretion} every disk considered in our model undergoes the same photoevaporative evolution albeit over different timescales set by $t_{\rm{LT}}$. 

When the disk evolution time $t = t_{\rm{LT}}$ we assume the disk rapidly clears, terminating planet formation and migration. This is motivated by photoevaporation models that show disks to rapidly clear once the photoevaporative mass-loss rate exceeds the disk accretion rate \citep{Owen2010, Haworth2016}. We emphasize that disk lifetimes are physically linked to the amount of UV and X-ray emission from the protostar, as stars with higher UV-excesses will have higher photoevaporative mass-loss rates, and correspondingly shorter disk lifetimes \citep{Owen2011}. This motivates our inclusion of $t_{\rm{LT}}$ in the photoevaporation factor in equation \ref{ViscousAccretion}.
 
The \citet{Chambers2009} disk model is divided into two regions: an inner region heated through viscous dissipation, and an outer region where radiative heating from the protostar dominates any other heating mechanism. We note that within the dead zone where disk winds dominate angular momentum transport, the effective ``viscous'' dissipation is, in reality, due to non-ideal MHD heating effects (eg. Ohmic heating at the disk midplane). The heat transition ($r_t$), a planet trap in our model, exists at the boundary between the inner region heated through viscous dissipation or non-ideal MHD heating and the outer region heated via radiation from the protostar.

\begin{figure*}
\centering
\includegraphics[width = 2.2in]{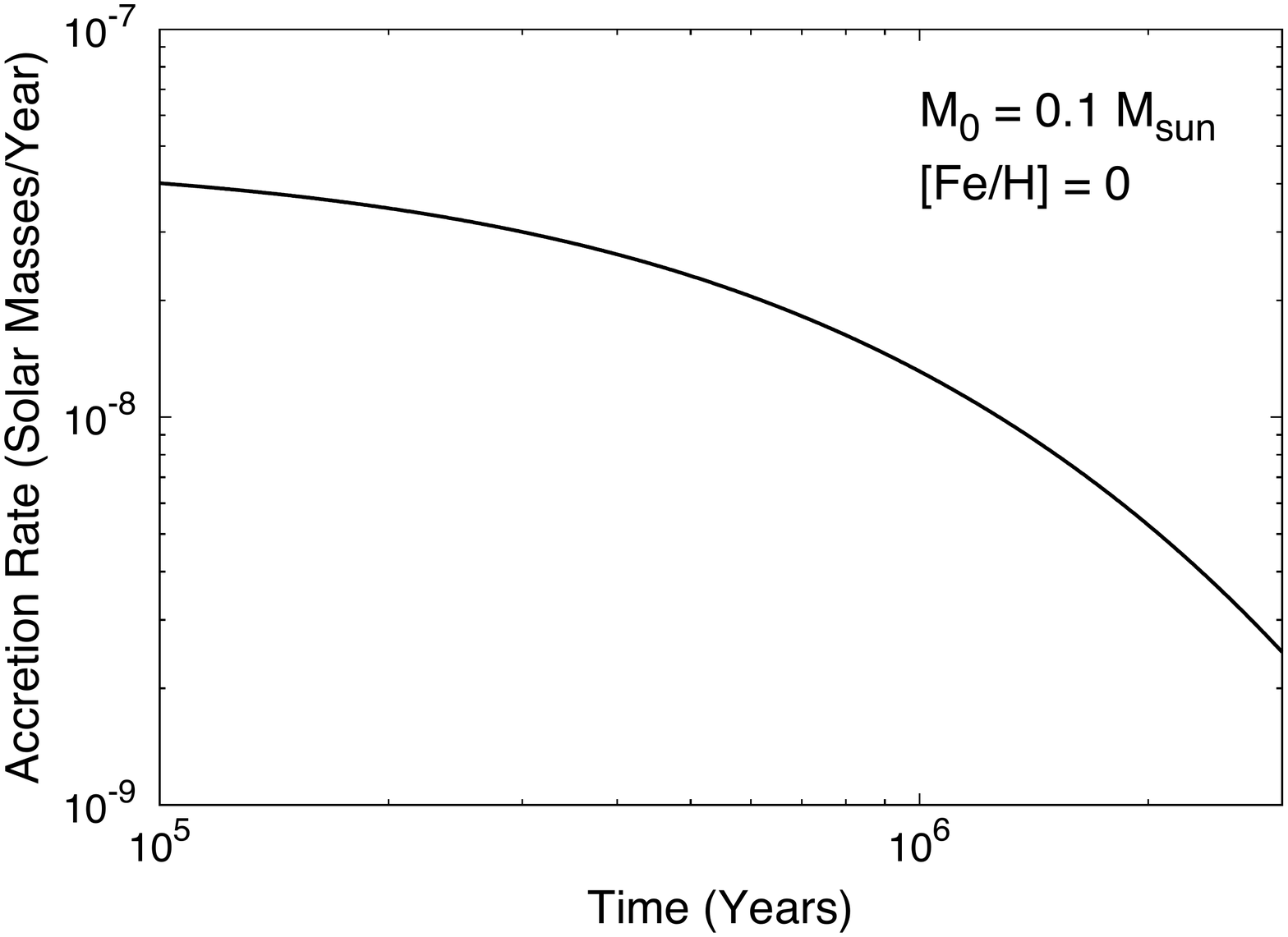} \includegraphics[width = 2.2 in]{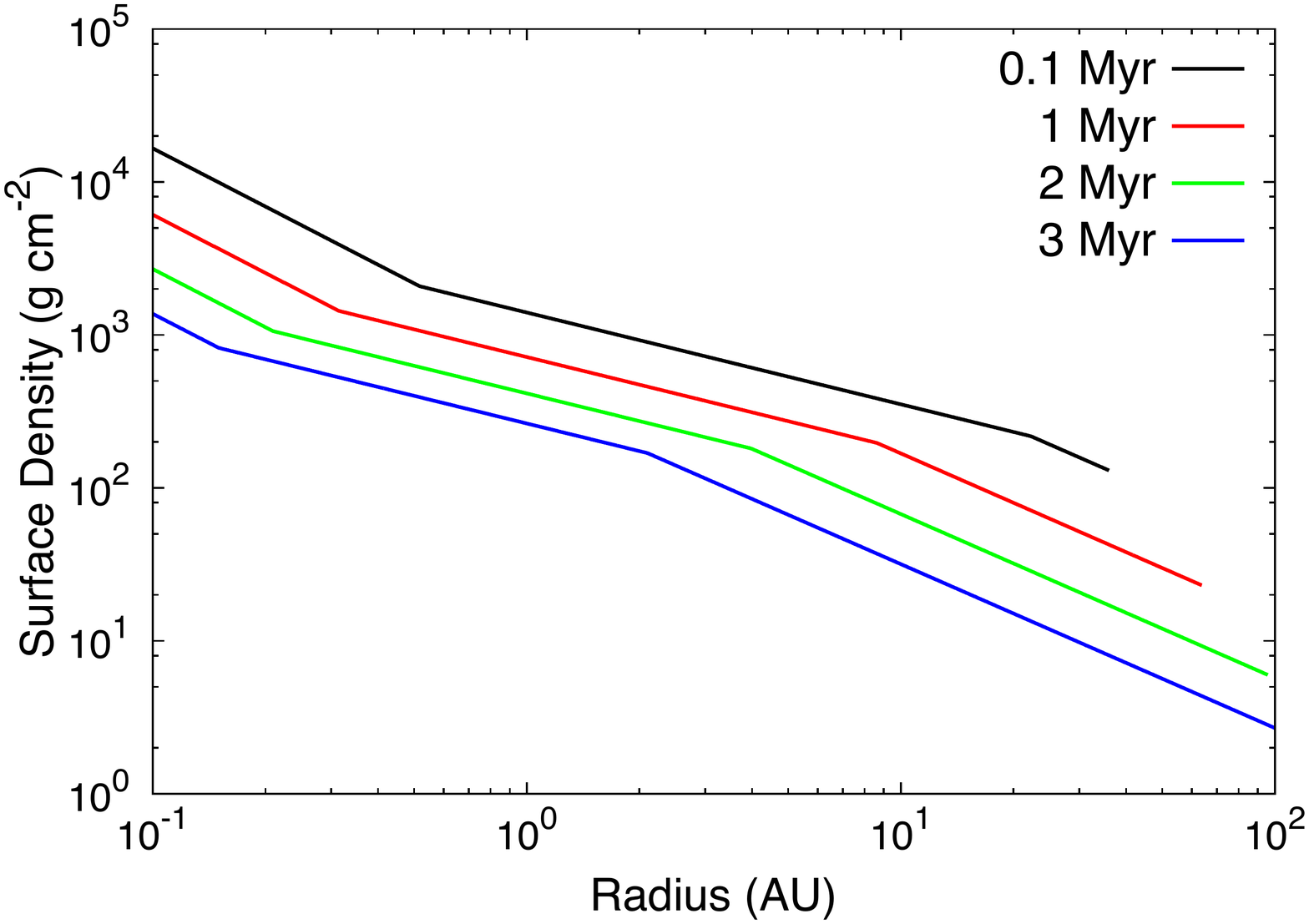} \includegraphics[width = 2.2 in]{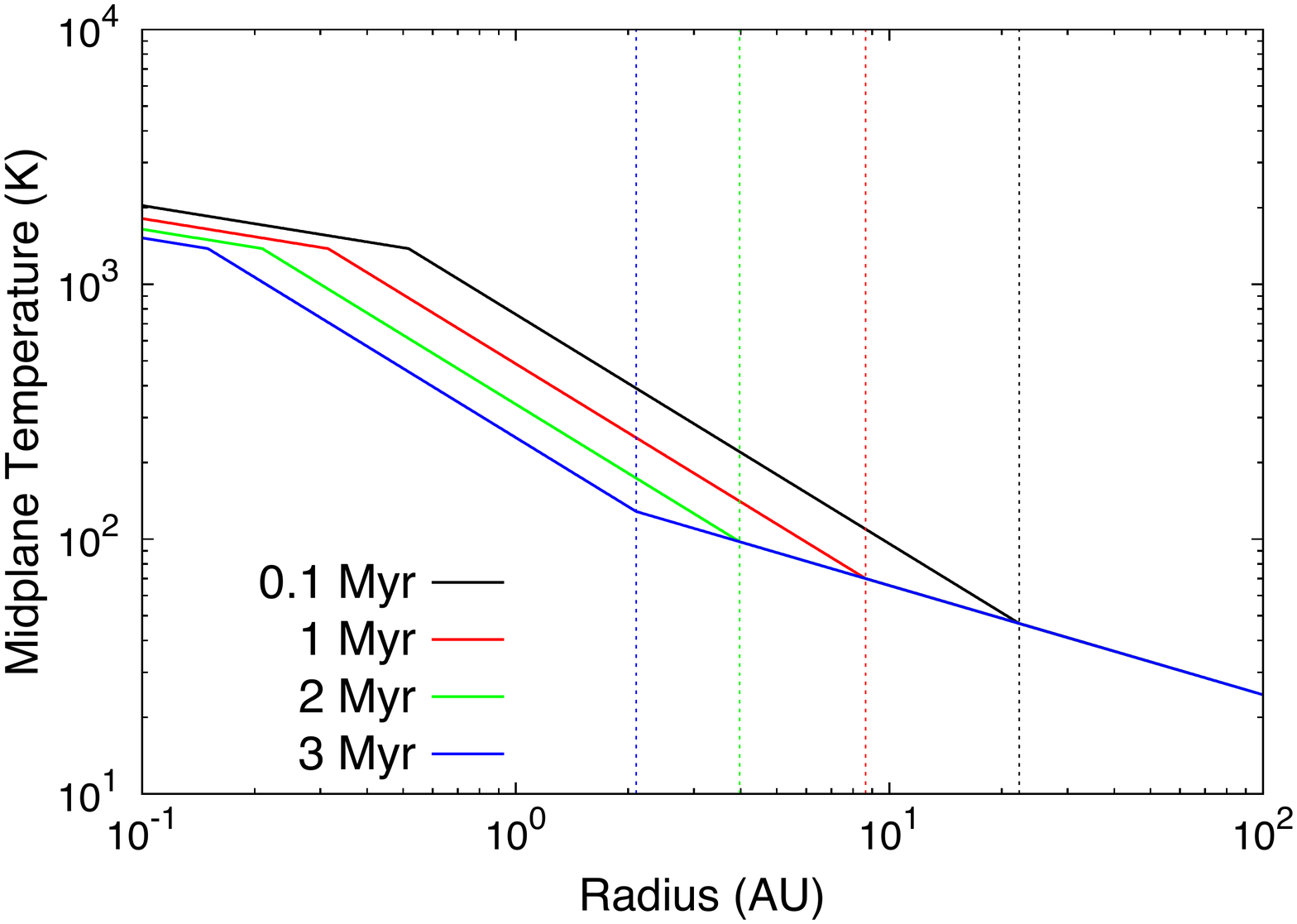}
\caption{The time evolution of the disk accretion rate, and radial profiles of surface density and midplane temperature are shown for a fiducial disk model. The vertical dashed lines in the right panel denote the location of the disk heat transition, existing at the boundary of the two different power law profiles pertaining to viscous and radiative heating. The innermost power law in the $\Sigma$ and $T$ plots within a few tenths of an AU corresponds to the region where the disk opacity is reduced to due evaporation of dust grains.}
\label{DiskPlot}
\end{figure*}

In figure \ref{DiskPlot}, we show the evolution of the disk accretion rate, as well as radial profiles of surface density and midplane temperature at several times throughout a fiducial disk's evolution. In table \ref{ChambersSigmaT}, we show the radius and accretion rate scalings of surface density and temperature profiles in both regions computed with our disk model. We refer the reader to section 2.1 of \citep{Alessi2017} for a complete description of our disk evolution model.
 
\begin{table}
\caption{Surface density ($\Sigma$) and temperature ($T$) dependencies on disk accretion rate and radius in both the viscous region ($r<r_t$) and the region heated by radiation from the host star ($r>r_t$).}
\centering
\begin{tabular} {|c|c|}
\hline
$r < r_t$ & $r > r_t$ \\
\hline
$\Sigma(r,t) \sim \dot{M}^{3/5} r^{-3/5}$ & $\Sigma(r,t) \sim \dot{M} r^{-15/14}$ \\
$T(r,t) \sim  \dot{M}^{2/5} r^{-9/10}$ & $T(r,t) \sim r^{-3/7}$\\
\hline
\end{tabular}
\label{ChambersSigmaT}
\end{table}

We calculate the dust to gas ratio throughout the disk as a function of disk metallicity [Fe/H] using,
\begin{equation} f_{\rm{dtg}} \equiv \frac{\Sigma_d}{\Sigma_g} = f_{\rm{dtg},0}10^{[\rm{Fe}/\rm{H}]} \;, \label{DustToGas} \end{equation}
where $f_{\rm{dtg},0} = 0.01$ is the fiducial dust to gas ratio for a Solar-metallicity disk, and $\Sigma_g \simeq \Sigma(r,t)$ calculated with our disk model. This simplified constant dust-to-gas ratio has no radial or time dependence, and thus does not include the effects of condensation fronts or dust evolution through radial drift, coagulation and fragmentation (which are considered, for example, in \citet*{Birnstiel2012} and \citet*{Cridland2017}). Models that include these effects have shown the dust-to-gas ratio has an abrupt increase across the ice line, and that the ionization structure of the disk is altered by including dust evolution effects, shifting the dead zone inward \citep*{Cridland2017}. By assuming a constant dust-to-gas ratio throughout the disk, we are neglecting the variations in the quantity that would arise across planet traps, and its affect on the solid accretion rates in our planet formation calculations. In a future paper (Alessi, Pudritz, \& Cridland, in prep.), we will include these effects in our population synthesis model.

Following \cite{Chambers2009}, our disk model assumes a constant opacity of $\kappa_0 = 3$ cm$^2$ g$^{-1}$ for Solar metallicity disks, except in a small ($\sim 0.1$ AU) inner region with temperatures exceeding 1380 K, where dust grains evaporate. We scale our assumed average disk opacity with metallicity as \citep{Remy2014},
\begin{equation} \kappa = \kappa_0 10^{\rm{[Fe/H]}} \;,\end{equation}
where the disk opacity is dominated by grain opacities over the disk temperature and metallicity range considered, -0.6 $\leq$ [Fe/H] $\leq$ 0.6. By assuming a constant disk opacity, we are neglecting the variation that would occur across ice lines. This transition in disk opacity is the physical cause for the ice line planet trap in our model, however this variation in $\kappa$ is not necessary for our model as we do not directly compute the planet-disk torques. Including a detailed opacity structure, as was done in \citet{Stepinski1998} would also affect the disk surface density and temperature profiles. However, this would be a small effect as the \citet{Chambers2009} disk model is only weakly sensitive to disk opacity, so variations in $\kappa$ would not greatly alter the global disk structure.

\subsection{Planet Migration \& Traps} 

The local density and composition of material in protoplanetary disks are heavily dependent on disk radius. Therefore, including the effects of planetary migration in core accretion scenarios is crucial to properly track a planet's time-dependent accretion rate and composition (e.g. \cite*{Cridland2016}). Migration of forming planets occurs due to the gravitational interaction, and resulting angular momentum exchange, between a forming planet and the surrounding disk material. 

\subsubsection{Trapped Type-I Migration}

The theory of type-I migration applies to low-mass ($\lesssim 10$ M$_\oplus$) forming planets that have only a small influence on surrounding disk material. To determine type-I migration rates, one must account for the Lindblad and corotation torques exerted on the planet, which depend on the planet and local disk conditions \citep{Lyra2010, HellaryNelson2012, Dittkrist2014, Baillie2016, Coleman2016b}. In the absence of a corotation torque, Lindblad torques exerted on a forming planet lead to rapid inward migration timescales of $\sim 10^5$ years \citep{GoldreichTremaine1980}. The discrepancy between this inward Lindblad migration timescale and the $\gtrsim 10^6$ year core accretion timescales (i.e. \citet{Pollack1996}) is known as the \emph{type-I migration problem}. 

The corotation torque, which often increases the planet's angular momentum, is a means for slowing the rapid inward migration caused by Lindblad torques and offers a solution to the type-I migration problem. The caveat with this mechanism is that the corotation torque's operation depends sensitively on the local disk conditions. In many cases, the corotation torque will saturate, exerting no positive torque on the planet which will then only be subjected to the strong negative Lindblad torques \citep{Masset2001, Masset2002}. The corotation torque has been shown to be sustained in disks when there is sufficient thermal and viscous diffusion within the corotation region \citep*{Paardekooper2011}.

Near planet traps, or inhomogeneities in disk surface density and temperature profiles, the corotation torque has been shown to remain unsaturated \citep{Masset2006, HP11}. Planet traps are zero torque equilibrium radii resulting from the summed contributions from the negative Lindblad torque and strong corotation torque operating locally near the planet trap. Moreover, planet traps are stable equilibria as planets on nearby orbits migrate into the trap due to the planet-disk torque, as has been demonstrated in numerical simulations \citep{Lyra2010, Coleman2016b}. In trapped type-I migration, the planets thus form while migrating inwards following their host trap's evolution with the disk \citep{Paardekooper2010}. The planet's migration timescale is therefore comparable to the disk's viscous evolution timescale of a few Myr, which is comparable to the core accretion timescale \citep{HP12}.

Other authors, such as \citet{HellaryNelson2012} and \citet{Dittkrist2014} have included multiple type-I migration regimes during early stages of planet formation. In this work, we only consider the trapped type-I migration regime. It remains possible, however, for corotation torques to saturate prior to the planet entering the type-II migration regime (see equation \ref{GapOpeningMass}). As discussed in \citet{Hasegawa2016}, the mass at which the corotation torque saturates ($M_{\rm{sat}}$) is similar to the gap-opening mass ($M_{\rm{gap}}$) where type-II migration begins. Moreover, we have shown in \citet{Alessi2017} that planets enter type-II migration prior to corotation torque saturation for a wide range of model parameters. We therefore include trapped type-I migration and type-II migration as the only migration regimes in our model. 

The planet traps we include are the water ice line, the heat transition, and the outer edge of the dead zone. Determining the locations of the traps within disks is a key component of our model, as we assume planets to form within traps for the entirety of their type-I migration mass-regime, and do not directly calculate the planet-disk torques. There are additional inhomogeneities and transitions that can lead to trapping present in disks that we do not include. One example is the dead zone inner edge, which is located at $\sim$ 0.1 AU \citep{Gammie1996}. Since this trap is restricted to the inner regions of the disk, it is unclear if such high temperatures ($\gtrsim$ 1000 K) will allow forming planets to accrete substantial amounts of gas.

There are also volatile ice lines in addition to water whose ability to act as traps has not yet been established and are not included here. An example of a prominent volatile in disks with an observed ice line is carbon monoxide \citep{Qi2013}. The CO condensation front will lie in the outer portion of the disk (roughly 30 AU), within the radiatively-heated region where the temperature is determined by the flux from the star, and not heat generated by the disk. Therefore, the temperature in this region has no time (or $\dot{M}$) dependence (see table \ref{ChambersSigmaT}). Thus, the trap will not evolve inward if included in our model, and planet formation at this condensation front would be confined to large orbital radii where the planetesimal accretion rate would be exceedingly low. 

In contrast to planetesimal accretion (considered in this paper's model), pebble accretion models predict higher solid accretion rates in the outer disk at the location of the CO ice line, resulting in shorter planet formation timescales within a typical disk lifetime \citep{AliDib2017}. However, recent works have shown that planet cores can only form up to $\lesssim$ 1 M$_\oplus$ by direct pebble accretion due to ablation in the forming cores' envelopes \citep{Alibert2017, Brouwers2018}.

Given this theoretical uncertainty, our model only considers planetesimal accretion which predicts low accretion rates in the outer disk. We therefore do not include the CO ice line as a trap in our model, as it will not contribute to the planet populations seen in the data, but rather would only produce low mass planets at large orbital radii.

While the CO$_2$ ice line lies in the disk's viscous region, and therefore will evolve inwards as the disk evolves, it is not included as a trap in this model as it is a factor of $\sim 10^6$ times less abundant than water \citep{Cridland2016}, and would not cause a sufficient opacity transition to dynamically trap planets (Cridland, Pudritz, \& Alessi, submitted). We rather follow \citet{HP12,HP13}, who showed that the three traps we do include in our model are sufficient to reproduce observed classes of planets. We note that the traps we do include are the general type of possibilities for traps within the body of the disk.

\begin{figure*}
\centering
\includegraphics[width = 2.2in]{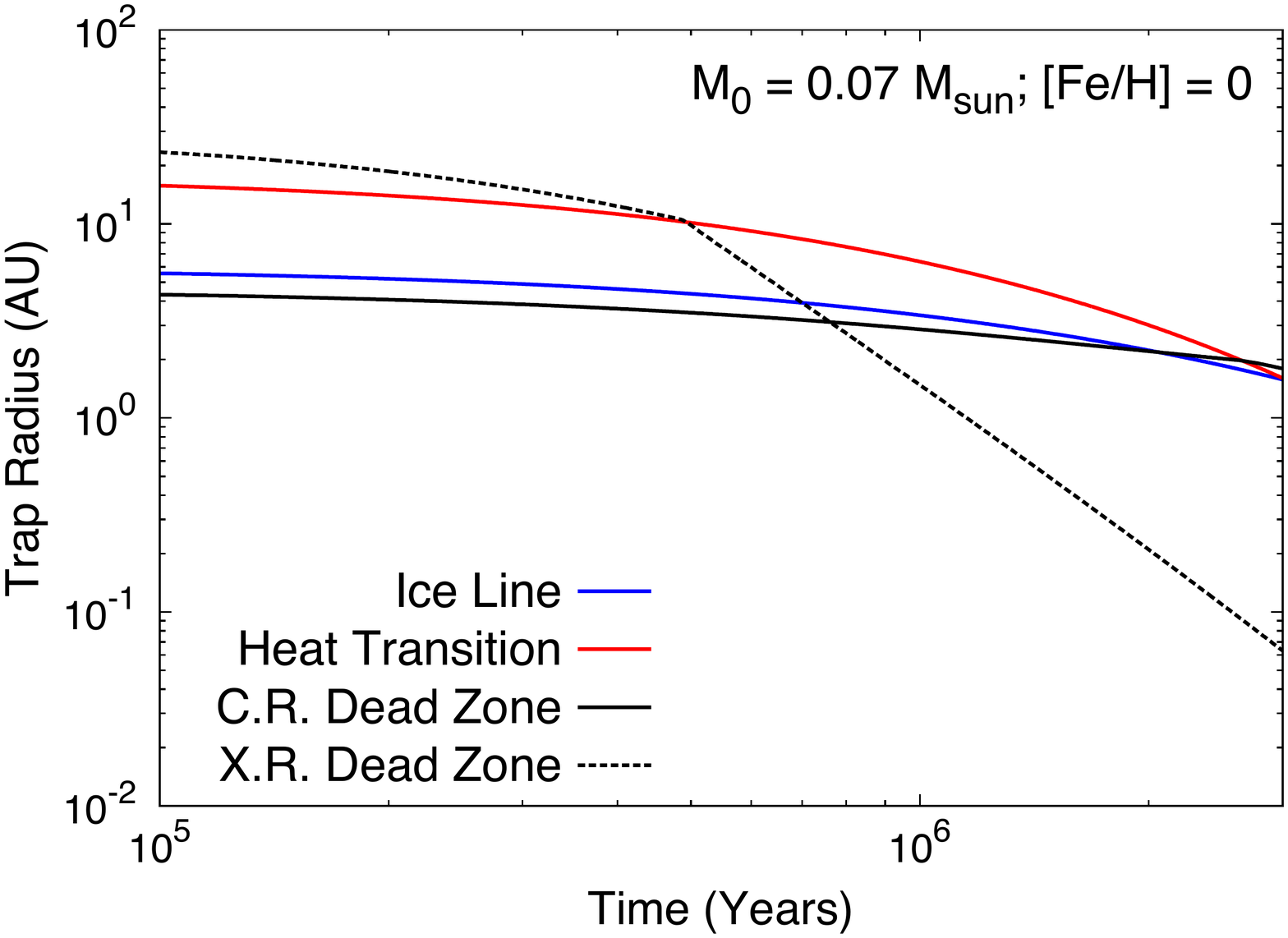} \includegraphics[width = 2.2 in]{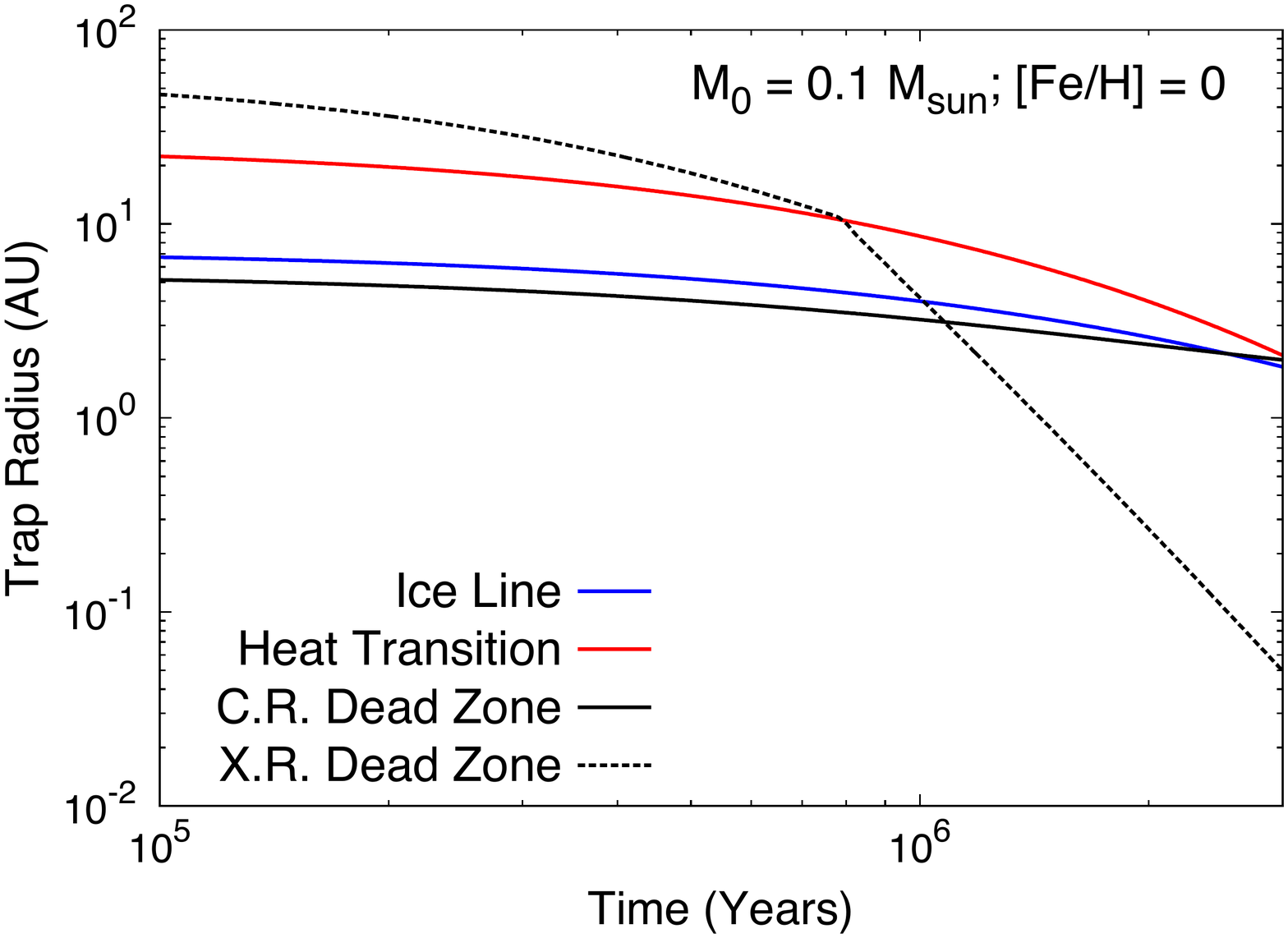} \includegraphics[width = 2.2 in]{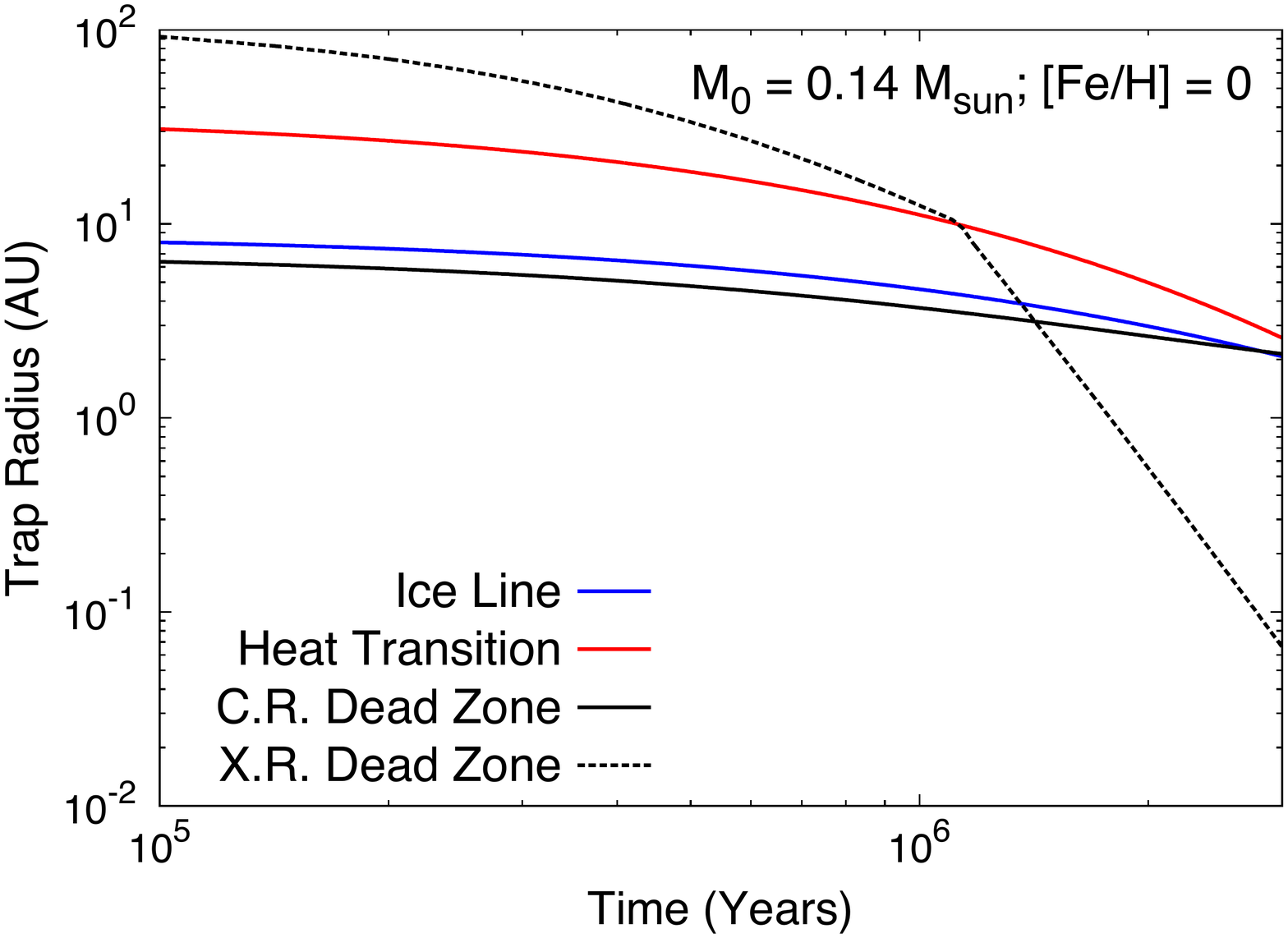}
\caption{Time evolution of planet trap locations is shown for disks of initial masses 0.07 M$_\oplus$, 0.1 M$_\oplus$ (a fiducial disk mass), and 0.14 M$_\oplus$. The fiducial disk mass corresponds to the  average value $\mu_m$ from the disk mass distribution used in our population synthesis model, while the light and heavy disk masses correspond to  $\pm 1\sigma_m$ variation from the mean.}
\label{Traps}
\end{figure*} 

%Ice Line
The location of the water ice line along the disk's midplane, $r_{il}$, is determined using an equilibrium chemistry calculation over the corresponding range of temperatures and pressures of the disk midplane throughout its evolution. We use the equilibrium chemistry software ChemApp to perform these calculations (distributed by GTT Technologies; http://www.gtt-technologies.de/newsletter) over the full range of metallicities considered in this work. We find that the location of the water ice line scales with the disk accretion rate as $r_{il} \sim \dot{M}^{4/9}$. This is the same scaling as was found in \citet{HP11} who tracked the position of the water condensation temperature, 170 K, through the disk's evolution.

%heat transition
The heat transition represents the midplane boundary between the region of the disk heated through viscous dissipation and the outer region heated via irradiation from the central star. Its location is defined in the \citet{Chambers2009} model at the disk radius where the midplane temperature due to viscous heating is equal to the temperature caused by radiation from the host star (i.e. equating the two temperature power laws whose scalings are shown in table \ref{ChambersSigmaT}). Its location scales with disk accretion rate as $r_t \sim \dot{M}^{28/33}$.

%dead zone

Within the dead zone, the disk ionization fraction will be too low for the MRI to drive turbulence. The outer edge of the dead zone, $r_{dz}$, thus separates an inner, turbulently inactive region from an outer, turbulent region. This location has been shown to be a planet trap due to the abrupt increase in scale height of the dust, whose radiation creates a thermal barrier to planet migration \citep{HP10}. In \citet{Alessi2017} (section 2.3.3) we describe our calculation of $r_{dz}$ in detail, which we summarize below.

To determine whether MRI-turbulence will be generated at a particular disk radius, we equate the MRI growth timescale to the Ohmic diffusion damping timescale over all vertical scales in the disk, the largest being the disk scale height \citet{Gammie1996}. This results in a condition for the MRI to be inactive, and for the disk radius in question to be within the disk dead zone, written in terms of the magnetic Elsasser number \citep{Blaes1994, Simon2013},
\begin{equation} \Lambda_0 = \frac{V_A^2}{\eta\Omega_K} \lesssim 1 \;,\label{MagneticReynolds} \end{equation}
where $V_A \simeq \alpha_{\rm{turb}}^{1/2}c_{\rm{s}}$ is the Alfv\'en speed and,
\begin{equation} \eta = \frac{234}{x_e}T^{1/2}\,\rm{cm}^2 \,\rm{s}^{-1}\;,\end{equation}
is the magnetic diffusivity, which depends on the electron fraction $x_e$.

The equilibrium electron fraction at a particular disk radius results from a balance between ionization and recombination rates. We calculate this electron fraction to be the solution of the following equation \citep{Oppenheimer1974},
\begin{equation} x_{\rm{e}}^3 + \frac{\beta_{\rm{t}}}{\beta_{\rm{d}}} x_{\rm{M}}x_{\rm{e}} - \frac{\zeta}{\beta_{\rm{d}}n}x_{\rm{e}} \frac{\zeta\beta_{\rm{t}}}{\beta_{\rm{d}}\beta_{\rm{r}}n}x_{\rm{M}} = 0\;,\label{polynomial} \end{equation}
where $\zeta$ is the ionization rate, $n$ is the local number density of disk material, and $x_{\rm{M}}$ is the metal fraction. There are three recombination processes accounted for with associated rate coefficients in equation \ref{polynomial}: dissociative recombination of electrons with molecular ions ($\beta_{\rm{d}} = 2\times 10^{-6} T^{-1/2}\,\textrm{cm}^3 \,\textrm{s}^{-1}$), radiative recombination of electrons with metal ions ($\beta_{\rm{r}} = 3\times10^{-11} T^{-1/2}\,\textrm{cm}^3\,\textrm{s}^{-1}$), and charge transfer from molecular ions to metal ions ($\beta_{\rm{t}} = 3\times10^{-9} \textrm{cm}^3 \,\textrm{s}^{-1}$) \citep{MP2003}.

We separately consider X-rays generated through magnetospheric accretion and interstellar cosmic rays as ionizing sources. Following \citep{Sano2000}, we calculate the cosmic ray ionization rate at the disk midplane by considering the interstellar cosmic ray ionization rate, $10^{-17}$ s$^{-1}$, attenuated over a length of 96 g cm$^{-2}$ \citep{Umebayashi1981},
\begin{equation} \zeta_{CR} = \frac{10^{-17}\,\textrm{s}^{-1}}{2}\exp \left(-\frac{\Sigma}{96\,\textrm{g}\;\textrm{cm}^{-2}}\right). \label{CRIonization} \end{equation}

Following \citet{MP2003}, the X-ray ionization rate is calculated using,
\begin{equation} \zeta_X = \left[\left(\frac{L_X}{E_X4\pi d^2}\right)\sigma(E_X)\right]\left(\frac{E_X}{\Delta \epsilon}\right)J(\tau,x_0)\;, \label{XRayIonization} \end{equation}
where $L_X \simeq 10^{30}$ ergs s$^{-1}$ is the X-ray luminosity of the protostar, $E_X=4$ keV is the X-ray energy considered, $d$ is the distance between the X-ray source and the midplane radius considered, and $\Delta \epsilon$ is the energy required to make an ion pair. The absorption cross section for X-rays of energy $E$ is \citep{Glassgold1997},
\begin{equation} \sigma(E) = 8.5\times10^{-23} \,\textrm{cm}^2\left(\frac{E}{\textrm{keV}}\right)^{-2.81}, \end{equation}
and the optical depth is,
\begin{equation} \tau(E) = N_H\sigma(E)\;,\end{equation}
where $N_H$ is the number density measured over the X-ray's path through the disk to the midplane radius considered. Lastly the X-ray attenuation factor in equation \ref{XRayIonization} is,
\begin{equation} J(\tau, x_0) = \int_{x_0}^\infty x^{-n}\exp(-x-\tau(E_X)x^{-n}) dx \,, \label{RadiativeTransfer}\end{equation}
written in terms of a dimensionless photon energy $x \equiv E/E_X$.

To calculate the location of the outer edge of the dead zone, we input the ionization rate corresponding to cosmic rays or X-rays (equation \ref{CRIonization} or \ref{XRayIonization}) into equation \ref{polynomial} to determine the electron fraction along the disk midplane. We then use equation \ref{MagneticReynolds} to determine the midplane radius where the critical magnetic Elsasser number condition is met, corresponding to the boundary between the MRI active and inactive regions of the disk.

%traps figure
In figure \ref{Traps}, we plot the time-dependent locations of planet traps in our model for three different initial disk masses, otherwise using fiducial disk parameters. For a range of disk masses corresponding to $\pm \sigma_m$ in our population synthesis calculations, the heat transition lies outside of the ice line for the entirety of a typical disk's lifetime of 3 Myr, while the cosmic ray dead zone exists $\lesssim$ 1 AU inside of the ice line. 

We note that the ice line, heat transition, and cosmic ray dead zone all evolve to $\simeq$ 1 AU at the end of a typical disk lifetime, as was found in \citet{Hasegawa2016}. Conversely, the X-ray dead zone exists outside of the ice line for the first $\sim$ Myr of disk evolution before evolving to within 0.1 AU after 3 Myr. This comparatively rapid evolution is a result of the X-ray ionization along the midplane being extremely sensitive to the disk surface density. 

We emphasize that the results of our planet formation models are connected to the location and evolution of the planet traps, since planets form within the traps throughout their type-I migration regime. For example, a planet forming within the heat transition forms outside $\sim$ 5 AU during oligarchic growth, accreting solids from a lower surface density region than the inner $\sim 0.1-0.5$ AU regions encountered by planets forming in the X-ray dead zone. The effects of the traps on our planet population synthesis models will be discussed in detail in section \ref{Results1}. 

%gap-opening
\subsubsection{Type-II Migration}
Type-II migration applies to planets that are massive enough to alter the local disk structure through the formation of an annular gap. Gap formation, and resulting type-II migration, allows planets to migrate away from the traps they were forming within during type-I migration. The gap-opening mass is reached when the planet's torque on the local disk material exceeds that of disk viscosity, or when the planet's Hill radius exceeds the disk's pressure scale height $H$. The gap-opening mass is written as \citep{MP2006},
\begin{equation} M_{\textrm{gap}} = M_*\;\textrm{min}\left[3h^3(r_p), \sqrt{40\alpha h^5(r_p)}\right]\,, \label{GapOpeningMass} \end{equation}
where $r_p$ is the planet's radius, and $h =H/r_p$ is the disk aspect ratio at the planet's location. We note that, when calculated this way, we are predicting the planet to open a gap in the disk when it overcomes the suppressing effects of either the disk viscosity \emph{or} the disk pressure (which are considered simultaneously, for example, in \citet{Crida2006}).

During type-II migration, the planet migrates following the disk viscous timescale of $\sim 10^6$ years, having a migration speed of,
\begin{equation} v_{\rm{mig,II}} \simeq -\nu/r_p \;. \end{equation}
When the planet's mass becomes comparable to the total disk mass within the planet's orbit, exceeding a critical mass of $M_{\rm{crit}} = \pi r_p^2 \Sigma$, it will resist migrating with the disk evolution \citep{Ivanov1999}. In this case, the slowed type-II migration speed becomes \citep{HP12},
\begin{equation} v_{\rm{mig,slow II}} \simeq -\frac{\nu}{r(1+M_p/M_{\rm{crit}})} \;.\end{equation}

\subsection{Core Accretion Model} \label{CoreAccretion}

We consider planet formation to take place through the core accretion scenario in this work, whereby an initially small planetary core accretes solids from the disk, building up its mass before accreting large amounts of gas. Our model considers an initial condition of a 0.01 M$_\oplus$ oligarch forming at an early stage of 10$^5$ years into disk evolution. 

The first stage of planet formation in our model is oligarchic growth, whereby growth of the planetary core takes place via accretion of planetesimals. The accretion rate in this regime is \citet{KokuboIda2002},
\begin{equation} \begin{aligned} \tau_{\rm{c,acc}}  \simeq &1.2\times10^5\;\textrm{yr}\;  \left(\frac{\Sigma_d}{10\;\textrm{g cm}^{-2}}\right)^{-1}
\\ & \times \left(\frac{r}{r_0}\right)^{1/2}\left(\frac{M_p}{M_\oplus}\right)^{1/3}\left(\frac{M_*}{M_\odot}\right)^{-1/6} 
\\ & \times\left[\left(\frac{b}{10}\right)^{-1/5}\left(\frac{\Sigma_g}{2.4\times10^3\;\textrm{g cm}^{-2}}\right)^{-1/5} \right.
\\ & \left. \times \left(\frac{r}{r_0}\right)^{1/20}\left(\frac{m}{10^{18} \;\textrm{g}}\right)\right]^2\;,\;\label{Solid_Accretion} \end{aligned}\end{equation}
where $m\simeq10^{18}$ g is the mass of accreted planetesimals and $b\simeq10$ is a parameter that defines the core's feeding zone. The corresponding accretion rate is $\dot{M} = M_p/\tau_{\rm{c,acc}}$.  

During the oligarchic growth phase, accreted planetesimals heats gas surrounding the planet, keeping it in hydrostatic balance. The transition from the oligarchic growth phase to gas accretion phases takes place when the planetesimal accretion decreases to the point that the heat released is insufficient to maintain pressure support of the surrounding gas, which then accretes onto the planet. The critical core mass, $M_{\rm{c,crit}}$ that separates these stages of formation depends on the planetesimal accretion rate and the envelope opacity $\kappa_{\rm{env}}$ as \citep{Ikoma2000, IdaLin2008, HP14},
\begin{equation} \begin{aligned} M_{\rm{c,crit}}  & \simeq f_{\rm{c,crit}} \left(\frac{1}{10^{-6} M_\oplus \;\textrm{yr}^{-1}}\frac{dM_p}{dt}\right)^{1/4} M_\oplus
\\&\simeq 10\,M_\oplus\left(\frac{1}{10^{-6} M_\oplus \;\textrm{yr}^{-1}}\frac{dM_p}{dt}\right)^{1/4}\left(\frac{\kappa_{\rm{env}}}{1\,\rm{cm}^2\,\rm{g}^{-1}}\right)^{0.3} \;. \label{CoreCrit} \end{aligned} \end{equation}
We note that this scaling is an update on the model presented in \citet{Alessi2017}, which ignored the dependence of the parameter $f_{\rm{c,crit}}$ on envelope opacity. We include the dependence of $M_{\rm{c,crit}}$ on $\kappa_{\rm{env}}$ in equation \ref{CoreCrit} and we fully explore the role of envelope opacity in our core accretion model.

For masses $M_p > M_{\rm{c,crit}}$ planet formation occurs via gas accretion, which takes place on the Kelvin-Helmholtz timescale \citep{Ikoma2000},
\begin{equation} \tau_{KH} \simeq 10^c \, \textrm{yr}\left(\frac{M_p}{M_\oplus}\right)^{-d}\;.\label{Gas_Accretion}\end{equation}
The values of parameters $c$ and $d$ in the Kelvin-Helmholtz timescale are physically linked to $\kappa_{\rm{env}}$. We include this effect in our model by using the fits shown in \citet{Mordasini2014}, that relate results of a numerical model of gas accretion to the Kelvin-Helmholtz parameters for a range of envelope opacities of $10^{-3}-10^{-1}$ cm$^2$ g$^{-1}$. The fit given for the Kelvin-Helmholtz $c$ parameter is,
\begin{equation} c = 10.7 + \log_{10}\left(\frac{\kappa_{\rm{env}}}{1\,\rm{cm}^2\,\rm{g}^{-1}}\right)\;.\label{KHc} \end{equation}
The Kevlin-Helmholtz $d$ parameter has a more complicated dependence on envelope opacity, ranging from $\approx$1.8-2.4 over the range of $\kappa_{\rm{env}}$ considered. The following piecewise-linear function reproduces the outputs shown in \citet{Mordasini2014},
\begin{equation} d = \left\{
\begin{array}{ll}
0.994 - 0.335\log\left(\frac{\kappa_{\rm{env}}}{\rm{cm}^2\,\rm{g}^{-1}}\right) & 0.001 \leq \frac{\kappa_{\rm{env}}}{\rm{cm}^2\,\rm{g}^{-1}} < 0.003 \\
1.954 + 0.045\log\left(\frac{\kappa_{\rm{env}}}{\rm{cm}^2\,\rm{g}^{-1}}\right) & 0.003\leq \frac{\kappa_{\rm{env}}}{\rm{cm}^2\,\rm{g}^{-1}} < 0.005\\
3.093 + 0.54\log\left(\frac{\kappa_{\rm{env}}}{\rm{cm}^2\,\rm{g}^{-1}}\right) & 0.005 \leq \frac{\kappa_{\rm{env}}}{\rm{cm}^2\,\rm{g}^{-1}} < 0.05
\end{array} \right.
\end{equation}
We do not consider envelope opacity values greater than 0.05 cm$^2$ g$^{-1}$ in this work. Our approach here is an updated treatment of the gas accretion model presented in \citet{Alessi2017}, that considered the Kelvin-Helmholtz $c$ and $d$ as separate parameters, ignoring their dependence on envelope opacities. 

Hence, we have reduced our model's parameter set ($f_{\rm{c,crit}}$ and the Kelvin-Helmholtz $c$ and $d$ parameters) to just one; $\kappa_{\rm{env}}$.

The disk lifetime plays a crucial role in core accretion calculations, setting an upper limit to the amount of time planets have to form and hence to the amount of material that they can accrete. In our model, planets can accrete up until the point where the disk is photoevaporated, but at the disk lifetime when the disk is dissipated, their formation is truncated. We find that oligarchic growth typically takes place on a $\lesssim$ Myr timescale. Conversely, gas accretion initially takes place on a long $>$ Myr timescale, which is comparable to the disk lifetime, before reaching $\sim 30$ M$_\oplus$ necessary to undergo runaway growth. The comparable timescale of the initially slow gas accretion phase and the disk lifetime reveals why super Earths form in the core accretion model; these are planets whose natal disks photoevaporate during their slow gas accretion phase.

During all stages of gas accretion, including the runaway growth phase, we only consider the Kelvin Helmholtz timescale, modified to include its $\kappa_{\rm{env}}$ dependence. We choose our treatment following \citet{HP12, HP13}, as these models also consider a trapped type-I migration phase and showed that such a gas accretion model can reproduce observed planet populations. The transition between the slow to runaway gas accretion phases in our model is simply a consequence of the decreasing Kelvin-Helmholtz timescale and the correspondingly increasing accretion rate.

Other works such as \citet{Machida2010, Dittkrist2014}, and \citet{Bitsch2015} have considered a disk limited accretion phase during late stages of planet formation which is in agreement with hydrostatic simulations (e.g. \citet{Lubow1999}. Both approaches are sensitive to $\kappa_{\rm{env}}$, and can produce similar results depending on the envelope opacities considered. 

Termination of gas accretion is related to gap opening, which depletes material in the planet's feeding zone \citep{Lissauer2009}. It is, however, unclear how soon a planet's accretion will terminate after opening a gap, with previous models finding that a substantial amount of material can flow through the gap and be accreted by the planet \citep{Lubow2006, Morbidelli2014}. We therefore parameterize the maximum masses of planets in our model in terms of their gap-opening masses, truncating formation when planets reach a mass of,
\begin{equation} M_{\textrm{max}} = f_{\textrm{max}}M_{\textrm{gap}}\;, \label{MaximumMass} \end{equation}
where $f_{\rm{max}}$ is a parameter in our model. We consider a range of values of 1-500 for $f_{\rm{max}}$, where a value of 1 corresponds to a planet whose accretion is terminated abruptly after opening a gap. Conversely, larger values of $f_{\rm{max}}$ correspond to planets that are able to continue accreting well beyond reaching their gap-opening masses (e.g. \citet{Kley1999}). \citet{HP13} considered the same method of terminating planet growth, and found that Jovian planet formation frequencies were insensitive to the parameter's setting when $f_{\rm{max}} \gtrsim 5$, while super Earth formation frequencies were sensitive to $f_{\rm{max}}$. We find that a range of $f_{\rm{max}}$ values is necessary in our population synthesis approach to obtain a range of Jovian masses seen in the M-a diagram.

\begin{figure}
\centering
\includegraphics[width = 3.1 in]{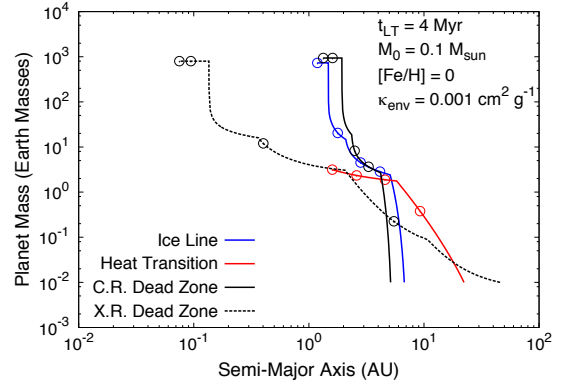}
\caption{Planet formation tracks are shown corresponding to each of the planet traps in our model within a disk with 4 Myr lifetime, 0.1 M$_\odot$ initial mass, and Solar metallicity. Open circles along the tracks denote the location of the planets at 1 Myr intervals. The ice line and cosmic ray dead zone each produce $\sim 10^3$ M$_\oplus$ Jupiters near 2 AU, while the X-ray dead zone produces a hot Jupiter orbiting within 0.1 AU. Lastly, the heat transition forms a $\sim$ 3 M$_\oplus$ super Earth orbiting near 2 AU.}
\label{Tracks}
\end{figure}

In figure \ref{Tracks}, we show planet formation tracks resulting from each of the traps in our model within a disk with 4 Myr lifetime, fiducial mass and metallicity. We initialize our calculations with a 0.01 M$_\oplus$ core beginning to form at $\tau_{\rm{int}} = 10^5$ years into the disk's lifetime, situated at an orbital radius coinciding with a planet trap. We place our embryos on traps because rapid typeI migration will quickly bring embryos from other initial locations into the traps\footnote{Our approach is different than \citet{Mordasini2009} who assume an initial distribution of starting positions distributed between 0.1-20 AU.}. With this set of disk parameters, both the ice line and cosmic ray dead zone produce Jupiters orbiting near 2 AU, while the X-ray dead zone forms a hot Jupiter, and the heat transition forms a super Earth. 

\subsection{Population Synthesis}

%Population synthesis: methods and statistical distributions
We employ the technique of planet population synthesis to account for the range of disk conditions suggested by observations and their affect on planet formation. Three of the four parameters we vary, being properties of the host star and disk, are external parameters - namely the disk lifetime, mass, and metallicity, whose distributions are discussed in section \ref{Observations}. Core accretion results depend sensitively on these quantities which are by the results of star formation in turbulent molecular clouds. Through varying these quantities, we aim to connect the observed ranges of host star and disk properties (which are vital to the planet formation process) with resulting planet populations. 

In addition to the disk lifetime, initial mass, and metallicity, the fourth parameter we stochastically vary is the $f_{\rm{max}}$ parameter discussed in section \ref{CoreAccretion} that sets the mass at which planet formation is terminated. For this parameter, we consider a log-uniform distribution with minimum 1 and maximum 500. We emphasize that this is the only parameter intrinsic in our model that is stochastically varied.

In our population synthesis calculations, we employ a Monte-Carlo routine whereby we stochastically sample each of the four varied parameters' distributions prior to each individual planet formation calculation. In each run, we calculate formation tracks for 1000 planets within each planet trap, for a total of 3000 planets in each population\footnote{We separately consider the cosmic ray and X-ray dead zone in each population run.}. For each population, we apply a synthetic observation described in Appendix \ref{Appendix} that allows us to filter out planets that have low probabilities of being observed with current technologies. This allows for a better comparison with the observed distributions. %Each population takes roughly 65 hours of computing on one core of a personal computer to complete all necessary calculations. 

For a subset of the population runs, we do not vary the disk metallicity, but rather choose a constant [Fe/H] value over the entire population. In these cases (specified in section \ref{Results}) the remaining three parameters' distributions are sampled to compute the population. For all populations, we consider a central star mass of 1 M$_\odot$, stellar radius 3 R$_\odot$, and effective temperature 4200 K that correspond with pre-main sequence tracks for a G-type star as shown in \citet{Siess2000}. In doing so, we are neglecting any consequences of deviating host-star masses from the Solar value on our resulting populations, as was investigated in \citet{IdaLin2005} and \citet{Alibert2011}. Lastly, we consider an effective disk $\alpha=0.001$ for all populations in this work. This setting of $\alpha$ is consistent with the recently observed upper limits of the TW-Hya disk's $\alpha_{\rm{turb}}$ of 0.007 \citep{Flaherty2018}.

In each population run, we specify the chosen $\kappa_{\rm{env}}$ value, which remains constant over the entire population. In this work, we do not consider the connection between the disk metallicity and $\kappa_{\rm{env}}$ and treat these a independent parameters. Envelope opacities of forming planets will depend on the time-dependent size distribution and composition of dust grains in the planet's envelope, and it is currently unknown how these are related to disk metallicity - if at all \citep{Mordasini2014b}.

\section{Results} \label{Results}

\begin{figure*} \ContinuedFloat*
\centering \includegraphics[width = 0.97\linewidth]{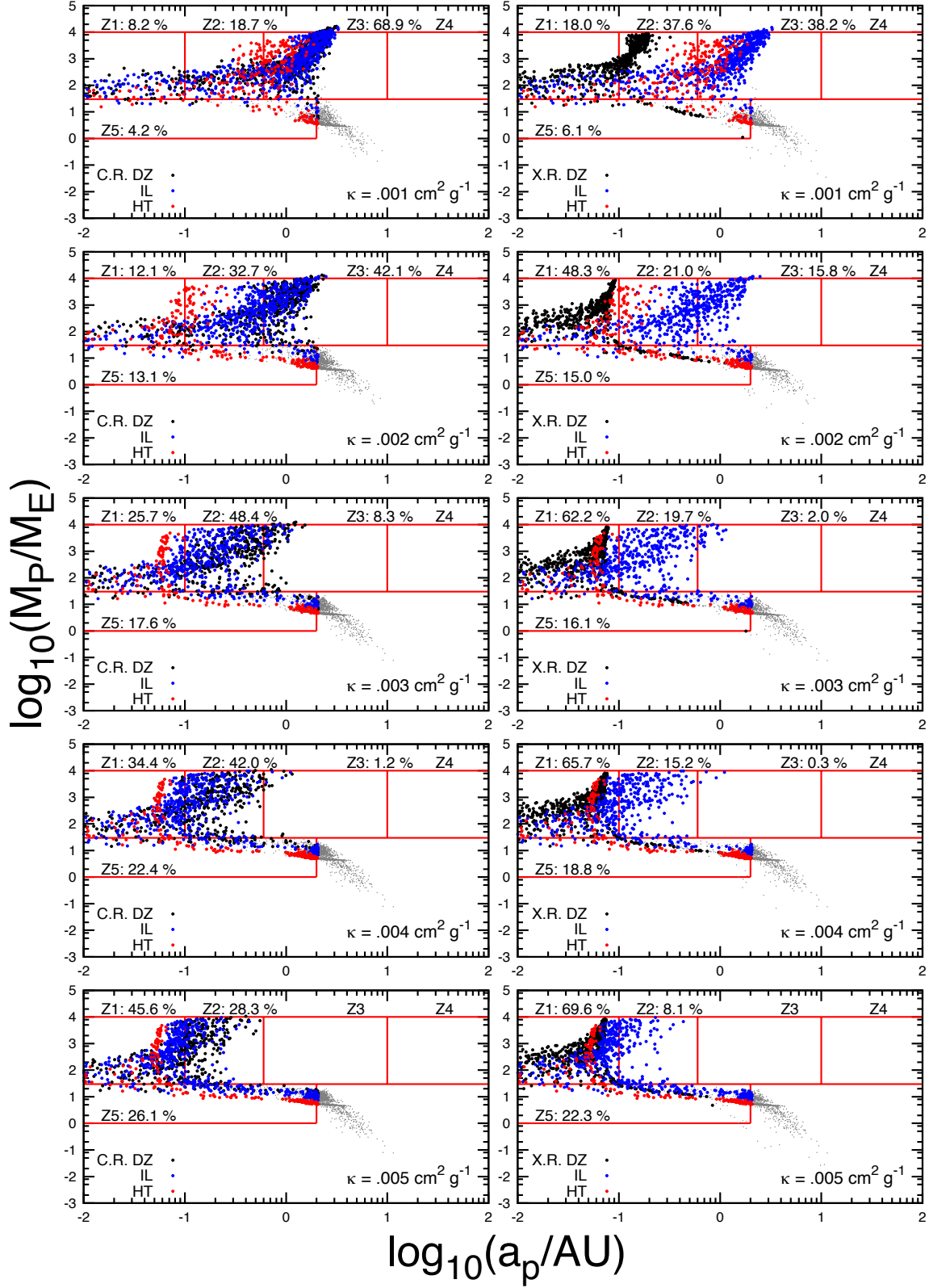}
\caption{Mass semi-major axis distributions of computed planet populations in Solar metallicity disks are shown. The left column of panels shows populations where cosmic ray ionization is considered to calculate the dead zone trap, and the right column corresponds to X-ray dead zone models. We increase the envelope opacity $\kappa_{\rm{env}}$ values from 0.001 cm$^2$ g$^{-1}$ to 0.005 cm$^2$ g$^{-1}$ from the top to bottom row of panels. Planets with low observation probabilities have been removed from the population and shown as small grey points. The remaining planets are coloured based on the trap they formed in. }
\label{SolarZ_Pops}
\end{figure*}
\begin{figure*} \ContinuedFloat
\centering \includegraphics[width=0.45\textwidth]{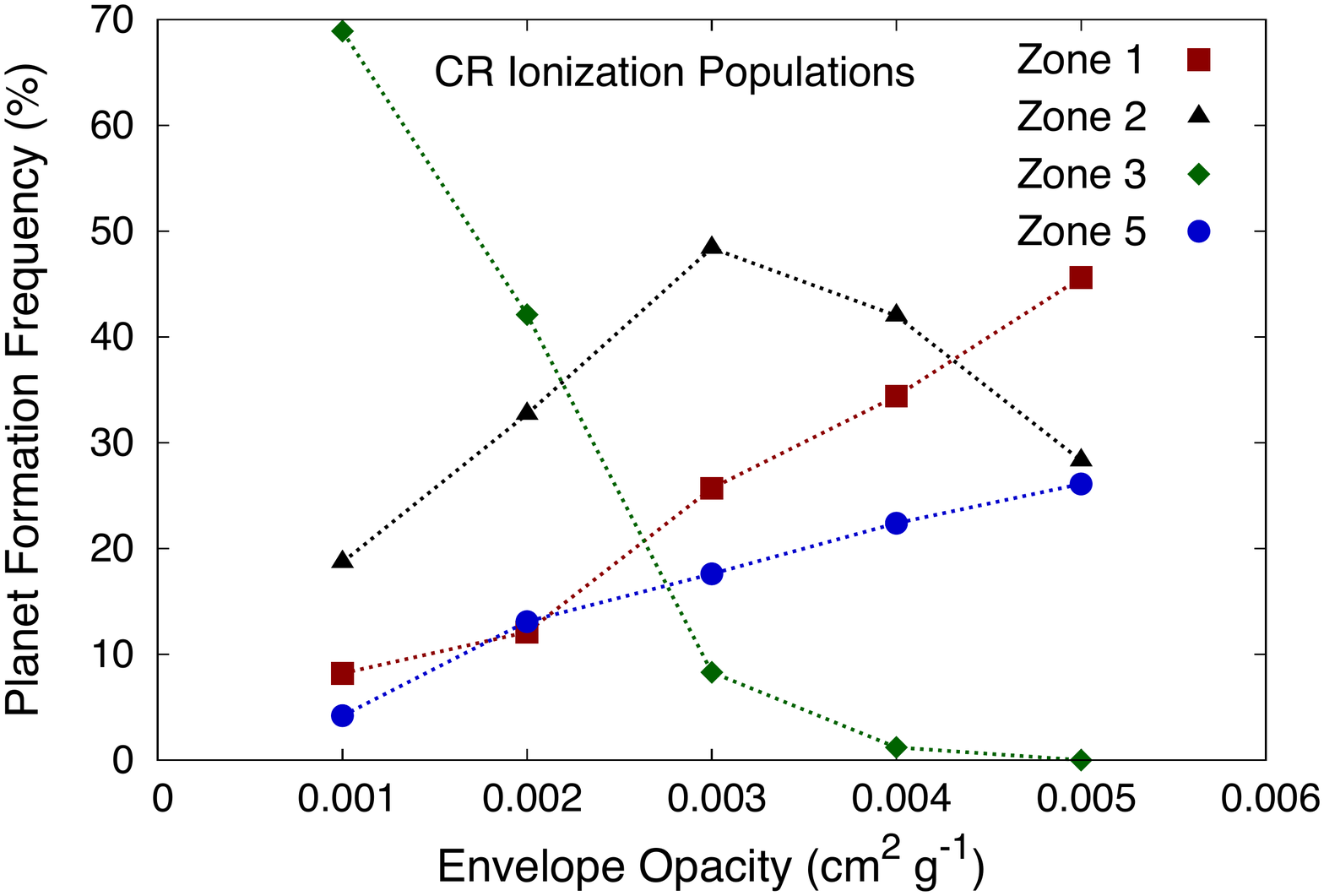} \includegraphics[width=0.45\textwidth]{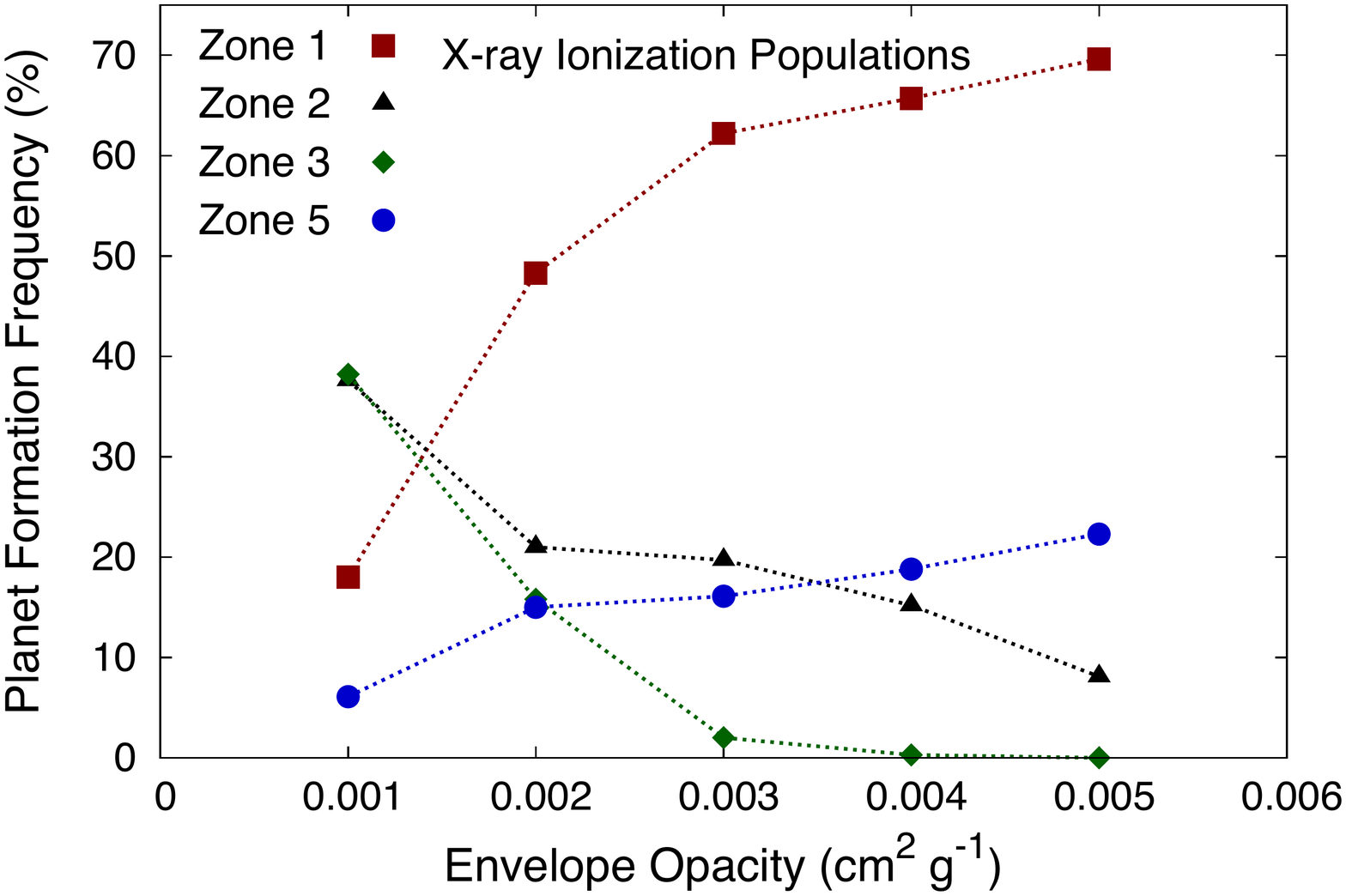}
\caption{Planet population frequencies for each zone are plotted for each envelope opacity considered in figure \ref{SolarZ_Pops}. The left panel corresponds to the cosmic ray dead zone runs, and the right panel to X-ray dead zone runs. Planets with low observation probabilities (the small grey points in figure \ref{SolarZ_Pops}) are not included when calculating zone frequencies.}
\label{SolarZ_Freqs}
\end{figure*}

We perform three separate sets of population synthesis calculations. First, in section \ref{Results1}, we consider only Solar metallicity disks and vary the values of $\kappa_{\rm{env}}$ to determine its affect on population structure. Next, in section \ref{Results2}, we compute populations with constant disk metallicities at non-Solar values. Lastly, in section \ref{Results3}, we compute populations with stochastically varied disk metallicities. In all populations, the disk lifetime and initial masses are stochastically sampled using the log-normal distributions for external parameters summarized in equation \ref{LogNormal}, unless otherwise specified. In all of these sections, two separate sets of populations are run; one that considers cosmic ray-ionized disks, and one considering X-ray ionized disks.

\subsection{Solar Metallicity Populations} \label{Results1}

In figure \ref{SolarZ_Pops}, we plot the outcomes of our population synthesis calculation for Solar metallicity disks, while considering different values of $\kappa_{\rm{env}}$. Populations generated in cosmic ray ionized disks are shown in the left column, and those for X-ray ionized disks in the right column. Each point on the diagrams represents the mass and orbital radius of a planet at the end of its natal disk's lifetime. The percentage included for each zone corresponds to the fraction of remaining planets that populate that region of the diagram. All populations shown have been observationally filtered using our method described in Appendix \ref{Appendix}, and planets with low estimated observation probabilities are shown in figure \ref{SolarZ_Pops} as small grey points. These planets are not included when calculating the percentage of planets populating zones in the diagram.

\subsubsection{Populations Arising From Planet Traps} 

The Jovian planets formed in our populations with the largest semi-major axes typically form within the cosmic ray dead zone or ice line traps. These are the two innermost traps in our model for the first $\sim 10^6$ years of disk evolution, each situated within 8 AU for typical disk masses (see figure \ref{Traps}) and evolving inwards. Compared to the outer traps, planets forming within the C.R. dead zone and the ice line therefore accrete from higher surface density regions of the disk, and have correspondingly lower accretion timescales during oligarchic growth. 

In contrast, Jovian planets with the smallest orbital radii tend to arise from formation within the X-ray dead zone or heat transition traps. In the case of the heat transition, the large orbital radii ($\gtrsim 10$ AU for the first 2 Myr of disk evolution) of the trap leads to planets accreting from lower surface density regions, leading to longer oligarchic growth timescales. This causes planets forming within this trap to have slightly lower critical core masses at the onset of gas accretion, and migrating within $\sim$ 2 AU before reaching this stage of formation (see equation \ref{CoreCrit}). The lower $M_{\rm{c,crit}}$ values in turn lead to subsequent gas accretion to take place on longer timescales, causing heat transition planets to migrate in further prior to reaching runaway growth (which only takes place in long-lived disks, with lifetimes $\gtrsim 5-6$ Myr).

The X-ray dead zone trap is initially the outermost trap in our model having orbital radii of $\gtrsim 20$ AU. However, this trap evolves inwards the fastest, and is located at $\sim$ 1-2 AU after 1 Myr of disk evolution, and is within 0.1 AU after 3 Myr, which is an average disk lifetime in our population synthesis models. Once the trap has quickly evolved to small orbital radii, planets forming within the trap accrete from high-density regions, and have low formation timescales (typically the shortest out of all the traps in our model). The trap's rapid evolution to small orbital radii leads to Jovians population zones 1 or 2 in our population runs. 

Our model produces a vast number of cores that fail to reach Jovian masses within their disks' lifetimes. In all of our population runs, more than $\sim$ half of our 3000 individual runs result in planets whose formation is terminated during the slow gas accretion phase or the oligarchic growth phase. However, the fraction of these failed cores that end up as super Earths or Neptunes within the observed zone 5 region of the M-a diagram remains low. Many of them are situated just outside the outer boundary of the zone (where our estimated observable limit is defined). Additionally, a large portion of the planets that do end up within zone 5 have low observation probabilities and are filtered out of the population. Our zone 5 population fractions are therefore substantially lower ($\sim 5\% - 25\%$) than what is found in observations ($78\%$). We believe that this is a consequence of our constant dust to gas model, and discuss this further in section 5.

Planet formation within each zone contributes to each zone 5 population. However, the heat transition forms most of the super Earths and Neptunes due to planet formation in this trap having the longest timescale. There is a large region of zone 5 that our planet formation model fails to populate; specifically the low-mass, low-period region of the diagram. At 1-2 AU, our lowest mass super Earths in zone 5 are $\sim$ 5 M$_\oplus$, and within 0.1 AU all zone 5 planets are Neptunes with masses exceeding 10 or 20 M$_\oplus$. This "zone of avoidance" within zone 5 in our model can be attributed to the inward migration of our formed planets being restricted to the rate at which the traps themselves migrate. The ice line, heat transition, and cosmic ray dead zone traps all converge near 1-2 AU after 3 Myr of disk evolution, so planets forming within these traps that have not reached their gap-opening masses will not be found at smaller orbital radii. The X-ray dead zone, having the most rapid inward migration, results in efficient planet formation. By the time the trap has evolved to 0.1 AU, a planet forming within the trap will exceed 10 M$_\oplus$. 

\subsubsection{Effects of Envelope Opacities; Cosmic Ray vs. X-Ray Dead Zone}

In figure \ref{SolarZ_Pops}, we consider a range of $\kappa_{\rm{env}}$ values between 0.001-0.005 cm$^2$ g$^{-1}$, the population pertaining to each value in a different row. The main effect of increasing envelope opacity is that it increases gas accretion timescales, as planet envelopes take longer to cool and contract. This causes the Jovian planet distribution in each population to shift to smaller orbital radii, as planets take longer to reach their runaway growth phase and in turn have more time to migrate inwards throughout their formation. The longer gas accretion timescales also favour the formation of zone 5 planets as the slow gas accretion phase takes longer, causing more planets' accretion to be terminated in this phase. In the case of the largest $\kappa_{\rm{env}}$ shown in figure \ref{SolarZ_Pops}, the resulting population has the largest frequency of zone 5 planets, and the gas giants are all within 0.6 AU.

Based on these conclusions, if we were to extend our investigated $\kappa_{\rm{env}}$ values towards unity, our populations would consist entirely of hot Jupiters (zone 1) and super Earths (zone 5). The increased formation timescales would result in a larger population of zone 5 planets, while also causing planets that do undergo runaway growth to migrate to even smaller orbital radii before doing so.

%Contrast XR & CRDZ jupiter distribution
%The populations corresponding to the cosmic ray dead zone show that this trap, in combination with the ice line and the heat transition, shows a gradual increase in the frequency of Jovian planets with orbital radius out to $\sim$3 AU, which represents the largest semi-major axis of Jupiters formed in our model due to the convergence of the planet traps near, or inside of 1 AU within disk evolution timescales (\textbf{connect with Cummings results}). 

%In contrast, the X-ray dead zone has a double-peaked radial distribution of Jovian planets, with the inner peak corresponding to the pile up of Jovians formed within the X-ray dead zone and the outer peak to planets formed (typically) in the ice line. The observed distributions suggest that zone 2 (between 0.1 and 0.6 AU) comprises a region where the frequency of Jovian planets is reduced. Populations using an X-ray dead zone, and the resulting double-peaked radial distribution of Jupiters, offer the only explanation for this feature of the observations in our model.

Comparing the left- and right-columns of figure \ref{SolarZ_Pops}, we find that the main difference between the CR and X-ray ionized disk models is that the CR models never result in a clear separation between the hot Jupiter and warm Jupiter populations. In contrast, the X-ray ionized disks have a distinct separation of these populations for low opacity models ($\kappa_{\rm{env}}$ = .001 - .002 cm$^2$ g$^{-1}$). Since this separation is a clear property of the observations we can conclude that based on our model that considers planetesimal accretion, (i) X-ray ionization dominates CR ionization effects in disks, a point supported by disk astrochemistry calculations \citep{Cleeves2013, Cleeves2015}, and that (ii) population separation between the hot Jupiter and warm Jupiter populations demands low envelope opacities (in agreement with results found in \citet{Mordasini2014}). \citet*{AliDib2017b} also obtained such a separation between the hot and warm Jupiter population using a pebble accretion planet formation model without considering a disk ionization model.

In figure \ref{SolarZ_Freqs}, we summarize these results by plotting each zone's frequency as a function of the envelope opacity considered in each population. For both cosmic ray dead zone and X-ray dead zone populations, zone 3 represents a reasonably large fraction of the population only for the lowest $\kappa_{\rm{env}}$ settings. At higher settings, this zone's frequency diminishes due to planet migration having a larger effect due to planets forming over longer timescales. For the same reason, the hot Jupiter population increases with envelope opacity. As was previously discussed, higher $\kappa_{\rm{env}}$ settings favour zone 5 planet formation as well, which is shown in figure \ref{SolarZ_Freqs}. 

\subsection{Effects of Disk Metallicity} \label{Results2}

In figure \ref{ConstantZ_Pops}, we show the results of our calculations that consider different disk metallicities while stochastically varying disk lifetimes and masses using equation \ref{LogNormal}. For all populations, we only consider an envelope opacity value of $\kappa_{\rm{env}} = 0.001$ cm$^2$ g$^{-1}$ for the reasons outlined in the previous subsection. Each population shown has filtered out planets with low probabilities of being observed using our method discussed in Appendix \ref{Appendix}. The populations shown in figure \ref{ConstantZ_Pops} reproduce the basic trends shown in section \ref{Results1} (i.e. the zones typically populated by the different traps), but highlight the effects of the setting of disk metallicity on our population results.

Disk metallicity effects the global dust to gas ratio throughout the disk (equation \ref{DustToGas}) and impacts each planet formation track during the oligarchic growth stage. Lower disk metallicities and dust-to-gas ratios result in longer oligarchic growth timescales. This causes planets to migrate further inward during this first phase of their formation following the evolution of the planet trap they are forming within. Additionally, critical core masses (equation \ref{CoreCrit}) are lower for planets forming in low-metallicity disks as well. This mass is the boundary between the oligarchic growth and slow gas accretion phases of formation. The low $M_{\rm{c,crit}}$ values achieved in low-metallicity disks result in longer gas accretion timescales, and in turn further inward migration. 

\begin{figure*} \ContinuedFloat*
\centering \includegraphics[width = 0.97\linewidth]{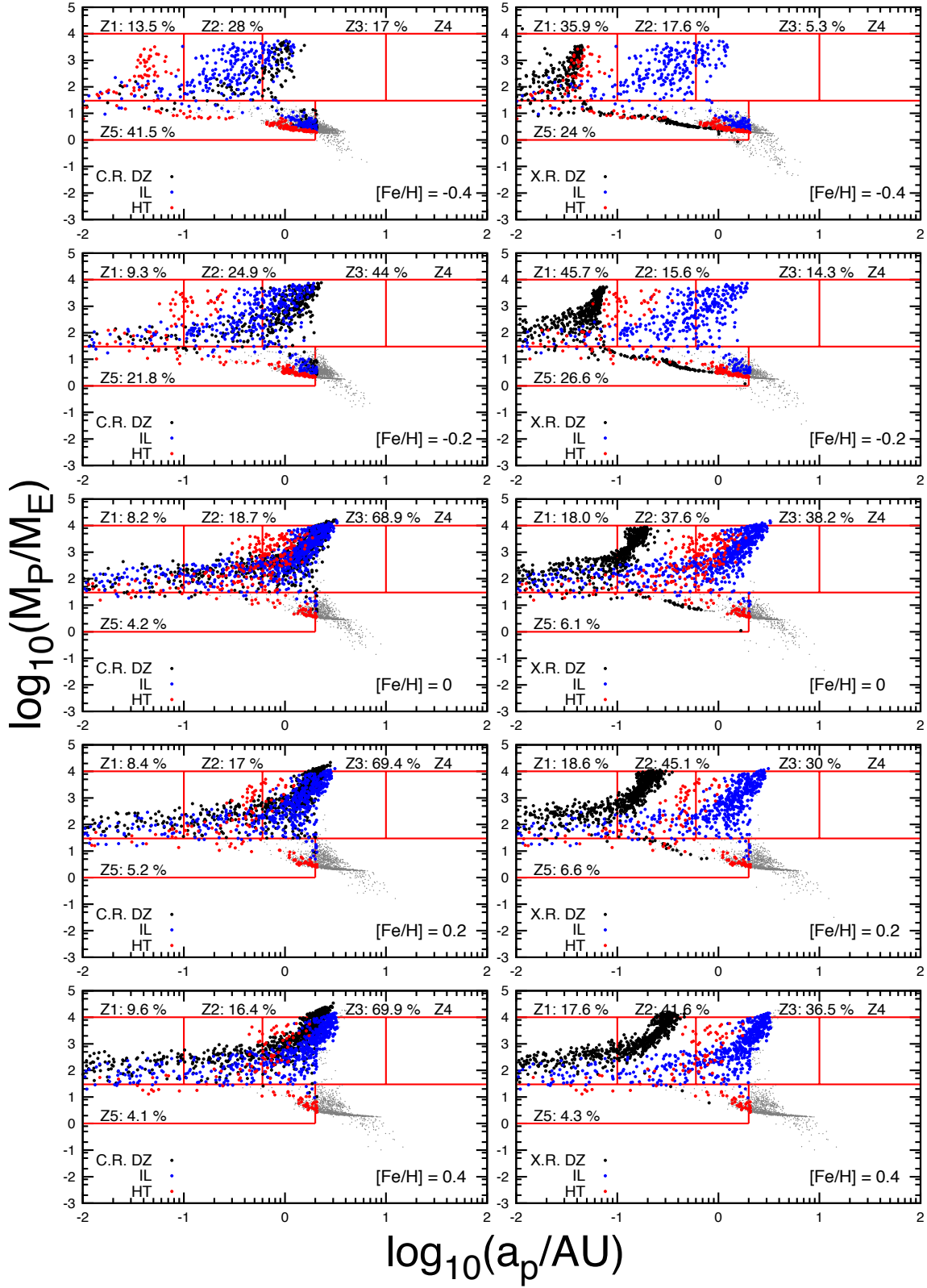}
\caption{Planet populations for constant disk metallicities are shown. The value of [Fe/H] is indicated for each population. In all runs, we consider an envelope opacity $\kappa_{\rm{env}} = 0.001$ cm$^2$ g$^{-1}$. Cosmic ray dead zone populations are shown in the left columns, and X-ray dead zone populations in the right. The plotted populations have been observationally filtered with small grey points indicating planets with low observation probabilities.}
\label{ConstantZ_Pops}
\end{figure*}
\begin{figure*} \ContinuedFloat
\centering
\includegraphics[width=0.45\textwidth]{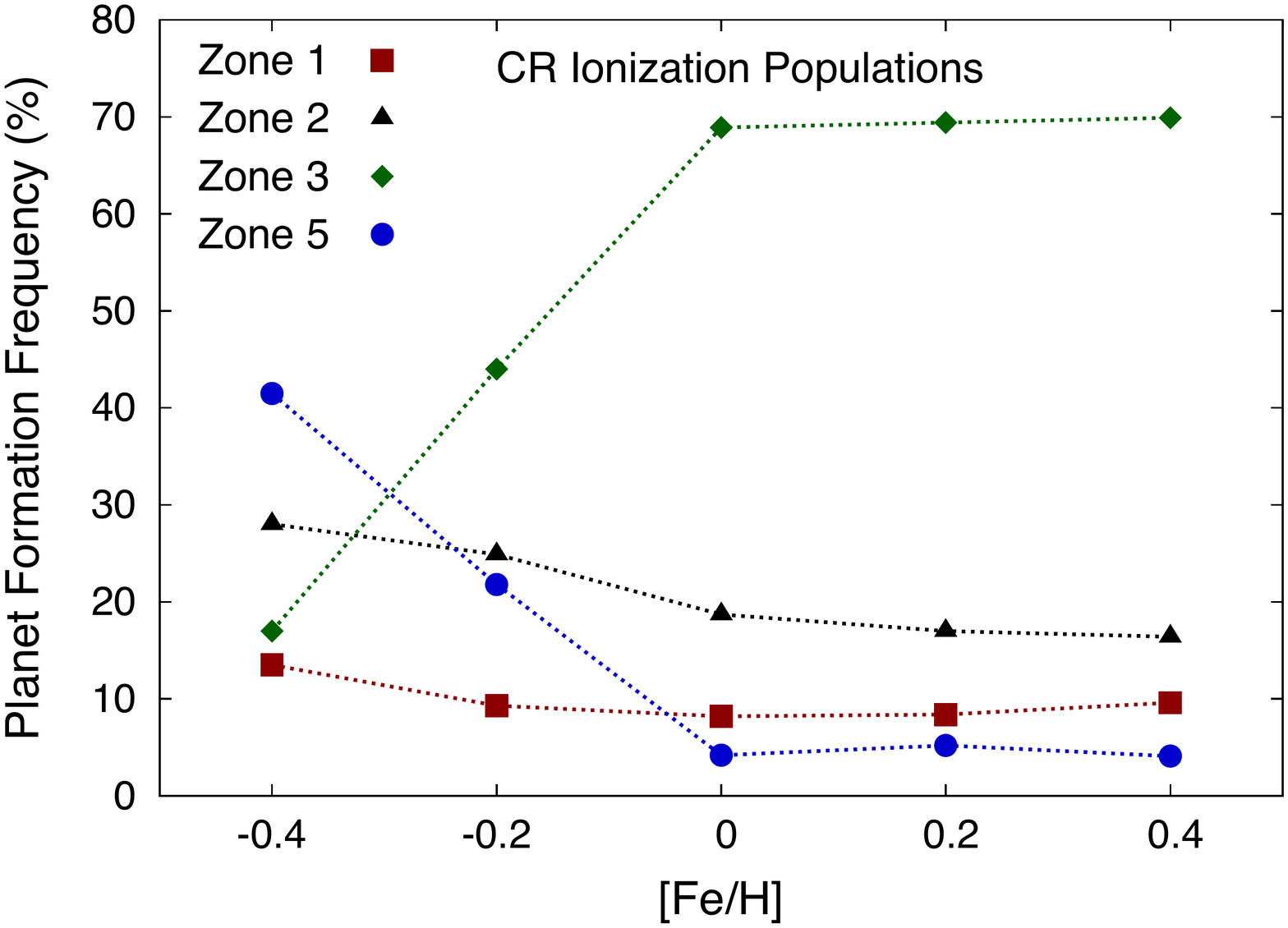} \includegraphics[width=0.45\textwidth]{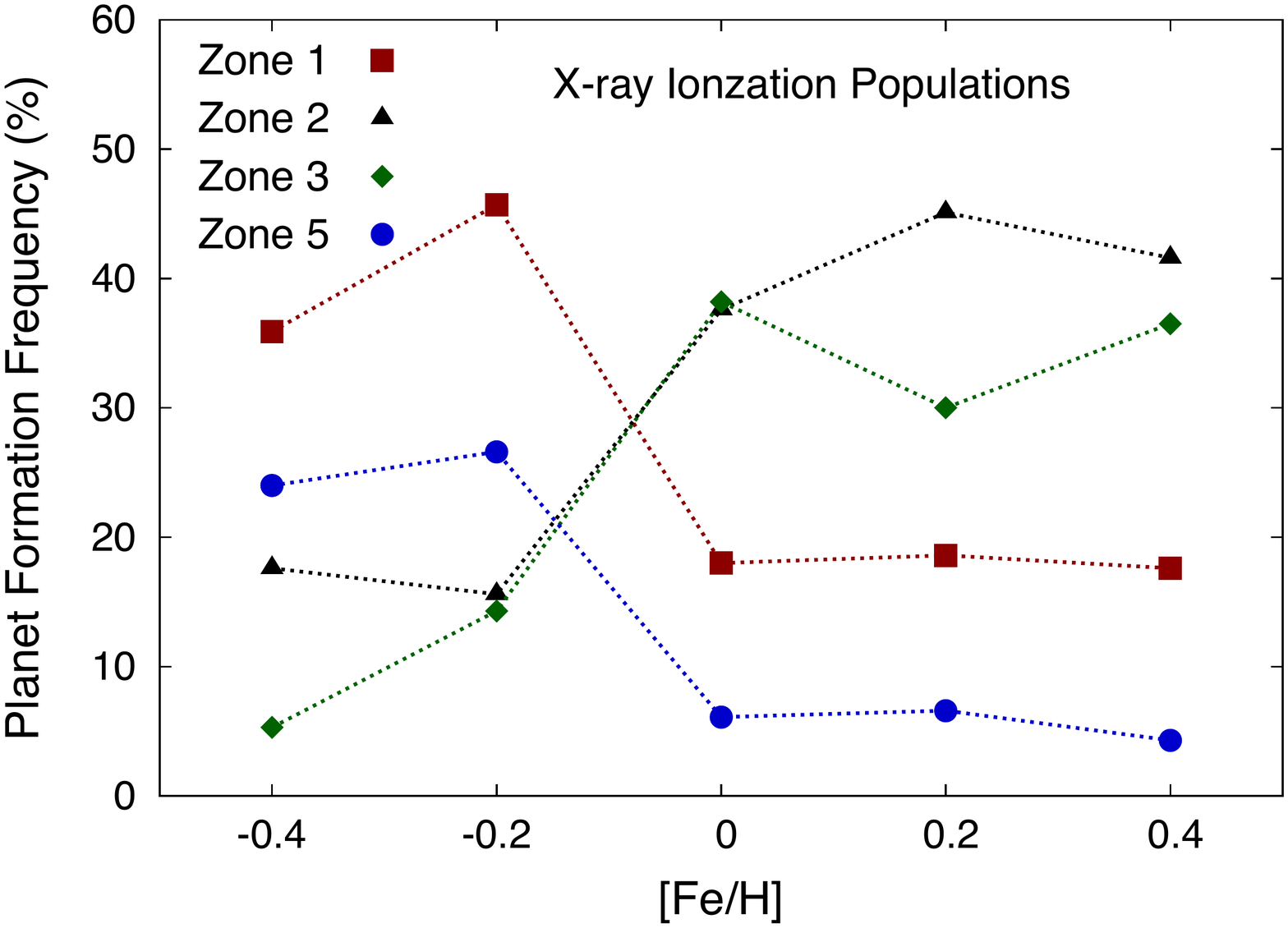}
\caption{Planet population frequencies from figure \ref{ConstantZ_Pops} are plotted for each zone as a function of disk metallicity.}
\label{ConstantZ_Freqs}
\end{figure*}

%Zone of avoidance
Due to the resulting low critical core masses, and the long formation timescales exposing planets to migration for the longest, the low-metallicity populations are particularly important for populating the zone of avoidance within zone 5 discussed in section \ref{Results1}. In the case of Solar metallicity, quite a large region of zone 5, corresponding to super Earths within 1 AU, remains unpopulated. Low metallicity settings do aid in populating this region, but only act to decrease the region's size, and there still remains a substantial region of zone 5 that our model is not able to populate.

In the upper panels of figure \ref{ConstantZ_Pops} where a metallicity of [Fe/H] = -0.4 is considered, the lowest mass planet that forms near 0.01 AU is roughly 10 M$_\oplus$, and at 0.1 AU is roughly 6 M$_\oplus$. Even with the optimal set-up to form low-mass zone 5 planets, our model cannot produce planets at these orbital radii with smaller masses, which are seen abundantly in the observed distribution. 

The increased formation timescales resulting from low disk metallicities favours the formation of super Earths and Neptunes, as was shown in previous metallicity studies such as \citet{Fischer2005} and \citet{Valenti2008}. In our low-metallicity populations, Jovian planets often require longer disk lifetimes than the average 3 Myr lifetimes considered in the $t_{\rm{LT}}$ distribution, and therefore form less frequently. The Jovian planets that do form are subject to migration for longer due to their increased formation timescale, which favours the formation of zone 1 and zone 2 Jupiters as opposed to zone 3. 

We note also that a variation of disk metallicity does not affect the double peaked structure of the Jovian populations, indicating that the low value of $\kappa_{\rm{env}}$ is still the key parameter (along with X-ray ionization), that controls this.

In figure \ref{ConstantZ_Freqs}, we show the frequencies by which planets populate various zones in figure \ref{ConstantZ_Pops} and their trends with disk metallicity. For both the cosmic ray and X-ray ionized models, the expected trend of low-metallicity disks favouring the formation of zone 5 planets is found. As disk metallicity is increased, the frequency of zone 5 planets decreases to less than 10 \% while the frequency of Jovian planets (sum of zones 1, 2, and 3) increases, in agreement with the observed planet-metallicity relation \citep{Fischer2005}. The distribution of Jovian planets in the populations in \ref{ConstantZ_Pops} is seen to shift outwards as disk metallicity is increased due to the shorter formation timescales.

For the case of cosmic ray populations, the majority of the Jovian population consists of zone 3 planets for metallicities [Fe/H] $\geq$ 0. The zone 1 and zone 2 frequencies decrease as metallicity is increased due to the distribution of Jovian planets shifting outwards.

The X-ray populations are capable of forming hot Jupiters even at the lowest metallicity settings considered. Since the X-ray dead zone trap evolves to small orbital radii early in the disk's evolution, planets forming within this trap accrete from higher surface density regions and have correspondingly lower formation timescales. Thus, the X-ray dead zone is the trap in our model that forms Jovian planets the most efficiently in low-metallicity disks. In figure \ref{ConstantZ_Freqs}, the hot Jupiter and the failed core populations (zone 1 and 5) comprise the majority of the sub-Solar metallicity populations, and their frequencies decrease as the metallicity increases beyond Solar. In the high metallicity cases, zones 2 and 3 are the highest frequency zones as the ice line and heat transition are able to form Jovians more often, and the distribution of Jovian planets shifts outwards due to shorter formation timescales.

\subsection{Stochastically Varied Disk Metallicity} \label{Results3}

\begin{figure*}
\centering
\includegraphics[width = 0.97\linewidth]{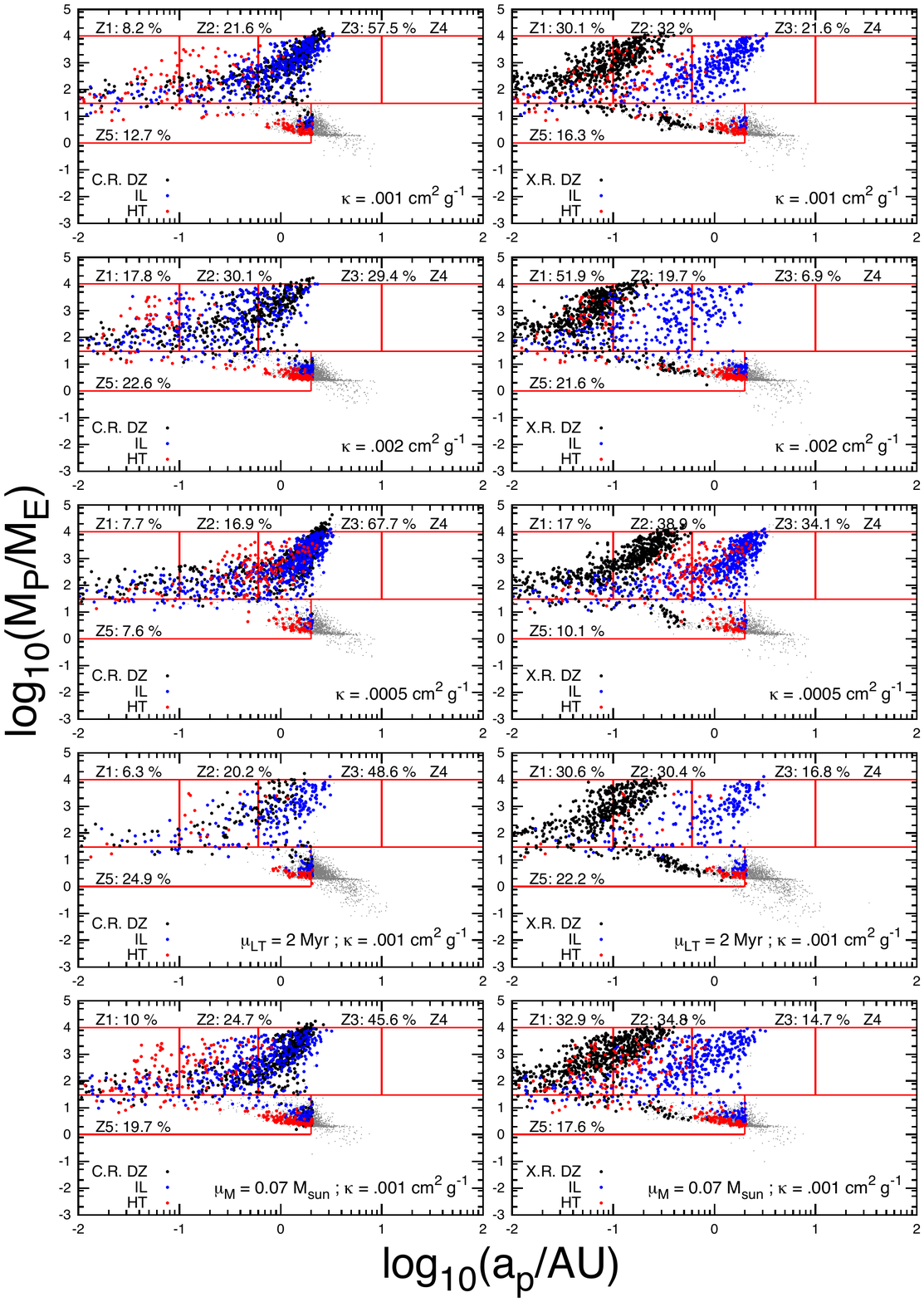}
\caption{Observationally-filtered planet populations are shown whereby disk metallicity is stochastically selected, along with disk lifetime and initial mass. In the top three rows, we consider different envelope opacity values of 0.001, 0.002, and 5$\times10^{-4}$ cm$^2$ g$^{-1}$ as indicated in each panel. In the bottom two rows, we shift the mean of the disk lifetime and initial mass distribution, respectively, to -1$\sigma$ of the mean.}
\label{VarZ_Pops}
\end{figure*}

In figure \ref{VarZ_Pops}, we show populations resulting from a stochastic variation of disk metallicity, lifetime, and initial mass corresponding to their observationally constrained distributions. In the upper three rows of this figure, we consider different envelope opacities; the fiducial value of 0.001 cm$^2$ g$^{-1}$ (top row), as well as a higher setting of $\kappa_{\rm{env}}=0.002$ cm$^2$ g$^{-1}$  (second row), and a lower setting of $5\times10^{-4}$ cm$^2$ g$^{-1}$ (third row). 

Similar to section \ref{Results1}, we find that the best setting of $\kappa_{\rm{env}}$ is 0.002 cm$^2$ g$^{-1}$ for models using a cosmic ray dead zone. Lower settings of $\kappa_{\rm{env}}$ result in a very high fraction of zone 3 planets as well as a reduced amount of zone 5 planets due to shorter gas accretion timescales. In all cases, our model forms a significantly smaller fraction of zone 5 planets than are present in the observed distribution.

\begin{figure*}
\centering
\includegraphics[width=0.45\textwidth]{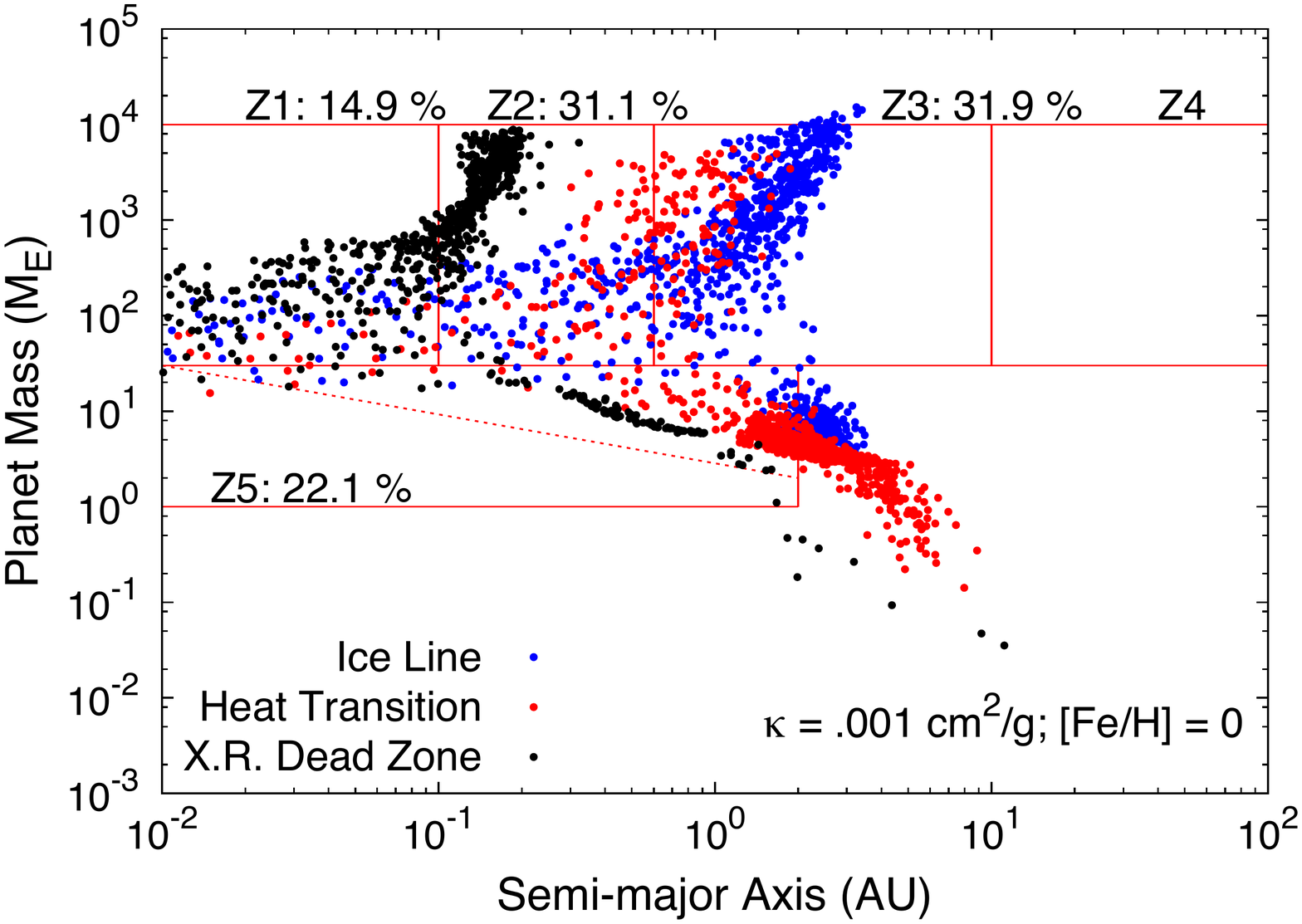} \includegraphics[width=0.45\textwidth]{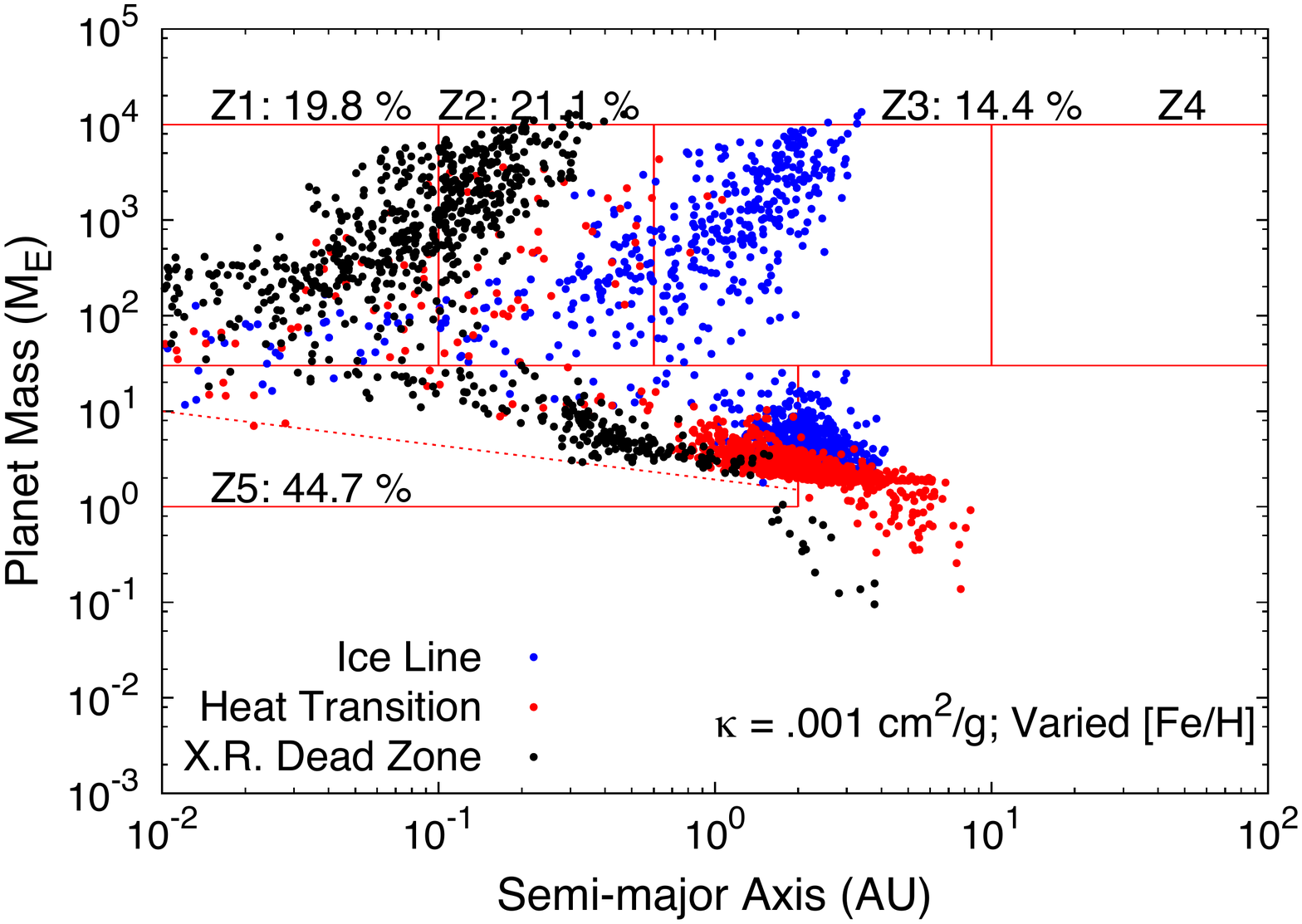}
\caption{Our populations that best correspond to the data are shown, without any observational filtering. Both populations use an X-ray ionization model and resulting dead zone structure, as well as a low envelope opacity of 0.001 cm$^2$ g$^{-1}$. The left panel corresponds to a Solar-metallicity disk for the entire population, and the right panel corresponds to stochastically-sampled disk metallicities. In both cases, our model does not populate a low-mass, low period region of zone 5 that we have identified for each population. The corresponding populations that have been corrected for observational biases are shown in figure \ref{SolarZ_Pops}, top right panel, and figure \ref{VarZ_Pops}, top right panel, respectively.}
\label{RawPops}
\end{figure*}

%Best setting for X-ray
In our populations that stochastically vary the disk metallicity, we again find that low settings of envelope opacity produce the best results when we consider an X-ray ionized disk. In the case of $\kappa_{\rm{env}}$ = 0.002 cm$^2$ g$^{-1}$, our model produces too few zone 3 Jupiters. Lower opacity settings are more optimal for X-ray dead zone populations as they result in a larger fraction of the Jupiter-mass planets having orbital radii larger than 1 AU. 

However, as was the case for the cosmic ray dead zone models, our populations form a much smaller fraction of super Earths and Neptunes compared to the observations. In the case of our best-fit $\kappa_{\rm{env}}$ = 0.001 cm$^2$ g$^{-1}$ setting for X-ray dead zone models from section \ref{Results1}, we only obtain a zone 5 fraction of 16.1 \%. For the lowest opacity setting considered, $\kappa_{\rm{env}} = 5 \times 10^{-4}$ cm$^2$ g$^{-1}$, the fraction of zone 5 planets further diminishes to roughly 10 \% due to short gas accretion timescales that favour the formation of Jupiters. We therefore consider $\kappa_{\rm{env}}$ = 0.001 cm$^2$ g$^{-1}$ to remain our optimal setting for X-ray dead zone populations that stochastically vary the disk metallicity.

In the bottom two rows of figure \ref{VarZ_Pops}, we show populations resulting from stochastically sampling modified disk lifetime and mass distributions. In the fourth row, we change the mean disk lifetime to 2 Myr, and in the bottom row, we reduce the mean initial disk mass to 0.07 M$_\odot$ (corresponding to -1$\sigma$ in its distribution). In both cases, we consider an envelope opacity of 0.001 $cm^2$ g$^{-1}$, and make no modification to the disk metallicity distribution we consider. 

Our motivation for modifying the disk lifetime and mass distributions is two-fold. First, since these distributions are somewhat poorly constrained observationally, we want to determine how robust our population results are to  $\sim 1 \sigma$ changes in the distributions' means which may result from future observations. Second, both these changes favour the formation of zone 5 planets (see discussion below), and we want to determine if such changes improve the fraction of zone 5 planets our model is able to form.

In the case of a reduced mean disk lifetime of 2 Myr (figure \ref{VarZ_Pops}, fourth row), the planets on average have less time to form, and this favours the formation of zone 5 planets. Comparing with the top row, we see that the zone 5 frequency has increased by roughly 6 \% which is not a drastic change. The frequencies within all of the zones do not change significantly, the total number of planets that populate zones has reduced. Decreasing the mean disk lifetime to 2 Myr results in an even more significant pile-up of planets just outside of zone 5 - masses between 1 and 20 M$_\oplus$ orbiting between 2 and 5 AU. Since these planets have low observation probabilities, they get filtered out of the population. 

When considering a reduced mean initial disk mass of 0.07 M$_\odot$ (figure \ref{VarZ_Pops}, bottom row), the resulting populations do have a slightly larger fraction of zone 5 planets. However, as was the case with a reduced mean disk lifetime, the fraction of planets within all of the zones does not change drastically when compared to the fiducial distributions. This shows that the results of our calculations, in terms of formation frequencies among the various zones, are robust to $\sim 1 \sigma$ changes in the mean of the disk lifetime and initial disk mass distributions.

%Best fit populations unfiltered. zone 5 fraction and distribution. zone of avoidance
In figure \ref{RawPops}, we highlight results that best approximate the planetary populations in the observed M-a diagram. Specifically, we show two populations that have no corrections for observational biases, and are direct outputs from our planet population synthesis model. The populations chosen give the best comparison to observations, and both use the X-ray dead zone, and an envelope opacity of 0.001 cm$^2$ g$^{-1}$. The left panel considers a Solar metallicity, while the right panel stochastically varies the disk metallicity within the population in accordance with equation \ref{NormalDist}.

These populations show that, before correcting our populations for observational biases, our model is capable of forming a significant fraction of zone 5 planets. The majority of our failed cores are situated between 1 and 5 AU and typically have low observation probabilities and are therefore likely to get filtered out. These fractions are still lower than the observed distribution's zone 5 fraction, but are much greater than the fractions that are obtained for populations that are corrected for observational biases.

In the case of a Solar metallicity disk (figure \ref{RawPops}, left panel), all zone 5 planets formed within $\sim$ 0.3 AU are $\gtrsim$ 20 M$_\oplus$, and are typically $\gtrsim$ 3 M$_\oplus$ near 1-2 AU. These represent the lowest mass planets our model can form that populate zone 5 using a Solar-metallicity disk. 

When considering the observed disk metallicity distribution (figure \ref{RawPops}, right panel), the resulting planet population appears less clustered and more scattered. In this case, we can form slightly lower-mass planets due to the sub-Solar metallicity portion of this distribution resulting in planets with lower critical core masses during their formation. The low-metallicity portion of the distribution results in planets that are $\gtrsim$ 8 M$_\oplus$ within 0.3 AU and $\gtrsim$ 2 M$_\oplus$ between 1-2 AU. In either case, there remains a large portion of zone 5 that is a zone of avoidance for super Earths in our core accretion - planet trap picture.  We address this and other general results below.

\section{Discussion} \label{Discussion}

%Distribution discussion: Talk about your computed distributions; X-ray vs CR Jupiter distribution & Cummings 2008? Why do we think X-rays are more appropriate. Shielding mentioned by Michigan group
Our population synthesis results show that the planet formation model used in this work is able to form the range of different planetary classes seen in observations. There are, however, regions of the planet mass semi-major axis diagram where our model does not produce planets. 

Our models are capable of forming Jovian planets for semi-major axes $\lesssim$ 2-3 AU, but do not produce any at larger orbital radii. The observed radial frequency distribution of Jovian planets, which peaks at roughly 3-10 AU, indicates that gas giants are less common at larger orbital radii \citep{Cumming2008, Bryan2016}. Thus, our populations agree with this result, as our model produces Jovian planets with $a_p \lesssim$ 3 AU, but does not produce longer period gas giants.

We separately consider X-rays (originating due to magnetospheric accretion) and interstellar cosmic rays as disk ionization sources. The different corresponding ionization rates result in notably different dead zone radii and evolution, leading to significantly different planet populations. The gas giants produced in the cosmic ray dead zone trap display a continuously increasing frequency with orbital radius out to 2 or 3 AU. The X-ray dead zone populations, conversely, show a double-peaked period distribution of gas giants as the X-ray dead zone tends to form hot Jupiters while the ice line forms gas giants at larger orbital radii near 1 AU. These models that include an X-ray dead zone offer an explanation to the observed reduction in frequency of gas giants between 0.1-0.6 AU \citep{Cumming2008}.

As discussed, for example, in \citet{Matt2005} and \citet{Frank2014}, young stars are associated with both disk and stellar (accretion powered) MHD winds that produce shocks that scatter cosmic rays and prevent them from reaching the disk \citep*{Cleeves2013}. Moreover, \citet{Cleeves2015} show that reduced cosmic ray ionization (a consequence of exclusion by disk winds) produce disk chemistry features in agreement with TW Hya observations. Our population results showing that X-ray ionized models produce populations that better reproduce observational features are in agreement with these results.
%There is a general point of physics here ... MHD disk winds will have shocks that scatter CRs... prventing them from reaching the disk.  All young stars are associated with MHD disk winds (see reviews Pudritz+2007, Frank+2014).   Moreover, the disk chemistry result (Cleeves + ) show that X-ray ionization gives better match to the disk chemistry.   So our results show that this does a better job on the populations, too. 

% lack of zone 5 planets - not picked up by observations
One key difference in our super Earth populations and the observations is that our model produces fewer close-in super Earths.  When comparing the direct population outputs (figure \ref{RawPops}) with the populations that account for observational biases (figure \ref{VarZ_Pops}), we see that a portion of the zone 5 planets our model are filtered out as they have low observation probabilities. Additionally, our model does produce a large number of failed cores, but a significant fraction of these orbit just outside the defined outer observable limit of 2 AU, and are situated on orbits out to 5 AU. The range of radii between 1-5 AU is the most common region for super Earths to form in our model as this is the region where the ice line, heat transition, and cosmic ray dead zone traps converge to. Thus, our model predicts there to be a significant population of super Earths with orbital radii of 1-5 AU.

%Zone of avoidance in population connection with HP16 - physical reason for this and implications. 
Our planet formation model does not produce low mass super Earths that migrate into short-period orbits. We noted this several times throughout section \ref{Results} and refer to this as a zone of avoidance of our model within zone 5 of the mass-semimajor axis diagram. The lowest mass super Earths formed in our model that populate zone 5 are $\sim$ 3 M$_\oplus$ and orbit outside of 1 AU. This feature of our populations is tied to the evolution of the traps, that evolve from orbital radii outside $\sim$ 8 AU to within $\sim$ 1 AU on disk evolution timescales. This prevents $\sim$ Earth-mass planets from migrating to short period orbits in our model as they evolve with the trap they are forming within. 

Planets can, however, become liberated from their host planet traps at the onset of type-II migration. In this case, planets must exceed their gap opening mass (equation \ref{GapOpeningMass}) of $\sim$ 10 M$_\oplus$ to type-II migrate separately from their trap. Initially, type-II migration takes place on shorter timescales than trapped type-I migration, so these planets can, in some cases migrate to short period orbits during their slow gas accretion phase before their accretion is terminated at the disk lifetime. Thus, the lowest mass zone 5 planets orbiting between 0.01-0.1 AU formed in our model have masses $\gtrsim$ 10 M$_\oplus$.

\citet{Hasegawa2016} also found that a core accretion model involving trapped type-I migration could not form a failed core population consistent with observations. In particular, this model could only form super Earths more massive than $\sim 4-5$ M$_\oplus$, similar to the results found in this work. 

Based on the results of our constant dust-to-gas ratio model, one might suggest that formation scenarios in addition to the failed core scenario are necessary to explain the observed super Earth population. In particular, \citet{Hasegawa2016} suggested that embryo assembly (similar to the mechanism by which the Solar System terrestrials formed) may produce super Earth populations that, in combination with the failed core scenario, compare well with observations.

We emphasize, however, that we have not included the effects that radial drift of dust may have on our populations. \citet{Birnstiel2012} and \citet{Cridland2017} have shown that dust evolution models result in depleted dust-to-gas ratios outside of the ice line. Based on these results, we predict that planet formation in the outer disk (for example, in the heat transition trap) will result in longer oligarchic growth timescales and correspondingly smaller $M_{\rm{c,crit}}$ values (see equation \ref{CoreCrit}). As discussed in section \ref{Results2}, this will in turn result in longer gas accretion timescales allowing migration to transport these planets to smaller orbital radii prior to disk dissipation. We expect that including the effects of radial drift and dust evolution will populate the zone of avoidance by moving the region of planet formation inward. We will fully explore these effects in an upcoming publication.

%How saturation of type I migration could change this. Connect to mordasini stuff
In this work, we only consider two migration regimes: trapped type-I migration and type-II migration. However, it is possible that there is an additional migration regime for which the corotation torque saturates prior to the onset of type-II migration \citep{Dittkrist2014}. In this case, the planet would no longer be trapped in the disk inhomogeneity, and would type-I migrate to inner regions of the disk on short timescales before reaching its gap opening mass. If this phase were applicable to planets in our model, it would greatly affect our population results.

In \citet{Alessi2017}, we perform an analysis to confirm that planets reach their gap-opening masses prior to their saturation-masses in our model, thereby confirming that the trapped phase is applicable throughout the entirety of the type-I migration regime in our model. There is however, a notable similarity in the gap-opening masses of planets in our model and typical saturation masses. As both of these quantities depend sensitively on the planet and local disk conditions, a model that considers in detail the planet-disk interactions is a prospect for future work (which is considered, for example, in \citet{HellaryNelson2012}, and Cridland, Pudritz, \& Alessi (2018), in prep.).

%Could dynamics change our results? How? Early stages (Hellary & Nelson) along with detailed planet-disk dynamics (Cridland + 2018). Late stages between the traps themselves (ie. Ida & Lin 2013? Paper from planets course).
We restrict our model to consider only the case of planet formation in isolation, and do not account for the dynamical interaction between planetary cores. We do this to provide a clear bench mark against which we can compare future calculations that include the effects of planet-planet interaction. We note that the traps in our model converge and intersect throughout disk evolution, and thus dynamic interactions would take place between forming planets. Planet-planet dynamics during planet formation (as was previously considered, for example, by \citet{HellaryNelson2012} and \citet*{Ida2013}) is another prospect for future work to determine the effects of resonances and scattering on our computed planet populations.

%Could dust evolution change this? How?
%Disk alpha - effect on our results
An additional parameter we have not varied that could affect our results is the disk $\alpha$, which we held to a constant value of 10$^{-3}$. If we were to consider larger estimates of $\alpha$ up to 10$^{-2}$, the following effects would take place: (i) disk evolution would take place on a shorter timescale, reducing the surface density quicker, causing planet formation to be less efficient; and (ii) inward migration rates would be faster. \citet{HP13} included a parameter study of $\alpha$ in their population synthesis work including planet traps. This work indeed showed that gas giants formed less efficiently when larger $\alpha$ values were considered. Additionally, zone 3 Jovians formed less frequently in populations when higher values of $\alpha$ were considered, while the zone 1 hot Jupiter population frequency increased with $\alpha$ - both as a result of the increased migration rates. Notably, the zone 5 population frequency was found to be insensitive to the particular setting of $\alpha$ within its estimated range of $10^{-3}-10^{-2}$.

%Opacity result (reduced ISM) dating back to Pollack but even more reduced like in Mordasini 2014. Factors affecting this. Connect with Mordasini 2014b result. How will future dust evolution help us with this?

We obtained the best comparison between computed populations and the observed distribution when considering low envelope opacities $\sim$ 10$^{-3}$ cm$^2$ g$^{-1}$. Our focus in this paper has not been to calculate $\kappa_{\rm{env}}$ values as they are acquired from the disk and processed in the forming planet's envelope, but rather to constrain its value by comparing resulting planet populations with the observed distribution. Calculating this directly would require one to track the dust grain sizes and compositions as they are accreted onto a planet, as well as how they are processed in the forming planet's envelope, which is beyond the scope of this paper. We note that one value of $\kappa_{\rm{env}}$ will not describe a planet's atmosphere throughout the entirety of its formation (as its opacity will change as a result of its accretion history), let alone describe an entire population of planets. However, by comparing our computed populations to observations we have constrained this parameter that sets gas accretion timescales.

The earliest core accretion models found that an envelope opacity $\sim$ 1 \% of the ISM opacity was necessary to achieve reasonable planet formation timescales \citep{Pollack1996}. Our results indicate  that an even further reduced envelope opacity of $\sim$ 0.1 \% the ISM opacity produces the best results, which is similar to the more recent result of \citet{Mordasini2014}, who found a reduction factor of 0.3 \% of the ISM opacity gave the best comparison of their computed populations with observations. It is notable that this result indicates that the best fit $\kappa_{\rm{env}}$ value is a factor of 1000 lower than the typical opacity assumed in our disk model for the case of Solar metallicity, which assumes a $\kappa_0$ of 3 cm$^2$ g$^{-1}$ \citep{Chambers2009}. One possible explanation is that significant grain growth takes place within envelopes of forming planets to achieve a much lower opacity than the surrounding disk material. This was shown to be the case in \citet{Mordasini2014b}, who found that efficient grain growth takes place over a significant portion of a planet atmosphere, leading to low optical depths consistent of a low $\kappa_{\rm{env}}$ value of 0.3 \% of the ISM opacity.

%%%%%%%%%%%%%%%%%%%%%%%%%%%%%%%%%%%%%%%%%%%%%%%%%%%%%%%%%%%%%%%

\section{Conclusions} \label{Conclusion}

In this paper, we have computed a comprehensive suite of planet population synthesis calculations, each involving three thousand model evolutionary tracks, that incorporate the observationally constrained distributions of disk lifetimes, masses, and metallicities in outcomes of a core accretion model with trapped type-I migration. Planets formed in our model are not free to migrate into the inner disk, but are rather tied to the radial locations of the planet traps they form within for all stages prior to type-II migration. The main focus of this paper has been to study the effects of forming planets' envelope opacities, as well as disk metallicities, on our results. We list our main conclusions here:

\begin{itemize}

\item We separately considered X-ray and cosmic ray disk ionization and resulting dead zones in our populations. These two traps produce significantly different populations as the X-ray dead zone tends to form massive planets within 0.5 AU while the cosmic ray dead zone forms Jovian planets near or outside 1 AU. When combined with the planets formed in the ice line and heat transition, populations involving a cosmic ray dead zone form a distribution of Jovian planets whose frequency increases with orbital radii out to $\sim$ 2 AU. Conversely, populations including an X-ray dead zone features a double-peaked distribution of Jupiters, with an inner peak within 0.3 AU consisting mainly of planets formed in the X-ray dead zone and an outer peak near 1 AU corresponding to planets formed within the ice line. Populations using an X-ray dead zone are the only ones that produce such a double-peaked feature in the Jovian planets' radial distribution that is seen in observations.

%X-Ray vs. CR DZ Populations

%Metallicity Dist.
\item We find that Jovian planets form less frequently in disks with sub-Solar metallicity. In such disks, failed cores (super Earths or Neptunes) are much more common. Our populations show that the formation frequency of Jovian planets is correlated with disk metallicity, in agreement with the planet-metallicity relation \citep{Fischer2005}.

%Zone of avoidance
\item Our planet formation model does not produce low-mass, low-period super Earths. The lowest mass planets formed in our model orbiting near 0.01 AU are $\sim$ 10 M$_\oplus$, and the lowest mass planets orbiting near 0.1 AU are $\sim$ 6 M$_\oplus$. The traps are located outside of 5 AU in the earliest stages of disk evolution, and converge to $\sim$ 1 AU within a typical 3 Myr disk lifetime, with the exception of the X-ray dead zone which evolves to $\sim 0.05$ AU. We suggest that this is a consequence of our assumption of a constant dust to gas ratio.

\item Since planet migration to small orbital radii are limited to large, disk evolution timescales, accretion throughout this process prevents our model from forming low-mass and low-period super Earths that have been observed. Based on the results of our constant dust-to-gas ratio model, we suggest these planets to have formed via a different process such as collisional growth, similar to the process that formed the Solar System's terrestrial planets. However, the radial drift of dust grains may result in a super Earth population that is much more concentrated towards small disk radii.  We defer this analysis to our next paper.

%Opacity conclusion
\item We find that low envelope opacities of $\sim 10^{-3}$ cm$^2$ g$^{-1}$ are necessary Jovian planets with orbital radii $\geq$ 1 AU in our populations. Higher envelope opacities (greater than 0.003 cm$^2$ g$^{-1}$) cause for longer formation timescales, whereby Jovian planets typically migrate within 0.6 AU prior to undergoing runaway growth. In populations considering an X-ray dead zone, we find an optimal setting of 0.001 cm$^2$ g$^{-1}$.

\end{itemize}

In our next paper in this series, we will examine the effect of radial drift of dust in the disk on planet formation and on the structure of planetary populations. 

\section*{Acknowledgements}
The authors thank Alex J. Cridland for fruitful discussions regarding this work. We also thank the anonymous referee for their helpful and insightful  comments. M.A. acknowledges funding from the National Sciences and engineering Research Council (NSERC) through the Alexander Graham Bell CGS/PGS Doctoral Scholarship. R.E.P. is supported by an NSERC Discovery Grant.This research has made use of the Exoplanet Orbit Database and the Exoplanet Data Explorer at exoplanets.org.

%%%%%%%%%%%%%%%%%%%%%%%%%%%%%%%%%%%%%%%%%%%%%%%%%%

%%%%%%%%%%%%%%%%%%%% REFERENCES %%%%%%%%%%%%%%%%%%

\bibliographystyle{mnras}
\bibliography{research}

\begin{thebibliography}{}
\makeatletter
\relax
\def\mn@urlcharsother{\let\do\@makeother \do\$\do\&\do\#\do\^\do\_\do\%\do\~}
\def\mn@doi{\begingroup\mn@urlcharsother \@ifnextchar [ {\mn@doi@}
  {\mn@doi@[]}}
\def\mn@doi@[#1]#2{\def\@tempa{#1}\ifx\@tempa\@empty \href
  {http://dx.doi.org/#2} {doi:#2}\else \href {http://dx.doi.org/#2} {#1}\fi
  \endgroup}
\def\mn@eprint#1#2{\mn@eprint@#1:#2::\@nil}
\def\mn@eprint@arXiv#1{\href {http://arxiv.org/abs/#1} {{\tt arXiv:#1}}}
\def\mn@eprint@dblp#1{\href {http://dblp.uni-trier.de/rec/bibtex/#1.xml}
  {dblp:#1}}
\def\mn@eprint@#1:#2:#3:#4\@nil{\def\@tempa {#1}\def\@tempb {#2}\def\@tempc
  {#3}\ifx \@tempc \@empty \let \@tempc \@tempb \let \@tempb \@tempa \fi \ifx
  \@tempb \@empty \def\@tempb {arXiv}\fi \@ifundefined
  {mn@eprint@\@tempb}{\@tempb:\@tempc}{\expandafter \expandafter \csname
  mn@eprint@\@tempb\endcsname \expandafter{\@tempc}}}

\bibitem[\protect\citeauthoryear{{Alessi}, {Pudritz}  \& {Cridland}}{{Alessi}
  et~al.}{2017}]{Alessi2017}
{Alessi} M.,  {Pudritz} R.~E.,   {Cridland} A.~J.,  2017, \mn@doi [\mnras]
  {10.1093/mnras/stw2360}, \href
  {http://adsabs.harvard.edu/abs/2017MNRAS.464..428A} {464, 428}

\bibitem[\protect\citeauthoryear{{Ali-Dib}}{{Ali-Dib}}{2017}]{AliDib2017}
{Ali-Dib} M.,  2017, \mn@doi [\mnras] {10.1093/mnras/stw2651}, \href
  {http://adsabs.harvard.edu/abs/2017MNRAS.464.4282A} {464, 4282}

\bibitem[\protect\citeauthoryear{{Ali-Dib}, {Johansen}  \& {Huang}}{{Ali-Dib}
  et~al.}{2017}]{AliDib2017b}
{Ali-Dib} M.,  {Johansen} A.,   {Huang} C.~X.,  2017, \mn@doi [\mnras]
  {10.1093/mnras/stx1272}, \href
  {http://adsabs.harvard.edu/abs/2017MNRAS.469.5016A} {469, 5016}

\bibitem[\protect\citeauthoryear{{Alibert}}{{Alibert}}{2017}]{Alibert2017}
{Alibert} Y.,  2017, \mn@doi [\aap] {10.1051/0004-6361/201630051}, \href
  {http://adsabs.harvard.edu/abs/2017A%26A...606A..69A} {606, A69}

\bibitem[\protect\citeauthoryear{{Alibert}, {Mordasini}  \& {Benz}}{{Alibert}
  et~al.}{2004}]{Alibert2004}
{Alibert} Y.,  {Mordasini} C.,   {Benz} W.,  2004, \mn@doi [\aap]
  {10.1051/0004-6361:20040053}, \href
  {http://adsabs.harvard.edu/abs/2004A%26A...417L..25A} {417, L25}

\bibitem[\protect\citeauthoryear{{Alibert}, {Mordasini}, {Benz}  \&
  {Winisdoerffer}}{{Alibert} et~al.}{2005}]{Alibert2005}
{Alibert} Y.,  {Mordasini} C.,  {Benz} W.,   {Winisdoerffer} C.,  2005, \mn@doi
  [\aap] {10.1051/0004-6361:20042032}, \href
  {http://adsabs.harvard.edu/abs/2005A%26A...434..343A} {434, 343}

\bibitem[\protect\citeauthoryear{{Alibert}, {Mordasini}  \& {Benz}}{{Alibert}
  et~al.}{2011}]{Alibert2011}
{Alibert} Y.,  {Mordasini} C.,   {Benz} W.,  2011, \mn@doi [\aap]
  {10.1051/0004-6361/201014760}, \href
  {http://adsabs.harvard.edu/abs/2011A%26A...526A..63A} {526, A63}

\bibitem[\protect\citeauthoryear{{Bai}}{{Bai}}{2016}]{Bai2016}
{Bai} X.-N.,  2016, \mn@doi [\apj] {10.3847/0004-637X/821/2/80}, \href
  {http://adsabs.harvard.edu/abs/2016ApJ...821...80B} {821, 80}

\bibitem[\protect\citeauthoryear{{Bai} \& {Stone}}{{Bai} \&
  {Stone}}{2013}]{BaiStone2013}
{Bai} X.-N.,  {Stone} J.~M.,  2013, \mn@doi [\apj]
  {10.1088/0004-637X/769/1/76}, \href
  {http://adsabs.harvard.edu/abs/2013ApJ...769...76B} {769, 76}

\bibitem[\protect\citeauthoryear{{Bailli{\'e}}, {Charnoz}  \&
  {Pantin}}{{Bailli{\'e}} et~al.}{2016}]{Baillie2016}
{Bailli{\'e}} K.,  {Charnoz} S.,   {Pantin} E.,  2016, \mn@doi [\aap]
  {10.1051/0004-6361/201528027}, \href
  {http://adsabs.harvard.edu/abs/2016A%26A...590A..60B} {590, A60}

\bibitem[\protect\citeauthoryear{{Bashi}, {Helled}, {Zucker}  \&
  {Mordasini}}{{Bashi} et~al.}{2017}]{Bashi2017}
{Bashi} D.,  {Helled} R.,  {Zucker} S.,   {Mordasini} C.,  2017, \mn@doi [\aap]
  {10.1051/0004-6361/201629922}, \href
  {http://adsabs.harvard.edu/abs/2017A%26A...604A..83B} {604, A83}

\bibitem[\protect\citeauthoryear{{Batalha} et~al.,}{{Batalha}
  et~al.}{2013}]{Batalha2013}
{Batalha} N.~M.,  et~al., 2013, \mn@doi [\apjs] {10.1088/0067-0049/204/2/24},
  \href {http://adsabs.harvard.edu/abs/2013ApJS..204...24B} {204, 24}

\bibitem[\protect\citeauthoryear{{Birnstiel}, {Klahr}  \&
  {Ercolano}}{{Birnstiel} et~al.}{2012}]{Birnstiel2012}
{Birnstiel} T.,  {Klahr} H.,   {Ercolano} B.,  2012, \mn@doi [\aap]
  {10.1051/0004-6361/201118136}, \href
  {http://adsabs.harvard.edu/abs/2012A%26A...539A.148B} {539, A148}

\bibitem[\protect\citeauthoryear{{Bitsch}, {Lambrechts}  \&
  {Johansen}}{{Bitsch} et~al.}{2015}]{Bitsch2015}
{Bitsch} B.,  {Lambrechts} M.,   {Johansen} A.,  2015, \mn@doi [\aap]
  {10.1051/0004-6361/201526463}, \href
  {http://adsabs.harvard.edu/abs/2015A%26A...582A.112B} {582, A112}

\bibitem[\protect\citeauthoryear{{Blaes} \& {Balbus}}{{Blaes} \&
  {Balbus}}{1994}]{Blaes1994}
{Blaes} O.~M.,  {Balbus} S.~A.,  1994, \mn@doi [\apj] {10.1086/173634}, \href
  {http://adsabs.harvard.edu/abs/1994ApJ...421..163B} {421, 163}

\bibitem[\protect\citeauthoryear{{Borucki} et~al.,}{{Borucki}
  et~al.}{2011}]{Borucki2011}
{Borucki} W.~J.,  et~al., 2011, \mn@doi [\apj] {10.1088/0004-637X/736/1/19},
  \href {http://adsabs.harvard.edu/abs/2011ApJ...736...19B} {736, 19}

\bibitem[\protect\citeauthoryear{{Brouwers}, {Vazan}  \& {Ormel}}{{Brouwers}
  et~al.}{2018}]{Brouwers2018}
{Brouwers} M.~G.,  {Vazan} A.,   {Ormel} C.~W.,  2018, \mn@doi [\aap]
  {10.1051/0004-6361/201731824}, \href
  {http://adsabs.harvard.edu/abs/2018A%26A...611A..65B} {611, A65}

\bibitem[\protect\citeauthoryear{{Bryan} et~al.,}{{Bryan}
  et~al.}{2016}]{Bryan2016}
{Bryan} M.~L.,  et~al., 2016, \mn@doi [\apj] {10.3847/0004-637X/821/2/89},
  \href {http://adsabs.harvard.edu/abs/2016ApJ...821...89B} {821, 89}

\bibitem[\protect\citeauthoryear{{Burke} et~al.,}{{Burke}
  et~al.}{2014}]{Burke2014}
{Burke} C.~J.,  et~al., 2014, \mn@doi [\apjs] {10.1088/0067-0049/210/2/19},
  \href {http://adsabs.harvard.edu/abs/2014ApJS..210...19B} {210, 19}

\bibitem[\protect\citeauthoryear{{Cassan} et~al.,}{{Cassan}
  et~al.}{2012}]{Cassan2012}
{Cassan} A.,  et~al., 2012, \mn@doi [\nat] {10.1038/nature10684}, \href
  {http://adsabs.harvard.edu/abs/2012Natur.481..167C} {481, 167}

\bibitem[\protect\citeauthoryear{{Chambers}}{{Chambers}}{2009}]{Chambers2009}
{Chambers} J.~E.,  2009, \mn@doi [\apj] {10.1088/0004-637X/705/2/1206}, \href
  {http://adsabs.harvard.edu/abs/2009ApJ...705.1206C} {705, 1206}

\bibitem[\protect\citeauthoryear{{Chiang} \& {Laughlin}}{{Chiang} \&
  {Laughlin}}{2013}]{ChiangLaughlin2013}
{Chiang} E.,  {Laughlin} G.,  2013, \mn@doi [\mnras] {10.1093/mnras/stt424},
  \href {http://adsabs.harvard.edu/abs/2013MNRAS.431.3444C} {431, 3444}

\bibitem[\protect\citeauthoryear{{Clarke}}{{Clarke}}{2007}]{Clarke2007}
{Clarke} C.~J.,  2007, \mn@doi [\mnras] {10.1111/j.1365-2966.2007.11547.x},
  \href {http://adsabs.harvard.edu/abs/2007MNRAS.376.1350C} {376, 1350}

\bibitem[\protect\citeauthoryear{{Cleeves}, {Adams}  \& {Bergin}}{{Cleeves}
  et~al.}{2013}]{Cleeves2013}
{Cleeves} L.~I.,  {Adams} F.~C.,   {Bergin} E.~A.,  2013, \mn@doi [\apj]
  {10.1088/0004-637X/772/1/5}, \href
  {http://adsabs.harvard.edu/abs/2013ApJ...772....5C} {772, 5}

\bibitem[\protect\citeauthoryear{{Cleeves}, {Bergin}, {Qi}, {Adams}  \&
  {{\"O}berg}}{{Cleeves} et~al.}{2015}]{Cleeves2015}
{Cleeves} L.~I.,  {Bergin} E.~A.,  {Qi} C.,  {Adams} F.~C.,   {{\"O}berg}
  K.~I.,  2015, \mn@doi [\apj] {10.1088/0004-637X/799/2/204}, \href
  {http://adsabs.harvard.edu/abs/2015ApJ...799..204C} {799, 204}

\bibitem[\protect\citeauthoryear{{Coleman} \& {Nelson}}{{Coleman} \&
  {Nelson}}{2016}]{Coleman2016b}
{Coleman} G.~A.~L.,  {Nelson} R.~P.,  2016, \mn@doi [\mnras]
  {10.1093/mnras/stw1177}, \href
  {http://adsabs.harvard.edu/abs/2016MNRAS.460.2779C} {460, 2779}

\bibitem[\protect\citeauthoryear{{Crida}, {Morbidelli}  \& {Masset}}{{Crida}
  et~al.}{2006}]{Crida2006}
{Crida} A.,  {Morbidelli} A.,   {Masset} F.,  2006, \mn@doi [\icarus]
  {10.1016/j.icarus.2005.10.007}, \href
  {http://adsabs.harvard.edu/abs/2006Icar..181..587C} {181, 587}

\bibitem[\protect\citeauthoryear{{Cridland}, {Pudritz}  \& {Alessi}}{{Cridland}
  et~al.}{2016}]{Cridland2016}
{Cridland} A.~J.,  {Pudritz} R.~E.,   {Alessi} M.,  2016, \mn@doi [\mnras]
  {10.1093/mnras/stw1511}, \href
  {http://adsabs.harvard.edu/abs/2016MNRAS.461.3274C} {461, 3274}

\bibitem[\protect\citeauthoryear{{Cridland}, {Pudritz}  \&
  {Birnstiel}}{{Cridland} et~al.}{2017}]{Cridland2017}
{Cridland} A.~J.,  {Pudritz} R.~E.,   {Birnstiel} T.,  2017, \mn@doi [\mnras]
  {10.1093/mnras/stw2946}, \href
  {http://adsabs.harvard.edu/abs/2017MNRAS.465.3865C} {465, 3865}

\bibitem[\protect\citeauthoryear{{Cumming}}{{Cumming}}{2004}]{Cumming2004}
{Cumming} A.,  2004, \mn@doi [\mnras] {10.1111/j.1365-2966.2004.08275.x}, \href
  {http://adsabs.harvard.edu/abs/2004MNRAS.354.1165C} {354, 1165}

\bibitem[\protect\citeauthoryear{{Cumming}, {Butler}, {Marcy}, {Vogt}, {Wright}
   \& {Fischer}}{{Cumming} et~al.}{2008}]{Cumming2008}
{Cumming} A.,  {Butler} R.~P.,  {Marcy} G.~W.,  {Vogt} S.~S.,  {Wright} J.~T.,
   {Fischer} D.~A.,  2008, \mn@doi [\pasp] {10.1086/588487}, \href
  {http://adsabs.harvard.edu/abs/2008PASP..120..531C} {120, 531}

\bibitem[\protect\citeauthoryear{{Dittkrist}, {Mordasini}, {Klahr}, {Alibert}
  \& {Henning}}{{Dittkrist} et~al.}{2014}]{Dittkrist2014}
{Dittkrist} K.-M.,  {Mordasini} C.,  {Klahr} H.,  {Alibert} Y.,   {Henning} T.,
   2014, \mn@doi [\aap] {10.1051/0004-6361/201322506}, \href
  {http://adsabs.harvard.edu/abs/2014A%26A...567A.121D} {567, A121}

\bibitem[\protect\citeauthoryear{{Ercolano} \& {Clarke}}{{Ercolano} \&
  {Clarke}}{2010}]{Ercolano2010}
{Ercolano} B.,  {Clarke} C.~J.,  2010, \mn@doi [\mnras]
  {10.1111/j.1365-2966.2009.16094.x}, \href
  {http://adsabs.harvard.edu/abs/2010MNRAS.402.2735E} {402, 2735}

\bibitem[\protect\citeauthoryear{{Fischer} \& {Valenti}}{{Fischer} \&
  {Valenti}}{2005}]{Fischer2005}
{Fischer} D.~A.,  {Valenti} J.,  2005, \mn@doi [\apj] {10.1086/428383}, \href
  {http://adsabs.harvard.edu/abs/2005ApJ...622.1102F} {622, 1102}

\bibitem[\protect\citeauthoryear{{Fischer}, {Marcy}  \& {Spronck}}{{Fischer}
  et~al.}{2014}]{Fischer2014}
{Fischer} D.~A.,  {Marcy} G.~W.,   {Spronck} J.~F.~P.,  2014, \mn@doi [\apjs]
  {10.1088/0067-0049/210/1/5}, \href
  {http://adsabs.harvard.edu/abs/2014ApJS..210....5F} {210, 5}

\bibitem[\protect\citeauthoryear{{Flaherty}, {Hughes}, {Teague}, {Simon},
  {Andrews}  \& {Wilner}}{{Flaherty} et~al.}{2018}]{Flaherty2018}
{Flaherty} K.~M.,  {Hughes} A.~M.,  {Teague} R.,  {Simon} J.~B.,  {Andrews}
  S.~M.,   {Wilner} D.~J.,  2018, \mn@doi [\apj] {10.3847/1538-4357/aab615},
  \href {http://adsabs.harvard.edu/abs/2018ApJ...856..117F} {856, 117}

\bibitem[\protect\citeauthoryear{{Frank} et~al.,}{{Frank}
  et~al.}{2014}]{Frank2014}
{Frank} A.,  et~al., 2014, \mn@doi [Protostars and Planets VI]
  {10.2458/azu_uapress_9780816531240-ch020}, \href
  {http://adsabs.harvard.edu/abs/2014prpl.conf..451F} {pp 451--474}

\bibitem[\protect\citeauthoryear{{Gammie}}{{Gammie}}{1996}]{Gammie1996}
{Gammie} C.~F.,  1996, \mn@doi [\apj] {10.1086/176735}, \href
  {http://adsabs.harvard.edu/abs/1996ApJ...457..355G} {457, 355}

\bibitem[\protect\citeauthoryear{{Glassgold}, {Najita}  \& {Igea}}{{Glassgold}
  et~al.}{1997}]{Glassgold1997}
{Glassgold} A.~E.,  {Najita} J.,   {Igea} J.,  1997, \mn@doi [\apj]
  {10.1086/303952}, \href {http://adsabs.harvard.edu/abs/1997ApJ...480..344G}
  {480, 344}

\bibitem[\protect\citeauthoryear{{Goldreich} \& {Tremaine}}{{Goldreich} \&
  {Tremaine}}{1980}]{GoldreichTremaine1980}
{Goldreich} P.,  {Tremaine} S.,  1980, \mn@doi [\apj] {10.1086/158356}, \href
  {http://adsabs.harvard.edu/abs/1980ApJ...241..425G} {241, 425}

\bibitem[\protect\citeauthoryear{{Gorti}, {Dullemond}  \& {Hollenbach}}{{Gorti}
  et~al.}{2009}]{Gorti2009}
{Gorti} U.,  {Dullemond} C.~P.,   {Hollenbach} D.,  2009, \mn@doi [\apj]
  {10.1088/0004-637X/705/2/1237}, \href
  {http://adsabs.harvard.edu/abs/2009ApJ...705.1237G} {705, 1237}

\bibitem[\protect\citeauthoryear{{Gorti}, {Hollenbach}  \& {Dullemond}}{{Gorti}
  et~al.}{2015}]{Gorti2015}
{Gorti} U.,  {Hollenbach} D.,   {Dullemond} C.~P.,  2015, \mn@doi [\apj]
  {10.1088/0004-637X/804/1/29}, \href
  {http://adsabs.harvard.edu/abs/2015ApJ...804...29G} {804, 29}

\bibitem[\protect\citeauthoryear{{Gressel} \& {Pessah}}{{Gressel} \&
  {Pessah}}{2015}]{Gressel2015b}
{Gressel} O.,  {Pessah} M.~E.,  2015, \mn@doi [\apj]
  {10.1088/0004-637X/810/1/59}, \href
  {http://adsabs.harvard.edu/abs/2015ApJ...810...59G} {810, 59}

\bibitem[\protect\citeauthoryear{{Gressel}, {Turner}, {Nelson}  \&
  {McNally}}{{Gressel} et~al.}{2015}]{Gressel2015}
{Gressel} O.,  {Turner} N.~J.,  {Nelson} R.~P.,   {McNally} C.~P.,  2015,
  \mn@doi [\apj] {10.1088/0004-637X/801/2/84}, \href
  {http://adsabs.harvard.edu/abs/2015ApJ...801...84G} {801, 84}

\bibitem[\protect\citeauthoryear{{Han}, {Wang}, {Wright}, {Feng}, {Zhao},
  {Fakhouri}, {Brown}  \& {Hancock}}{{Han} et~al.}{2014}]{Han2014}
{Han} E.,  {Wang} S.~X.,  {Wright} J.~T.,  {Feng} Y.~K.,  {Zhao} M.,
  {Fakhouri} O.,  {Brown} J.~I.,   {Hancock} C.,  2014, \mn@doi [\pasp]
  {10.1086/678447}, \href {http://adsabs.harvard.edu/abs/2014PASP..126..827H}
  {126, 827}

\bibitem[\protect\citeauthoryear{{Hasegawa}}{{Hasegawa}}{2016}]{Hasegawa2016}
{Hasegawa} Y.,  2016, \mn@doi [\apj] {10.3847/0004-637X/832/1/83}, \href
  {http://adsabs.harvard.edu/abs/2016ApJ...832...83H} {832, 83}

\bibitem[\protect\citeauthoryear{{Hasegawa} \& {Pudritz}}{{Hasegawa} \&
  {Pudritz}}{2010}]{HP10}
{Hasegawa} Y.,  {Pudritz} R.~E.,  2010, \mn@doi [\apjl]
  {10.1088/2041-8205/710/2/L167}, \href
  {http://adsabs.harvard.edu/abs/2010ApJ...710L.167H} {710, L167}

\bibitem[\protect\citeauthoryear{{Hasegawa} \& {Pudritz}}{{Hasegawa} \&
  {Pudritz}}{2011}]{HP11}
{Hasegawa} Y.,  {Pudritz} R.~E.,  2011, \mn@doi [\mnras]
  {10.1111/j.1365-2966.2011.19338.x}, \href
  {http://adsabs.harvard.edu/abs/2011MNRAS.417.1236H} {417, 1236}

\bibitem[\protect\citeauthoryear{{Hasegawa} \& {Pudritz}}{{Hasegawa} \&
  {Pudritz}}{2012}]{HP12}
{Hasegawa} Y.,  {Pudritz} R.~E.,  2012, \mn@doi [\apj]
  {10.1088/0004-637X/760/2/117}, \href
  {http://adsabs.harvard.edu/abs/2012ApJ...760..117H} {760, 117}

\bibitem[\protect\citeauthoryear{{Hasegawa} \& {Pudritz}}{{Hasegawa} \&
  {Pudritz}}{2013}]{HP13}
{Hasegawa} Y.,  {Pudritz} R.~E.,  2013, \mn@doi [\apj]
  {10.1088/0004-637X/778/1/78}, \href
  {http://adsabs.harvard.edu/abs/2013ApJ...778...78H} {778, 78}

\bibitem[\protect\citeauthoryear{{Hasegawa} \& {Pudritz}}{{Hasegawa} \&
  {Pudritz}}{2014}]{HP14}
{Hasegawa} Y.,  {Pudritz} R.~E.,  2014, \mn@doi [\apj]
  {10.1088/0004-637X/794/1/25}, \href
  {http://adsabs.harvard.edu/abs/2014ApJ...794...25H} {794, 25}

\bibitem[\protect\citeauthoryear{{Haworth}, {Clarke}  \& {Owen}}{{Haworth}
  et~al.}{2016}]{Haworth2016}
{Haworth} T.~J.,  {Clarke} C.~J.,   {Owen} J.~E.,  2016, \mn@doi [\mnras]
  {10.1093/mnras/stv3016}, \href
  {http://adsabs.harvard.edu/abs/2016MNRAS.457.1905H} {457, 1905}

\bibitem[\protect\citeauthoryear{{Hellary} \& {Nelson}}{{Hellary} \&
  {Nelson}}{2012}]{HellaryNelson2012}
{Hellary} P.,  {Nelson} R.~P.,  2012, \mn@doi [\mnras]
  {10.1111/j.1365-2966.2011.19815.x}, \href
  {http://adsabs.harvard.edu/abs/2012MNRAS.419.2737H} {419, 2737}

\bibitem[\protect\citeauthoryear{{Howard} et~al.,}{{Howard}
  et~al.}{2012}]{Howard2012}
{Howard} A.~W.,  et~al., 2012, \mn@doi [\apjs] {10.1088/0067-0049/201/2/15},
  \href {http://adsabs.harvard.edu/abs/2012ApJS..201...15H} {201, 15}

\bibitem[\protect\citeauthoryear{{Ida} \& {Lin}}{{Ida} \&
  {Lin}}{2004a}]{IdaLin2004}
{Ida} S.,  {Lin} D.~N.~C.,  2004a, \mn@doi [\apj] {10.1086/381724}, \href
  {http://adsabs.harvard.edu/abs/2004ApJ...604..388I} {604, 388}

\bibitem[\protect\citeauthoryear{{Ida} \& {Lin}}{{Ida} \&
  {Lin}}{2004b}]{IdaLin2004b}
{Ida} S.,  {Lin} D.~N.~C.,  2004b, \mn@doi [\apj] {10.1086/424830}, \href
  {http://adsabs.harvard.edu/abs/2004ApJ...616..567I} {616, 567}

\bibitem[\protect\citeauthoryear{{Ida} \& {Lin}}{{Ida} \&
  {Lin}}{2005}]{IdaLin2005}
{Ida} S.,  {Lin} D.~N.~C.,  2005, \mn@doi [\apj] {10.1086/429953}, \href
  {http://adsabs.harvard.edu/abs/2005ApJ...626.1045I} {626, 1045}

\bibitem[\protect\citeauthoryear{{Ida} \& {Lin}}{{Ida} \&
  {Lin}}{2008}]{IdaLin2008}
{Ida} S.,  {Lin} D.~N.~C.,  2008, \mn@doi [\apj] {10.1086/523754}, \href
  {http://adsabs.harvard.edu/abs/2008ApJ...673..487I} {673, 487}

\bibitem[\protect\citeauthoryear{{Ida}, {Lin}  \& {Nagasawa}}{{Ida}
  et~al.}{2013}]{Ida2013}
{Ida} S.,  {Lin} D.~N.~C.,   {Nagasawa} M.,  2013, \mn@doi [\apj]
  {10.1088/0004-637X/775/1/42}, \href
  {http://adsabs.harvard.edu/abs/2013ApJ...775...42I} {775, 42}

\bibitem[\protect\citeauthoryear{{Ikoma}, {Nakazawa}  \& {Emori}}{{Ikoma}
  et~al.}{2000}]{Ikoma2000}
{Ikoma} M.,  {Nakazawa} K.,   {Emori} H.,  2000, \mn@doi [\apj]
  {10.1086/309050}, \href {http://adsabs.harvard.edu/abs/2000ApJ...537.1013I}
  {537, 1013}

\bibitem[\protect\citeauthoryear{{Ivanov}, {Papaloizou}  \&
  {Polnarev}}{{Ivanov} et~al.}{1999}]{Ivanov1999}
{Ivanov} P.~B.,  {Papaloizou} J.~C.~B.,   {Polnarev} A.~G.,  1999, \mn@doi
  [\mnras] {10.1046/j.1365-8711.1999.02623.x}, \href
  {http://adsabs.harvard.edu/abs/1999MNRAS.307...79I} {307, 79}

\bibitem[\protect\citeauthoryear{{Kley}}{{Kley}}{1999}]{Kley1999}
{Kley} W.,  1999, \mn@doi [\mnras] {10.1046/j.1365-8711.1999.02198.x}, \href
  {http://adsabs.harvard.edu/abs/1999MNRAS.303..696K} {303, 696}

\bibitem[\protect\citeauthoryear{{Kokubo} \& {Ida}}{{Kokubo} \&
  {Ida}}{2002}]{KokuboIda2002}
{Kokubo} E.,  {Ida} S.,  2002, \mn@doi [\apj] {10.1086/344105}, \href
  {http://adsabs.harvard.edu/abs/2002ApJ...581..666K} {581, 666}

\bibitem[\protect\citeauthoryear{{Lissauer}, {Hubickyj}, {D'Angelo}  \&
  {Bodenheimer}}{{Lissauer} et~al.}{2009}]{Lissauer2009}
{Lissauer} J.~J.,  {Hubickyj} O.,  {D'Angelo} G.,   {Bodenheimer} P.,  2009,
  \mn@doi [\icarus] {10.1016/j.icarus.2008.10.004}, \href
  {http://adsabs.harvard.edu/abs/2009Icar..199..338L} {199, 338}

\bibitem[\protect\citeauthoryear{{Lubow} \& {D'Angelo}}{{Lubow} \&
  {D'Angelo}}{2006}]{Lubow2006}
{Lubow} S.~H.,  {D'Angelo} G.,  2006, \mn@doi [\apj] {10.1086/500356}, \href
  {http://adsabs.harvard.edu/abs/2006ApJ...641..526L} {641, 526}

\bibitem[\protect\citeauthoryear{{Lubow}, {Seibert}  \& {Artymowicz}}{{Lubow}
  et~al.}{1999}]{Lubow1999}
{Lubow} S.~H.,  {Seibert} M.,   {Artymowicz} P.,  1999, \mn@doi [\apj]
  {10.1086/308045}, \href {http://adsabs.harvard.edu/abs/1999ApJ...526.1001L}
  {526, 1001}

\bibitem[\protect\citeauthoryear{{Lynden-Bell} \& {Pringle}}{{Lynden-Bell} \&
  {Pringle}}{1974}]{LBP1974}
{Lynden-Bell} D.,  {Pringle} J.~E.,  1974, \mn@doi [\mnras]
  {10.1093/mnras/168.3.603}, \href
  {http://adsabs.harvard.edu/abs/1974MNRAS.168..603L} {168, 603}

\bibitem[\protect\citeauthoryear{{Lyra}, {Paardekooper}  \& {Mac Low}}{{Lyra}
  et~al.}{2010}]{Lyra2010}
{Lyra} W.,  {Paardekooper} S.-J.,   {Mac Low} M.-M.,  2010, \mn@doi [\apjl]
  {10.1088/2041-8205/715/2/L68}, \href
  {http://adsabs.harvard.edu/abs/2010ApJ...715L..68L} {715, L68}

\bibitem[\protect\citeauthoryear{{Machida}, {Kokubo}, {Inutsuka}  \&
  {Matsumoto}}{{Machida} et~al.}{2010}]{Machida2010}
{Machida} M.~N.,  {Kokubo} E.,  {Inutsuka} S.-I.,   {Matsumoto} T.,  2010,
  \mn@doi [\mnras] {10.1111/j.1365-2966.2010.16527.x}, \href
  {http://adsabs.harvard.edu/abs/2010MNRAS.405.1227M} {405, 1227}

\bibitem[\protect\citeauthoryear{{Masset}}{{Masset}}{2001}]{Masset2001}
{Masset} F.~S.,  2001, \mn@doi [\apj] {10.1086/322446}, \href
  {http://adsabs.harvard.edu/abs/2001ApJ...558..453M} {558, 453}

\bibitem[\protect\citeauthoryear{{Masset}}{{Masset}}{2002}]{Masset2002}
{Masset} F.~S.,  2002, \mn@doi [\aap] {10.1051/0004-6361:20020240}, \href
  {http://adsabs.harvard.edu/abs/2002A%26A...387..605M} {387, 605}

\bibitem[\protect\citeauthoryear{{Masset}, {Morbidelli}, {Crida}  \&
  {Ferreira}}{{Masset} et~al.}{2006}]{Masset2006}
{Masset} F.~S.,  {Morbidelli} A.,  {Crida} A.,   {Ferreira} J.,  2006, \mn@doi
  [\apj] {10.1086/500967}, \href
  {http://adsabs.harvard.edu/abs/2006ApJ...642..478M} {642, 478}

\bibitem[\protect\citeauthoryear{{Matsumura} \& {Pudritz}}{{Matsumura} \&
  {Pudritz}}{2003}]{MP2003}
{Matsumura} S.,  {Pudritz} R.~E.,  2003, \mn@doi [\apj] {10.1086/378846}, \href
  {http://adsabs.harvard.edu/abs/2003ApJ...598..645M} {598, 645}

\bibitem[\protect\citeauthoryear{{Matsumura} \& {Pudritz}}{{Matsumura} \&
  {Pudritz}}{2006}]{MP2006}
{Matsumura} S.,  {Pudritz} R.~E.,  2006, \mn@doi [\mnras]
  {10.1111/j.1365-2966.2005.09737.x}, \href
  {http://adsabs.harvard.edu/abs/2006MNRAS.365..572M} {365, 572}

\bibitem[\protect\citeauthoryear{{Matsumura}, {Pudritz}  \&
  {Thommes}}{{Matsumura} et~al.}{2007}]{MP2007}
{Matsumura} S.,  {Pudritz} R.~E.,   {Thommes} E.~W.,  2007, \mn@doi [\apj]
  {10.1086/513175}, \href {http://adsabs.harvard.edu/abs/2007ApJ...660.1609M}
  {660, 1609}

\bibitem[\protect\citeauthoryear{{Matt} \& {Pudritz}}{{Matt} \&
  {Pudritz}}{2005}]{Matt2005}
{Matt} S.,  {Pudritz} R.~E.,  2005, \mn@doi [\apjl] {10.1086/498066}, \href
  {http://adsabs.harvard.edu/abs/2005ApJ...632L.135M} {632, L135}

\bibitem[\protect\citeauthoryear{{Mayor} et~al.,}{{Mayor}
  et~al.}{2011}]{Mayor2011}
{Mayor} M.,  et~al., 2011, preprint, \href
  {http://adsabs.harvard.edu/abs/2011arXiv1109.2497M} {} (\mn@eprint {arXiv}
  {1109.2497})

\bibitem[\protect\citeauthoryear{{Morbidelli}, {Szul{\'a}gyi}, {Crida}, {Lega},
  {Bitsch}, {Tanigawa}  \& {Kanagawa}}{{Morbidelli}
  et~al.}{2014}]{Morbidelli2014}
{Morbidelli} A.,  {Szul{\'a}gyi} J.,  {Crida} A.,  {Lega} E.,  {Bitsch} B.,
  {Tanigawa} T.,   {Kanagawa} K.,  2014, \mn@doi [\icarus]
  {10.1016/j.icarus.2014.01.010}, \href
  {http://adsabs.harvard.edu/abs/2014Icar..232..266M} {232, 266}

\bibitem[\protect\citeauthoryear{{Mordasini}}{{Mordasini}}{2014}]{Mordasini2014b}
{Mordasini} C.,  2014, \mn@doi [\aap] {10.1051/0004-6361/201423702}, \href
  {http://adsabs.harvard.edu/abs/2014A%26A...572A.118M} {572, A118}

\bibitem[\protect\citeauthoryear{{Mordasini}, {Alibert}  \& {Benz}}{{Mordasini}
  et~al.}{2009a}]{Mordasini2009}
{Mordasini} C.,  {Alibert} Y.,   {Benz} W.,  2009a, \mn@doi [\aap]
  {10.1051/0004-6361/200810301}, \href
  {http://adsabs.harvard.edu/abs/2009A%26A...501.1139M} {501, 1139}

\bibitem[\protect\citeauthoryear{{Mordasini}, {Alibert}, {Benz}  \&
  {Naef}}{{Mordasini} et~al.}{2009b}]{Mordasini2009b}
{Mordasini} C.,  {Alibert} Y.,  {Benz} W.,   {Naef} D.,  2009b, \mn@doi [\aap]
  {10.1051/0004-6361/200810697}, \href
  {http://adsabs.harvard.edu/abs/2009A%26A...501.1161M} {501, 1161}

\bibitem[\protect\citeauthoryear{{Mordasini}, {Alibert}, {Georgy}, {Dittkrist},
  {Klahr}  \& {Henning}}{{Mordasini} et~al.}{2012}]{Mordasini2012c}
{Mordasini} C.,  {Alibert} Y.,  {Georgy} C.,  {Dittkrist} K.-M.,  {Klahr} H.,
  {Henning} T.,  2012, \mn@doi [\aap] {10.1051/0004-6361/201118464}, \href
  {http://adsabs.harvard.edu/abs/2012A%26A...547A.112M} {547, A112}

\bibitem[\protect\citeauthoryear{{Mordasini}, {Klahr}, {Alibert}, {Miller}  \&
  {Henning}}{{Mordasini} et~al.}{2014}]{Mordasini2014}
{Mordasini} C.,  {Klahr} H.,  {Alibert} Y.,  {Miller} N.,   {Henning} T.,
  2014, \mn@doi [\aap] {10.1051/0004-6361/201321479}, \href
  {http://adsabs.harvard.edu/abs/2014A%26A...566A.141M} {566, A141}

\bibitem[\protect\citeauthoryear{{Morton}, {Bryson}, {Coughlin}, {Rowe},
  {Ravichandran}, {Petigura}, {Haas}  \& {Batalha}}{{Morton}
  et~al.}{2016}]{Morton2016}
{Morton} T.~D.,  {Bryson} S.~T.,  {Coughlin} J.~L.,  {Rowe} J.~F.,
  {Ravichandran} G.,  {Petigura} E.~A.,  {Haas} M.~R.,   {Batalha} N.~M.,
  2016, \mn@doi [\apj] {10.3847/0004-637X/822/2/86}, \href
  {http://adsabs.harvard.edu/abs/2016ApJ...822...86M} {822, 86}

\bibitem[\protect\citeauthoryear{{Nakatani}, {Hosokawa}, {Yoshida}, {Nomura}
  \& {Kuiper}}{{Nakatani} et~al.}{2017}]{Nakatani2017}
{Nakatani} R.,  {Hosokawa} T.,  {Yoshida} N.,  {Nomura} H.,   {Kuiper} R.,
  2017, preprint, \href {http://adsabs.harvard.edu/abs/2017arXiv170604570N} {}
  (\mn@eprint {arXiv} {1706.04570})

\bibitem[\protect\citeauthoryear{{Ndugu}, {Bitsch}  \& {Jurua}}{{Ndugu}
  et~al.}{2018}]{Ndugu2018}
{Ndugu} N.,  {Bitsch} B.,   {Jurua} E.,  2018, \mn@doi [\mnras]
  {10.1093/mnras/stx2815}, \href
  {http://adsabs.harvard.edu/abs/2018MNRAS.474..886N} {474, 886}

\bibitem[\protect\citeauthoryear{{Oppenheimer} \& {Dalgarno}}{{Oppenheimer} \&
  {Dalgarno}}{1974}]{Oppenheimer1974}
{Oppenheimer} M.,  {Dalgarno} A.,  1974, \mn@doi [\apj] {10.1086/153030}, \href
  {http://adsabs.harvard.edu/abs/1974ApJ...192...29O} {192, 29}

\bibitem[\protect\citeauthoryear{{Ormel}}{{Ormel}}{2014}]{Ormel2014}
{Ormel} C.~W.,  2014, \mn@doi [\apjl] {10.1088/2041-8205/789/1/L18}, \href
  {http://adsabs.harvard.edu/abs/2014ApJ...789L..18O} {789, L18}

\bibitem[\protect\citeauthoryear{{Owen}, {Ercolano}, {Clarke}  \&
  {Alexander}}{{Owen} et~al.}{2010}]{Owen2010}
{Owen} J.~E.,  {Ercolano} B.,  {Clarke} C.~J.,   {Alexander} R.~D.,  2010,
  \mn@doi [\mnras] {10.1111/j.1365-2966.2009.15771.x}, \href
  {http://adsabs.harvard.edu/abs/2010MNRAS.401.1415O} {401, 1415}

\bibitem[\protect\citeauthoryear{{Owen}, {Ercolano}  \& {Clarke}}{{Owen}
  et~al.}{2011}]{Owen2011}
{Owen} J.~E.,  {Ercolano} B.,   {Clarke} C.~J.,  2011, \mn@doi [\mnras]
  {10.1111/j.1365-2966.2010.17818.x}, \href
  {http://adsabs.harvard.edu/abs/2011MNRAS.412...13O} {412, 13}

\bibitem[\protect\citeauthoryear{{Paardekooper}, {Baruteau}, {Crida}  \&
  {Kley}}{{Paardekooper} et~al.}{2010}]{Paardekooper2010}
{Paardekooper} S.-J.,  {Baruteau} C.,  {Crida} A.,   {Kley} W.,  2010, \mn@doi
  [\mnras] {10.1111/j.1365-2966.2009.15782.x}, \href
  {http://adsabs.harvard.edu/abs/2010MNRAS.401.1950P} {401, 1950}

\bibitem[\protect\citeauthoryear{{Paardekooper}, {Baruteau}  \&
  {Kley}}{{Paardekooper} et~al.}{2011}]{Paardekooper2011}
{Paardekooper} S.-J.,  {Baruteau} C.,   {Kley} W.,  2011, \mn@doi [\mnras]
  {10.1111/j.1365-2966.2010.17442.x}, \href
  {http://adsabs.harvard.edu/abs/2011MNRAS.410..293P} {410, 293}

\bibitem[\protect\citeauthoryear{{Pascucci} \& {Sterzik}}{{Pascucci} \&
  {Sterzik}}{2009}]{Pascucci2009}
{Pascucci} I.,  {Sterzik} M.,  2009, \mn@doi [\apj]
  {10.1088/0004-637X/702/1/724}, \href
  {http://adsabs.harvard.edu/abs/2009ApJ...702..724P} {702, 724}

\bibitem[\protect\citeauthoryear{{Pollack}, {Hubickyj}, {Bodenheimer},
  {Lissauer}, {Podolak}  \& {Greenzweig}}{{Pollack} et~al.}{1996}]{Pollack1996}
{Pollack} J.~B.,  {Hubickyj} O.,  {Bodenheimer} P.,  {Lissauer} J.~J.,
  {Podolak} M.,   {Greenzweig} Y.,  1996, \mn@doi [\icarus]
  {10.1006/icar.1996.0190}, \href
  {http://adsabs.harvard.edu/abs/1996Icar..124...62P} {124, 62}

\bibitem[\protect\citeauthoryear{{Qi} et~al.,}{{Qi} et~al.}{2013}]{Qi2013}
{Qi} C.,  et~al., 2013, \mn@doi [Science] {10.1126/science.1239560}, \href
  {http://adsabs.harvard.edu/abs/2013Sci...341..630Q} {341, 630}

\bibitem[\protect\citeauthoryear{{R{\'e}my-Ruyer} et~al.,}{{R{\'e}my-Ruyer}
  et~al.}{2014}]{Remy2014}
{R{\'e}my-Ruyer} A.,  et~al., 2014, \mn@doi [\aap]
  {10.1051/0004-6361/201322803}, \href
  {http://adsabs.harvard.edu/abs/2014A%26A...563A..31R} {563, A31}

\bibitem[\protect\citeauthoryear{{Rowe} et~al.,}{{Rowe}
  et~al.}{2014}]{Rowe2014}
{Rowe} J.~F.,  et~al., 2014, \mn@doi [\apj] {10.1088/0004-637X/784/1/45}, \href
  {http://adsabs.harvard.edu/abs/2014ApJ...784...45R} {784, 45}

\bibitem[\protect\citeauthoryear{{Sano}, {Miyama}, {Umebayashi}  \&
  {Nakano}}{{Sano} et~al.}{2000}]{Sano2000}
{Sano} T.,  {Miyama} S.~M.,  {Umebayashi} T.,   {Nakano} T.,  2000, \mn@doi
  [\apj] {10.1086/317075}, \href
  {http://adsabs.harvard.edu/abs/2000ApJ...543..486S} {543, 486}

\bibitem[\protect\citeauthoryear{{Shakura} \& {Sunyaev}}{{Shakura} \&
  {Sunyaev}}{1973}]{SS1973}
{Shakura} N.~I.,  {Sunyaev} R.~A.,  1973, \aap, \href
  {http://adsabs.harvard.edu/abs/1973A%26A....24..337S} {24, 337}

\bibitem[\protect\citeauthoryear{{Siess}, {Dufour}  \& {Forestini}}{{Siess}
  et~al.}{2000}]{Siess2000}
{Siess} L.,  {Dufour} E.,   {Forestini} M.,  2000, \aap, \href
  {http://adsabs.harvard.edu/abs/2000A%26A...358..593S} {358, 593}

\bibitem[\protect\citeauthoryear{{Simon}, {Bai}, {Stone}, {Armitage}  \&
  {Beckwith}}{{Simon} et~al.}{2013}]{Simon2013}
{Simon} J.~B.,  {Bai} X.-N.,  {Stone} J.~M.,  {Armitage} P.~J.,   {Beckwith}
  K.,  2013, \mn@doi [\apj] {10.1088/0004-637X/764/1/66}, \href
  {http://adsabs.harvard.edu/abs/2013ApJ...764...66S} {764, 66}

\bibitem[\protect\citeauthoryear{{Stepinski}}{{Stepinski}}{1998}]{Stepinski1998}
{Stepinski} T.~F.,  1998, \mn@doi [\icarus] {10.1006/icar.1998.5893}, \href
  {http://adsabs.harvard.edu/abs/1998Icar..132..100S} {132, 100}

\bibitem[\protect\citeauthoryear{{Umebayashi} \& {Nakano}}{{Umebayashi} \&
  {Nakano}}{1981}]{Umebayashi1981}
{Umebayashi} T.,  {Nakano} T.,  1981, \pasj, \href
  {http://adsabs.harvard.edu/abs/1981PASJ...33..617U} {33, 617}

\bibitem[\protect\citeauthoryear{{Valenti} \& {Fischer}}{{Valenti} \&
  {Fischer}}{2008}]{Valenti2008}
{Valenti} J.~A.,  {Fischer} D.~A.,  2008, \mn@doi [Physica Scripta Volume T]
  {10.1088/0031-8949/2008/T130/014003}, \href
  {http://adsabs.harvard.edu/abs/2008PhST..130a4003V} {130, 014003}

\bibitem[\protect\citeauthoryear{{Venturini}, {Alibert}  \& {Benz}}{{Venturini}
  et~al.}{2016}]{Venturini2016}
{Venturini} J.,  {Alibert} Y.,   {Benz} W.,  2016, \mn@doi [\aap]
  {10.1051/0004-6361/201628828}, \href
  {http://adsabs.harvard.edu/abs/2016A%26A...596A..90V} {596, A90}

\bibitem[\protect\citeauthoryear{{Wang} \& {Fischer}}{{Wang} \&
  {Fischer}}{2015}]{Wang2015}
{Wang} J.,  {Fischer} D.~A.,  2015, \mn@doi [\aj] {10.1088/0004-6256/149/1/14},
  \href {http://adsabs.harvard.edu/abs/2015AJ....149...14W} {149, 14}

\bibitem[\protect\citeauthoryear{{Zeng} \& {Sasselov}}{{Zeng} \&
  {Sasselov}}{2013}]{Zeng2013}
{Zeng} L.,  {Sasselov} D.,  2013, \mn@doi [\pasp] {10.1086/669163}, \href
  {http://adsabs.harvard.edu/abs/2013PASP..125..227Z} {125, 227}

\makeatother
\end{thebibliography}

%%%%%%%%%%%%%%%%% APPENDICES %%%%%%%%%%%%%%%%%%%%%

\appendix

\section{Observationally Filtering Computed Populations} \label{Appendix}

In this section, we describe our method of correction our computed populations for observational biases by estimating their observation probabilities. We obtain separate estimations for transit and radial velocity detection probabilities and assume that planets with high transit detection probabilities \emph{or} high radial velocity detection probabilities are observable and are therefore included in the populations. Planets that have low detection probabilities corresponding to both methods are filtered out of the populations.

To estimate detection probabilities for each computed planet, we use signal to noise ratios (SNR). We assume that cases where S/N $\geq$ 10 correspond to successful observations of a planet. This is a conservative estimate as a SNR of 10 corresponds to a detection at the 10$\sigma$ confidence level. For each population of 3000 planets this results in a false positive probability of 10$^{-18}$. This is the same SNR threshold used in \citet{Howard2012} to confirm detections of hot Jupiters in \emph{Kepler}'s first quarter.

In the case of a transit detection, the SNR ratio is \citep{Howard2012},
\begin{equation} \textrm{S/N}_{\rm{transit}} = \frac{R_p^2/R_*^2}{\sigma_{\textrm{CDPP}}}\sqrt{\frac{n_{\rm{tr}}t_{rm{dur}}}{3\,\textrm{hr}}}\;,\label{SNR}\end{equation}
where $R_p$ is the planet's radius, $R_* = R_\odot$ is the radius of the star, here $n_{\rm{tr}}$ is the number of detected transits during the observation time, $t_{\rm{dur}}$ is the transit duration, and $\sigma_{\textrm{CDPP}}$ is the differential photometric precision measured over a duration of 3 hours. The ``noise'' factor in equation \ref{SNR} is $\sigma_{\textrm{cdpp}}$ is related to errors associated with telescope stability, transit light curve, and stellar activity. As reported in \citet{Howard2012}, typical 3 hour $\sigma_{\textrm{cdpp}}$ values range from 30-300 ppm. 

%Planet radii
Our model does not directly calculate planet radii as we do not consider interior structure models. To estimate the radii of low mass planets $M_p < 30$ M$_\oplus$, where we compute solid abundances using our disk equilibrium chemistry model, we interpolate over results presented in \citet{Zeng2013} who performed an interior model of solid planets in order to calculate their radii as a function of planet mass and composition. For higher mass planets, we use the empirical relation from \citet{Bashi2017} to estimate planet radii. 

%T_dur and n_tr
Since this model does not consider dynamical effects during or after planet formation, we do not have an estimate of orbital eccentricities or inclinations of planets in our computed populations. We therefore make the simplifying assumption that both of these quantities are zero when calculating the transit durations $t_{\rm{dur}}$. To estimate the number of transits $n_{\rm{tr}}$, the total ``observation time'' needs to be defined, which we assume to be 3 years to roughly correspond to the \emph{Kepler} mission (prior to the \emph{K2} mission).

The radial velocity amplitude corresponding to a planet of mass $M_p$ orbiting a star $M_s$ with period $P$ is \citep{Cumming2004},
\begin{equation} K = \frac{28.4 \,\textrm{m/s}}{\sqrt{1-e^2}}\frac{M_p \sin{i}}{M_{\rm{Jup}}}\left(\frac{M_s+M_p}{M_\odot}\right)^{-2/3}\left(\frac{P}{1\,\textrm{year}}\right)^{-1/3}\;.\label{RV_amplitude}\end{equation}
In this work, $M_s = M_\odot$, and we assume $e=0$ for all planets for the reasons discussed previously. The corresponding SNR for a radial velocity detection is,
\begin{equation} \textrm{S/N}_{\rm{RV}} = \frac{K}{\sigma_{\rm{RV}}}\sqrt{n_{\rm{orb}}}\;,\label{SNR_RV}\end{equation}
where $n_{\rm{orb}}$ are the number of orbits during the observation time and $\sigma_{\rm{RV}}$ is the noise associated with a radial velocity detection. Sources of error related to the RV method are caused by a combination of measurement error and stellar ``jitter'' (pulsation and/or surface convection that adds noise to an RV signal). These two sources of error are reported in \citet{Cumming2004} to be 3-5 m/s each. Taking into account addition in quadrature, we estimate $\sigma_{RV}$ to range from 4-10 m/s. We consider a longer observation time of 6 years for transit detections when estimating $n_{\rm{orb}}$.

We assume uniform distributions for both $\sigma_{\textrm{cdpp}}$ and $\sigma_{RV}$ over the ranges previously mentioned and calculate the corresponding SNR distribution for each planet formed in the population. To estimate a detection probability associated with each method, we calculate the fraction of the SNR distributions that are $\geq$ 10, which we chose as our threshold. We then performed a simple Monte Carlo routine using each planet's detection probabilities to determine if each planet was observed or not. We note that the majority of planets in each population have detection probabilities calculated with this method of or 1 or 0, and in these cases this is a binary problem. However, there is always a portion of planets who have a non-binary detection probability, which correspond to planets that would be observed around stars with low noise values, or unlikely to be observed around stars with high noise values. It is these cases that necessitate the Monte Carlo calculation.

%%%%%%%%%%%%%%%%%%%%%%%%%%%%%%%%%%%%%%%%%%%%%%%%%%

% Don't change these lines
\bsp	% typesetting comment
\label{lastpage}
\end{document}